\documentclass[pdflatex,sn-mathphys-num]{sn-jnl}% Math and Physical Sciences Numbered Reference Style 
\usepackage{graphicx}%
\usepackage{multirow}%
\usepackage{amsmath,amssymb,amsfonts}%
\usepackage{amsthm}%
\usepackage{mathrsfs}%
\usepackage[title]{appendix}%
\usepackage{xcolor}%
\usepackage{textcomp}%
\usepackage{manyfoot}%
\usepackage{booktabs}%
\usepackage{algorithm}%
\usepackage{algorithmicx}%
\usepackage{algpseudocode}%
\usepackage{listings}%
\theoremstyle{thmstyleone}%
\theoremstyle{thmstyletwo}%

\theoremstyle{thmstylethree}%

\begin{document}

\title[Article Title]{The Impact of Solar -- Terrestrial Plasma and Magnetic Field on the Detection of Space-borne Gravitational Wave Detections}


\author*[1,2]{\sur{Wei} \fnm{Su}  }\email{suwei25@mail.sysu.edu.cn}



\affil*[1]{\orgdiv{School of Physics and Astronomy}, \orgname{MOE Key Laboratory of TianQin Mission, TianQin Research Center for Gravitational Physics, Frontiers Science Center for TianQin, Gravitational Wave Research Center of CNSA, Sun Yat-sen University (Zhuhai Campus)}, \orgaddress{\street{} \city{Zhuhai}, \postcode{519082}, \state{Guangdong}, \country{China}} }

% \affil[2]{\orgdiv{Department}, \orgname{Organization}, \orgaddress{\street{Street}, \city{City}, \postcode{10587}, \state{State}, \country{Country}}}

% \affil[3]{\orgdiv{Department}, \orgname{Organization}, \orgaddress{\street{Street}, \city{City}, \postcode{610101}, \state{State}, \country{Country}}}


\abstract{
% The abstract serves both as a general introduction to the topic and as a brief, non-technical summary of the main results and their implications. Authors are advised to check the author instructions for the journal they are submitting to for word limits and if structural elements like subheadings, citations, or equations are permitted.
Space-borne gravitational wave detections raise new questions for heliophysics: how the Sun-Terrestrial space environment affect gravitational wave detection, and to what extent? Space-borne gravitational wave detectors use laser interferometry to measure displacement variations between two free test masses caused by gravitational waves.  Space-borne gravitational wave detectors require extremely high measurement accuracy, making it necessary to take into account the effects of space plasma and magnetic field. On one hand, laser propagation through space plasma can induce optical path difference noise, affecting distance measurement accuracy. On the other hand, interactions between space magnetic field and the test masses can generate acceleration noise. This review introduces studies on laser propagation noise and space magnetic acceleration noise in space gravitational wave detection. And this review presents a method, time-delay interferometry, to suppress laser propagation noise.
}




\maketitle

\section{Introduction}\label{sec1}

In early 2016, the discovery of gravitational waves (GWs) was officially announced by the LIGO collaboration \citep{LIGO2016}, and since then, modern astronomy has entered the multi-messenger era. According to the GW signature we can get some parameters that are difficult to obtain from electromagnetic (EM) wave observation, such as the mass and spin of black holes and neutron stars. Until the beginning of 2025, more than 100 GW events and candidate events have been detected \citep{GW170817,GW170817CHN,LIGO-Virgo2019ApJ,LIGO-Virgo2019PhRvX,Bailes2021NatRP}, all of which were discovered by ground-borne GW detectors in the frequency band of about 10--1000 Hz. In order to expand the frequency band of GW detection, it is proposed to expand the laser interferometer for GW detection from the ground to the space, so that the interferometric arm can be $\sim 10^{5}$ to $\sim 10^{6}$ kilometers, and the corresponding sensitivity frequency expands to the low-frequency band from $10^{-4}$ Hz to 1 Hz. At present, the proposed space GW detection plans are Laser Interferometer Space Antenna (LISA) led by European Space Agency (ESA) \citep{LISA2017}, DECIGO of Japan \citep{DECIGO2011}, Taiji (TJ) \citep{TJ2017} and TianQin (TQ) \citep{Luo2016} of China, etc.

Deploying GW detectors in space can avoid the effects of ground vibrations, the Earth gravity gradient force, atmosphere and human activities. However, space is not a vacuum, for space-borne GW detection, it is necessary to fully consider the effects of the space environment, such as space magnetic field, plasma, high-energy particles, which will directly affect the success or failure of space-borne GW detection. Now the space-borne GW detection route is to use the laser interferometry to measure the displacement change of the free test mass (TM) caused by GW, and its core measurement indices are: 1. displacement ranging accuracy, the laser interferometry ranging accuracy of the space-borne GW detection is required to be of the order of $10^{-12}$ $\rm m~Hz^{-1/2}$ \citep{TJ2021,Mei2020ptep,LISA2024}; 2. acceleration noise, which is a parameter to evaluate whether the TM is ‘free’ or not, and the accuracy of acceleration measurements for space-borne GW detection is required to be of the order of $10^{-15}$ $\rm m~s^2~Hz^{-1/2}$ \citep{TJ2021,Mei2020ptep,LISA2024}. This review focuses on these two core indices to analyse the laser ranging noise and acceleration noise caused by the space plasma and magnetic field.

From the perspective of GW detection, firstly, space plasma can lead to laser propagation effect \citep{Smetana2020,Lu2020,Su2021,Jennrich2021}. There are dispersion and Faraday rotation effects when the laser propagates in space plasma, which will lead to time delay, optical path difference (OPD), wavefront distortion polarization angle change, and so on, affecting the laser ranging accuracy. Secondly, the space magnetic field will lead to acceleration noise \citep{Schumaker2003,Stebbins2004,Su2020}. Since the TM in the inertial sensor of GW detector has weak remanent magnet moment and residual charge, the interaction of remanent magnet moment and residual charge with the space magnetic field can generate magnetic moment force and Lorentz force, resulting in acceleration noise \citep{Hanson2003,Su2020}.

\begin{figure}[h]
\centering
\includegraphics[width=0.75\textwidth, trim = 120 50 120 50]{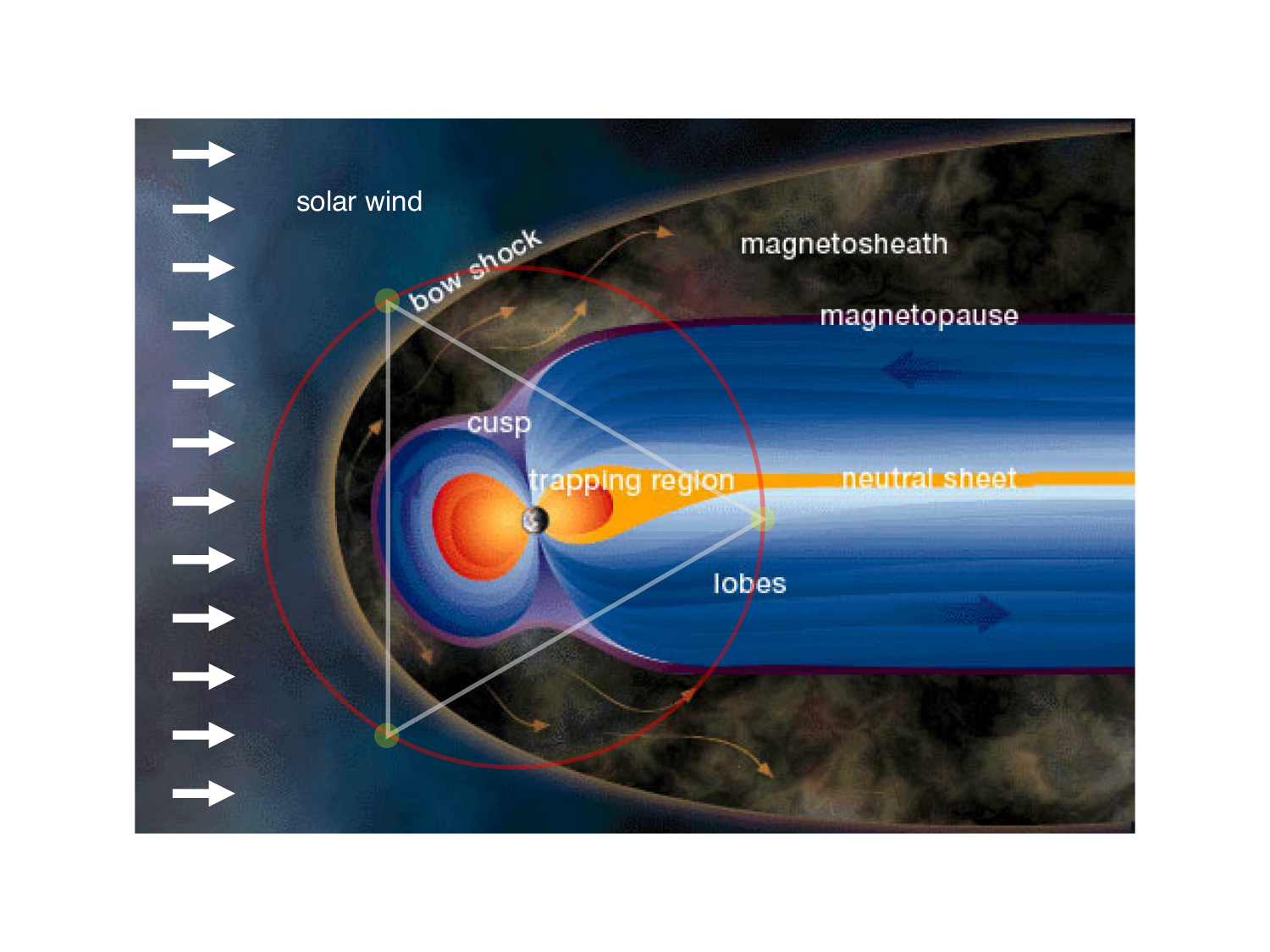}
\caption{The schematic of GW detectors in space environment \citep{Su2020}.}
\label{fig-overview}
\end{figure}

From the perspective of the space environment, solar-terrestrial physical processes on various time and space scales affect the detection of GW in space, the schematic is show in Figure \ref{fig-overview}. On the time scale, the sensitive frequency band for space-borne GW detection is from 0.1 mHz to 1 Hz, which corresponds to a time resolution of up to 1 s. On the other hand, The Sun is the source of the space plasma and magnetic field, and the solar activity has a periodicity of about 11 years \citep{Charbonneau2010LRSP}. Therefore, the time scale to be considered for the space-borne GW detection is from the order of 1 s to $10^8$ s (11 years $\approx 3 \times 10^8$ s), covering 8 orders of magnitude. On the space scale, the arm length of the GW detectors reach the order of $\sim 10^5$ to $\sim 10^6$ km, and the velocity of GW detectors will travel about 2 km (TQ) or about 30 km (LISA and TJ) in 1 s. Thus, the smallest space scale we need to consider reaches the order of $\sim 1$ km. From $\sim 1$ to $\sim 10^6$ km spans 6 orders of magnitude in space scale.
Due to the 8 orders interval in time scale and 6 orders interval in space scale, we need to consider multiple scales of heliophysics processes, e.g., plasma dynamics, magnetohydrodynamics (MHD), and global scales \citep{Zhang2007ApJ,Lin2015SSRv,LiG2017,Guo2024}. For the scale of plasma dynamics, e.g., magnetic reconnection\citep{Zhou2018,Zhou2019,LiWY2021}, plasma waves \citep{Zhao2017,Huang2020a,Huang2022}, and turbulence \citep{He2013,Wu2020} need to be considered. For the MHD scale, structures such as shocks \citep{Su2022,Trotta2025}, magnetic clouds \citep{Zhang1988,Bothmer1998}, and MHD instabilities \citep{Feng2013} need to be considered. For the global scale, the evolution of the global structure of the Earth's magnetosphere and the heliosphere \citep{Parker1958,Parker1958PoF}, and the corotating interaction regions (CIRs) \citep{Gosling1999,Allen2021,Wijsen2023} should to be considered. In addition, dramatic changes of space plasma and magnetic field caused by eruption phenomena of the Sun and Earth (e.g., flares \citep{Lin2000,Song2016,Yang2018}, coronal mass ejections (CME) \citep{ChenPF2011,Cheng2017}, magnetic storms \citep{Tsurutani1988,Gonzalez1994,Nagai1998}, etc.) should to be considered. All these heliophysics processes will have impacts on GW detection and need to be evaluated.

% In this review, the theory of space environment effects in GW detection, and the heliophysics methods, models, and data that used in the research are presented in Section $\ref{sec2}$. 
This review presents the theory of space environment effects on space-borne GW detection, along with the heliophysics methods, models, and data used in the research, which are detailed in Section $\ref{sec2}$.
Section $\ref{sec3}$ is the results of laser propagation noise (LPN) in space plasma on GW detections, and the methods and results of suppressing LPN. Section $\ref{sec4}$ gives the results of the effect of acceleration noise due to space magnetic fields on GW detection. Finally, there are discussion and conclusion.

\section{Theories, models, and data of space plasma and magnetic field effects on GW detection}\label{sec2}

\subsection{EM waves propagation in the space plasma}
\label{sec2-1}

% 电磁波传播在等离子体中的传播可以用**Appleton-Hartree（A-H）公式**来描述。利用麦克斯韦方程组和等离子体中的动力学方程来得到电磁波的在磁化等离子中的传播特性，最终得出折射率的表达式，即A-H公式。

% 在此，我们取磁场$\boldsymbol{B}$方向为$z$方向，波矢$\boldsymbol{k}$与磁场夹角方向为$\theta$，垂直$\boldsymbol{B}$和$\boldsymbol{k}$的方向为$y$ 方向。

The propagation of EM waves in plasma can be described by the Appleton--Hartree (A--H) Equation. By employing dynamic equations of plasma and Maxwell's Equations, the propagation characteristics of EM waves in magnetized plasma are derived, culminating in the expression for the refractive index, the A--H Equation \cite{Bittencourt2004, Hutchinson2002}.

Here, the coordinate is taken as follows: The magnetic field $\boldsymbol{B}_0$ is aligned with the $z$-axis, the wavevector $\boldsymbol{k}$ forms an angle $\theta$ with the magnetic field, and the $x$-axis is perpendicular to both $\boldsymbol{B}_0$ and $\boldsymbol{k}$.

% ### 运动方程

% #### 电子运动方程

% 从运动方程出发。电子在电场和磁场中收到电场力和洛伦兹力，其运动方程为
The motion equation of an electron in the collision plasma is as follow,
\begin{equation}
m \frac{\mathrm{d} \boldsymbol{u} }{\mathrm{d} t} = -e(\boldsymbol{E} + \boldsymbol{u} \times \boldsymbol{B}_0 ) - m\nu \boldsymbol{u}
\label{eq:motion}
\end{equation}
% 其中，$m_e$ 是电子质量，$e$ 是电子电荷，$\boldsymbol{u}$ 是电子速度，$\boldsymbol{B}_0$ 是背景磁场。假设背景磁场是静态且均匀的。
where, $m$ is electron mass, $e$ is electron charge, $\boldsymbol{u}$ is the velocity of the electron, $\boldsymbol{E}$ is the electric field, $\boldsymbol{B}_0$ is the background magnetic field, $\nu$ is the collision frequency.
% 假设背景磁场是静态且均匀的。

% 对于单色波，其电场可表示为，
For a harmonic monochromatic EM wave with angular frequency $\omega$, the electric field $\boldsymbol{E}$ can be written as,
\begin{equation}
    \boldsymbol{E}(\boldsymbol{r}, t) = \boldsymbol{E}_0 \mathrm{e}^{\mathrm{i} (\boldsymbol{k} \cdot \boldsymbol{r} - \omega t)}
\label{eq:E}
\end{equation}
% 以上式为例，对于简谐波有
Thus, for harmonic wave, there are,
\begin{equation}
    \frac{\partial}{\partial t} = -\mathrm{i} \omega , \quad \frac{\partial^2}{\partial t^2} = - \omega^2, \quad \nabla = \mathrm{i} \boldsymbol{k}
\label{eq:partial-t-s}
\end{equation}
% 假设其速度扰动是简谐的，对此式做傅立叶变换(将 Eq. $\eqref{eq:3}$ 带入运动方程 Eq. $\eqref{eq:1}$ )，有，
Substitute Equation ($\ref{eq:partial-t-s}$) into Equation ($\ref{eq:motion}$), the motion equation becomes,
\begin{equation}
    \left(1 + \mathrm{i}\frac{\nu}{\omega}\right)\boldsymbol{u} + \frac{\mathrm{i} e }{\omega m } (\boldsymbol{u} \times \boldsymbol{B}_0) = \frac{\mathrm{i}}{\omega m } e \boldsymbol{E}
\label{eq:motion1}
\end{equation}

% #### 求解运动方程

% 电场和速度扰动为矢量，在此写为分量形式，
% $$
% \begin{aligned}
% \boldsymbol{E} &= (E_x, E_y, E_z) \\
% \boldsymbol{v} &= (v_x, v_y, v_z)
% \end{aligned}
% $$
% % 在最开始选取的坐标系中有，
Since the magnetic field is along $z$-axis, in the Lorentz force term ($\boldsymbol{u} \times \boldsymbol{B_0}$) of motion Equation (\ref{eq:motion}),
\begin{equation*}
    \boldsymbol{u} \times \boldsymbol{B_0} = B_0 u_y \boldsymbol{e}_x - B_0 u_x \boldsymbol{e}_y
\end{equation*}
where, $\boldsymbol{e}_x$ and $\boldsymbol{e}_y$ are the unit vector in $x$- and $y$-axes, respectively.
% 将运动方程$\eqref{eq:5}$写为分量的形式，
% 将电子回旋频率标记为 $\omega_{\mathrm{B} } = \frac{e B_0}{m_e}$ ，上式整理为，
Labelling the electron cyclotron frequency $\omega_{\mathrm{B} } = e B_0/m$, and the plasma oscillation frequency $\omega_\mathrm{p}^2 = \frac{e^2 n_\mathrm{e} }{m \varepsilon}$, where $n_\mathrm{e}$ is the electron number density, and $\varepsilon$ is the electric permittivity.
% 令，
And introducing the notation,
\begin{equation}
\begin{aligned}
X &= \frac{\omega_{\mathrm{p}}^2 } {\omega^2}, \quad
Y = \frac{\omega_{\mathrm{B} } } {\omega}, \quad Z = \frac{\nu}{\omega} \\
U &= 1 + \mathrm{i}Z
\end{aligned}
\label{eq:UXY}
\end{equation}
% 得，
With the denotation of Equation ($\ref{eq:UXY}$), the motion Equation (\ref{eq:motion}) can be written in following form,
\begin{equation}
\left(\begin{array}{ccc}
U & \mathrm{i} Y & 0 \\
-\mathrm{i} Y & U & 0 \\
0 & 0 & U
\end{array}\right)\left(\begin{array}{l}
u_{x} \\
u_{y} \\
u_{z}
\end{array}\right)=-\frac{\mathrm{i} e}{m \omega}\left(\begin{array}{c}
E_{x} \\
E_{y} \\
E_{z}
\end{array}\right)
\label{eq:motion-matrix}
\end{equation}
% 对上式中的矩阵求逆，速度$\boldsymbol{u}$可以由电场$\boldsymbol{E}$表示为如下，
Inverting the matrix in the above Equation ($\ref{eq:motion-matrix}$), then the velocity $\boldsymbol{u}$ can be expressed by electric field $\boldsymbol{E}$ in the following form,
\begin{equation}
\begin{pmatrix} 
u_x  \\
u_y \\
u_z
\end{pmatrix} = 
-\frac{\mathrm{i} e}{m \omega U\left(U^{2}-Y^{2}\right)}
\begin{pmatrix} 
U^{2} & -\mathrm{i} U Y & 0 \\
\mathrm{i} U Y & U^{2} & 0 \\
0 & 0 & \left(U^{2}-Y^{2}\right)
\end{pmatrix}
\begin{pmatrix} 
E_x  \\
E_y \\
E_z
\end{pmatrix}
\label{eq:motion-matrix-I}
\end{equation}

% ### Maxwell方程组到色散方程

% #### 从 Maxwell 方程组到电磁波方程

% 有 Maxwell 方程组，
The Maxwell Equations in general form is as follow,
\begin{equation}
\begin{aligned}
\nabla \times \boldsymbol{E} &= - \frac{\partial \boldsymbol{B} }{\partial t} \\
\nabla \times \boldsymbol{B} &= \mu \boldsymbol{j} + \mu \varepsilon \frac{\partial \boldsymbol{E} }{\partial t}
\label{eq:Maxwell}
\end{aligned}
\end{equation}
% 我们可以得到，
here, $\mu$ is magnetic permeability.
According to Maxwell Equations, we can get the electromagnetic wave equation as follow,
\begin{equation}
\nabla \times (\nabla \times \boldsymbol{E}) + \omega^2 \mu \varepsilon \boldsymbol{E} + \mathrm{i} \omega \mu \boldsymbol{j} = 0 
\label{eq:EM-wave}
\end{equation}
% 考虑最简单的简谐波形式，其电场 $\boldsymbol{E}(\boldsymbol{r}, t) = \boldsymbol{E}_0 \mathrm{e}^{\mathrm{i} (\boldsymbol{k} \cdot \boldsymbol{r} - \omega t)}$ 。$\nabla$ 算符为对空间的导数：$\nabla = \boldsymbol{i} \frac{\partial}{\partial x} + \boldsymbol{j} \frac{\partial}{\partial y} + \boldsymbol{k} \frac{\partial}{\partial z} = \partial/\partial \boldsymbol{r}$ ，有，

% 又有欧姆定律，
There is Ohm's law,
\begin{equation}
    \boldsymbol{j} = \sigma \boldsymbol{E}
    \label{eq:Ohm}
\end{equation}
where, $\boldsymbol{j}$ is current density, $\mathbf{\sigma}$ is conductivity tensor.
% 将欧姆定律 Eq. $\eqref{eq:11}$ 带入 Eq. $\eqref{eq:10}$，
% 上式可以进一步写为，

In coordinate system that we have chosen in the beginning of Section $\ref{sec2-1}$,
$\boldsymbol{k} = \sin \theta k \boldsymbol{e_y} + \cos \theta k \boldsymbol{e_z}$.
And considering that $\nabla \times (\nabla \times \boldsymbol{E}) = \nabla(\nabla \cdot \boldsymbol{E}) - \nabla^2 \boldsymbol{E}$, and $\nabla = \mathrm{i} \boldsymbol{k}$ (Equation (\ref{eq:partial-t-s})), and combining Ohm's law (Equation (\ref{eq:Ohm})), the wave Equation ($\ref{eq:EM-wave}$) can be written as follow,
\begin{equation}
    \left(k^2 k_{ij} + \omega^2 \varepsilon \mu  \epsilon_{ij} \right) \boldsymbol{E} = 0
    \label{eq:dispersion-rela}
\end{equation}
where, $k$ is as follow, 
\begin{equation}
    k = \omega \sqrt{\varepsilon \mu} = \omega \frac{N}{c}
    \label{eq:k-N}
\end{equation}
here, $c$ is the light speed, $N$ is the phase refraction index. And $k_{ij}$ in Equation \eqref{eq:dispersion-rela},
\begin{equation}
k_{ij} = \begin{pmatrix} 
-1 & 0 & 0 \\
0 & \sin^2 \theta-1 & \sin \theta \cos \theta  \\
0 & \sin \theta \cos \theta & \cos^2 \theta-1
\end{pmatrix}
\end{equation}
and $\epsilon_{ij}$ is dielectric tensor,
\begin{equation}
    \epsilon_{ij} = \boldsymbol{I} + \frac{\mathrm{i}}{\omega \varepsilon} \sigma
    \label{eq:dielectic-conductivity}
\end{equation}

% #### 从电磁波方程到色散方程

% 为保证 Eq. $\eqref{eq:15}$在一般情况下为 0，有，
In order to ensure that Equation ($\ref{eq:dispersion-rela}$) is 0 in general, there is, 
\begin{equation}
    \omega^2 \varepsilon \mu  (k_{ij} + \epsilon_{ij}) = 0
    \label{eq:dispersion}
\end{equation}
% 至此，得到了 $k$ 与 $\omega$ 的关系，即**色散关系**。
At this stage, the relation between $k$ and $\omega$, i.e., the dispersion relation, is obtained.

% ### 从电流密度到电导率张量和介电张量

% #### 从电流密度到电导率张量

% 电流密度 $\mathbf{J}$ 由电子的运动引起，其定义式为：
The current density $\boldsymbol{j}$ is caused by the movement of charged particles, and it is defined as,
\begin{equation}
\boldsymbol{j} = -n_\mathrm{e} e \boldsymbol{u} 
\label{eq:current-def}
\end{equation}
% 将运动方程$\eqref{eq:7}$中的 $\boldsymbol{v}$ 带入 Eq. $\eqref{eq:17}$ (电流密度 $\boldsymbol{j}$ )中，
where, $n_\mathrm{e}$ is the number density of electron. Take the definition of current density $\boldsymbol{j}$ into motion Equation ($\ref{eq:motion-matrix-I}$)
\begin{equation}
\boldsymbol{j} = \frac{\mathrm{i} n_\mathrm{e} e^2}{m \omega U\left(U^{2}-Y^{2}\right)}
\begin{pmatrix} 
U^{2} & -\mathrm{i} U Y & 0 \\
\mathrm{i} U Y & U^{2} & 0 \\
0 & 0 & \left(U^{2}-Y^{2}\right)
\end{pmatrix}
\begin{pmatrix} 
E_x  \\
E_y \\
E_z
\end{pmatrix}
\end{equation}
% 根据 Eq $\eqref{eq:11}$ (欧姆定律，$\boldsymbol{j} = \mathbf{\sigma} \boldsymbol{E}$)，结合 Eq. $\eqref{eq:18}$将电导率张量 $\boldsymbol{\sigma}$ 写为张量的形式，
Considering Ohm's law, we can obtain the conductivity tensor $\sigma_{ij}$,
\begin{equation}
\sigma_{ij} = 
\frac{\mathrm{i} n_\mathrm{e} e^2}{m \omega U\left(U^{2}-Y^{2}\right)}
\begin{pmatrix} 
U^{2} & -\mathrm{i} U Y & 0 \\
\mathrm{i} U Y & U^{2} & 0 \\
0 & 0 & \left(U^{2}-Y^{2}\right)
\end{pmatrix}
\end{equation}

% #### 从电导率张量到介电张量

% 由 Eq. $\eqref{eq:22}$ (相对介电张量 $\boldsymbol{\epsilon}$ 与电导率张量 $\boldsymbol{\sigma}$ 的关系 $\boldsymbol{\epsilon} =  \mathbf{I} + \mathrm{i} \boldsymbol{\sigma}/(\omega \varepsilon_0)  $ )，且考虑到等离子体振荡频率 $\omega_\mathrm{p}$ 有 $\omega_\mathrm{p}^2 = \frac{n_0 e^2}{m_e \varepsilon_0}$ ，相对介电张量中 $\mathrm{i} \boldsymbol{\sigma}/(\omega \varepsilon_0) $ 为，
% 相对介电张量 $\boldsymbol{\epsilon}$ 为，
And considering the relationship between dielectric tensor $\epsilon_{ij}$ and conductivity tensor $\sigma_{ij}$ in Equation $\eqref{eq:dielectic-conductivity}$, $\epsilon_{ij}$ can be got as follow,
\begin{equation}
\epsilon_{ij} = \begin{pmatrix} 
1 - \frac{XU}{ U^2 - Y^2  } & \mathrm{i} \frac{XY}{ U^2 - Y^2  } & 0 \\
-\mathrm{i} \frac{XY}{ U^2 - Y^2  } & 1 - \frac{XU}{ U^2 - Y^2  } & 0 \\
0 & 0 & 1 - \frac{X }{U }
\end{pmatrix}
\end{equation}

% ### 求解折射率方程

% 相速度的定义为，
% The definition of the phase velocity is as follow,
% \begin{equation}
% v_{\mathrm{p}} = \frac{\omega}{k} = \frac{1}{\sqrt{\mu \varepsilon}}
% \end{equation}
% % 相折射率的定义为，
% Thus, 
% \textbf{The phase refraction index is as follow,}
% \begin{equation}
% N = \frac{c}{v_{\mathrm{p}}} = \frac{ck}{\omega} = c \sqrt{\varepsilon \mu}
% \label{eq:N-k}
% \end{equation}

% 将 Eqs. $\eqref{eq:33} \eqref{eq:35}$ 带入色散关系中 Eq. $\eqref{eq:24}$得，
Take the relation between $N$ and $k$ (Equation $\eqref{eq:k-N}$) into dispersion relation (Equation \eqref{eq:dispersion-rela}), we can obtain the following determinant,
\begin{equation}
\begin{vmatrix} 
-N^2 + 1 - \frac{XU }{ U^2 - Y^2 } & \mathrm{i}\frac{XY }{U^2-Y^2 } & 0 \\
- \mathrm{i}\frac{XY }{U^2-Y^2 } & -\cos^2 \theta ~N^2 + 1 - \frac{X }{ U^2 - Y^2 } & \sin \theta \cos \theta ~N^2 \\
0 & \sin \theta \cos \theta ~N^2 & -\sin^2 \theta ~N^2 + 1 - \frac{X}{U}
\end{vmatrix}
= 0
\label{eq:dispersion-det-N}
\end{equation}
% 在此，令，
With following denotation,
\begin{equation}
\begin{aligned}
S & =1-\frac{X U}{U^{2}-Y^{2}}, \quad D =-\frac{X Y}{U^{2}-Y^{2}}, \quad P =1-\frac{X}{U} \\
R &= S+D, \quad L= S-D
\end{aligned}
\end{equation}
% 展开行列式$\eqref{eq:39}$，
we can expand the determinant Equation ($\ref{eq:dispersion-det-N}$) as follow,
\begin{equation}
    \left(S \sin^2 \theta + P \cos^2 \theta \right) N^4 - \left[RL \sin^2 \theta + SP \left(1 + \cos^2 \theta \right) \right] N^2 + PRL = 0
    \label{eq:N2}
\end{equation}
% 现在令，
% Here, we introduce the denotation,
% \begin{equation}
% \begin{aligned}
% A &= S \sin^2 \theta + P \cos^2 \theta \\
% B &= RL \sin^2 \theta + SP \left(1 + \cos^2 \theta \right) \\
% C &= PRL 
% \end{aligned}
% \end{equation}
% % 式$\eqref{eq:41}$可以写为，
% thus Equation ($\ref{eq:N2}$) can be written as follow,
% \begin{equation}
%     A N^4 - B N^2 + C = 0
% \end{equation}
% 解$N^2$得，
Ultimately, we can obtain the solution of $N^2$ as follow \cite{Bittencourt2004},
\begin{equation}
    N^2 = 1 - \frac{X}{U - \frac{1}{2 (U - X)} Y^2 \sin^2 \theta \pm \sqrt{\frac{1}{4 (U - X)^2} Y^4 \sin^4 \theta + Y^2 \cos^2 \theta}}
    \label{eq:A-H}
\end{equation}
% 此式即为 **Appleton-Hartree（A-H）公式**。
It is A--H Equation.

% 在空间中，等离子体是无碰撞的，$\nu$ = 0，所以运动方程$\eqref{eq:1}$为，
For collisionless plasma with $\nu = 0$, the collision term in the motion Equation (\ref{eq:motion}) can be ignored. Thus $Z$ is neglected, and $U = 1$. 
In this case, the A--H equation of collisionless plasma as follow \citep{Hutchinson2002},
\begin{equation}
    N^{2}=1-\frac{X(1-X)}{1-X-\frac{1}{2} Y^{2} \sin ^{2} \theta \pm\left[\left(\frac{1}{2} Y^{2} \sin ^{2} \theta\right)^{2}+(1-X)^{2} Y^{2} \cos ^{2} \theta\right]^{1 / 2}} 
    \label{eq:A-H-cl}
\end{equation}
% 此式即为无碰撞等离子体的 **Appleton-Hartree（A-H）公式**，描述了电磁波在磁化等离子体中的传播，特别是其折射率随频率、磁场和等离子体参数的变化。

% ### 讨论

A--H equation describes the refractive index for EM waves in magnetized plasma. 
The $\pm$ sign in A--H equation is associated with $Y$, which is related to the magnetic cyclotron frequency $\omega_B$, and further to magnetic field $\boldsymbol{B}$.
The $\pm$ sign shows that the magnetic field makes the plasma an anisotropic medium. The anisotropy implies that the wave's propagation speed and attenuation depend not only on the frequency of EM waves but also on its direction of propagation relative to the magnetic field.

For space plasma is collisionless, the typical number density $n_\mathrm{e}$ is on the order of 1 cm$^{-3}$, the typical number temperature is on the order of $10^4$ K, and the typical magnetic field $B$ is on the order of 10 nT. 
Under the above conditions, $\omega_\mathrm{p} \sim 10^5$ Hz, $\omega_\mathrm{B} \sim 10^3$ Hz, and $\nu \sim 10^{-8}$ Hz, thus, $X \gg Y^2 \gg Z$. Thus, the space plasma can be assumed to be collisionless and un-magnetized plasma. And A--H Equation $\eqref{eq:A-H-cl}$ further degenerates as follow,
\begin{equation}
    N^2 = 1 - X
\label{eq:A-H-f_p}
\end{equation}

%  $\pm$ 号也说明磁场使得等离子体成为各向异性介质，即波的传播速度和衰减特性不仅取决于波的频率，还与传播方向和磁场的相对取向有关。

% 当 $\omega_p \gg \omega_B$ 时，$X \gg Y^2$，此时，含有 $Y$ 的项可以忽略，此时 A-H 公式退化为，

% 即各项同性的情况。比较 Eq. $\eqref{eq:38}$ 与 Eq. $\eqref{eq:39}$，我们可以看出，电磁波在磁化等离子体中的折射率的各项异性是由磁场导致的。

Comparing Equation $\eqref{eq:A-H-cl}$ and Equation $\eqref{eq:A-H-f_p}$, it reveals that without the magnetic field, the plasma changes from anisotropic medium to isotropic one, the propagation of EM waves in the isotropic plasma becomes simplified.

The group refractive index $N_\mathrm{g}$ can be derived as follow,
\begin{equation}
N_\mathrm{g} = \frac{\partial (N \omega)}{\partial \omega} = 
\frac{\partial \left[\omega \left(1 - \frac{\omega_{\mathrm{p}}^2}{\omega^2}\right)\right] }{\partial \omega} \approx 1 + \frac{X}{2} = 1 + \frac{K n_e}{2f^2}
\end{equation}
Here, $f$ is the EM wave frequency, $K = e^2/(4\pi m_{\mathrm{e} } \varepsilon_0 ) = 80.6~\rm m^3s^{-2}$.

The time ($\tau$) of EM waves propagating a distance $L$ in space plasma is,
\begin{equation}\label{Eq7}
\tau = \int_{L} \frac{\mathrm{d} s}{v_g} = \int_{L} \frac{\mathrm{d} s}{c/N_g}
\end{equation}
where $c$ is the speed of light in vacuum,
The time delay ($\Delta \tau$) of the EM waves propagation in space plasma relative to the vacuum case is,
\begin{equation}
\Delta \tau = \frac{1}{c} \int_{L} (1 + \frac{K n_{\mathrm{e}}}{2f^2}) \mathrm{d} s - \frac{L}{c} = \frac{K}{2 c f^2} \int_{L} n_{\mathrm{e}} \mathrm{d} s 
\label{eq:dt}
\end{equation}
% Here, $\int_{L} N \mathrm{d} s$ is called TEC. 
$\int_{L} n_{\mathrm{e}} \mathrm{d} s$ is the integrated electron number density along the laser link. According to Equation \eqref{eq:dt}, the OPD can be calculated as following,
\begin{equation}
\label{eq:dl}
\Delta l = c \Delta \tau = \frac{K}{2 f^2} \int_{L} n_{\mathrm{e}} \mathrm{d} s
\end{equation}
According to Equation \eqref{eq:dl}, we can obtain the OPD noise $\Delta l$ due to laser propagation in space plasma.

\subsection{The model of acceleration noise due to space magnetic field}\label{sec2-2}

% 带有磁矩的物体(检验质量)在磁场中受到的静磁力可以表示为，
An object with a magnetic moment in the magnetic field will be subjected to a magnetic moment force $\boldsymbol{F}_{\rm M}$, which can be expressed as follows,
\begin{equation}
\boldsymbol{F}_{\rm M} = \nabla (\boldsymbol{M} \cdot \boldsymbol{B}) \\
\label{eq:F-M}
\end{equation}
% 其中，$\boldsymbol{M}$为TM的磁矩，$\boldsymbol{B}$为背景磁场。在这里，$\boldsymbol{B}$由空间磁场$\boldsymbol{B}_{\mathrm{sp} }$和卫星自身磁场$\boldsymbol{B}_{\mathrm{sc} }$组成：$\boldsymbol{B} = \boldsymbol{B}_{\mathrm{sp} } + \boldsymbol{B}_{\mathrm{sc} } $；$\boldsymbol{M}$为检验质量上的磁矩，由剩磁矩$\boldsymbol{M}_{r}$和背景磁场导致的感应磁矩$\boldsymbol{M}_{i}$组成：$\boldsymbol{M} = \boldsymbol{M}_{\mathrm{r} } + \boldsymbol{M}_{\mathrm{i} }$。其中，$\boldsymbol{M}_{\mathrm{i} }$可以分解为由$\boldsymbol{B}_{\mathrm{sp} }$和$\boldsymbol{B}_{\mathrm{sc} }$导致的两部分：
where, $\boldsymbol{B}$ is the background magnetic field, and $\boldsymbol{M}$ is the magnetic moment of the TM.
Here, $\boldsymbol{B}$ is composed of the spacecraft magnetic field $\boldsymbol{B}_{\rm sc}$ and the space magnetic field $\boldsymbol{B}_{\rm sp}$, 
$\boldsymbol{B} = \boldsymbol{B}_{\rm sp} + \boldsymbol{B}_{\rm sc}$; And the 
$\boldsymbol{M}$ including the remanent magnetic moment $\boldsymbol{M}_{\rm r}$ 
and the inductive magnetic moment $\boldsymbol{M}_{\rm i}$, $\boldsymbol{M}_{\rm tm} = \boldsymbol{M}_{\rm r} + \boldsymbol{M}_{\rm i}$. And
both $\boldsymbol{B}_{\rm sp}$ and $\boldsymbol{B}_{\rm sc}$ can induce $\boldsymbol{M}_{\rm i}$,
\begin{equation}
\boldsymbol{M}_{\mathrm{i} } = \boldsymbol{M}_{\mathrm{isc} } + \boldsymbol{M}_{\mathrm{isp} } = \frac{\chi_{\mathrm{m} } V \boldsymbol{B}_{\mathrm{sp}} }{\mu_0} + \frac{\chi_{\mathrm{m} } V \boldsymbol{B}_{\mathrm{sc} } }{\mu_0} \\
\label{eq:mag-moment}
\end{equation}
% 其中$\chi_{\mathrm{m} }$是TM的磁化率，$V$是TM的体积，$\mu_0$为真空磁导率。
here, $\chi_{\rm m}$ is magnetic susceptibility, $V$ is the volume of the TM, and $\mu_0$ is the vacuum magnetic permeability.

% 把$\boldsymbol{B}$, $\boldsymbol{M}$和Eq. $\eqref{eq:mag-moment}$式带入Eq. $\eqref{eq:F-M}$式，可以得到磁矩力导致的加速度噪声为，
Take magnetic field $\boldsymbol{B}$, magnetic moment $\boldsymbol{M}$ and Equation $\eqref{eq:mag-moment}$ into Equation $\eqref{eq:F-M}$, the acceleration noise due to the magnetic moment force can be obtained as,
\begin{equation}
\begin{aligned}
a &= \frac{1}{m} \nabla [(\boldsymbol{M}_{\mathrm{r} } + \boldsymbol{M}_{\mathrm{isp} } + \boldsymbol{M}_{\mathrm{isc} } ) \cdot (\boldsymbol{B}_{\mathrm{sp} } + \boldsymbol{B}_{\mathrm{sc} } )] \\
& = \frac{1}{m} \nabla \left( \boldsymbol{M}_{\mathrm{r} } \cdot \boldsymbol{B}_{\mathrm{sp} } + \boldsymbol{M}_{\mathrm{r} } \cdot \boldsymbol{B}_{\mathrm{sc} }  + \frac{2 \chi_{\mathrm{m} }V}{\mu_0} \boldsymbol{B}_{\mathrm{sp} } \cdot \boldsymbol{B}_{\mathrm{sc} } + \frac{\chi_{\mathrm{m} }V}{\mu_0} B_{\mathrm{sp} }^2 + \frac{\chi_{\mathrm{m} }V}{\mu_0} B_{\mathrm{sc} }^2 \right) \\
\end{aligned}
\label{eq:mag-moment-expand}
\end{equation}
where, $m$ is the mass of the TM.
% 其中$m$为TM的质量。

% 在此，我们关注空间磁场导致的加速度噪声，即只考虑角标含有sp的项，在此忽略Eq. $\eqref{eq:mag-moment-expand}$式的第2项和第5项。可以上式可以简化为，
In this work, we focus only on the acceleration noise due to the space magnetic field, and neglect the terms without space magnetic field (second and fifth terms in Equation $\eqref{eq:mag-moment-expand}$). Thus, Equation $\eqref{eq:mag-moment-expand}$ can be simplified as follow,
\begin{equation}
a_{\mathrm{M} } = \frac{1}{m} \nabla \left( \boldsymbol{M}_{\mathrm{r} } \cdot \boldsymbol{B}_{\mathrm{sp} } + \frac{2 \chi_{\mathrm{m}}V}{\mu_0} \boldsymbol{B}_{\mathrm{sp} } \cdot \boldsymbol{B}_{\mathrm{sc} } + \frac{\chi_{\mathrm{m}}V}{\mu_0} B_{\mathrm{sp} }^2 \right) 
\label{eq:mag-moment-sp}
\end{equation}

% 根据矢量运算规则
According to the vector operation rules 
$\nabla(\boldsymbol{f}\cdot\boldsymbol{g}) = (\boldsymbol{f} \cdot \nabla) \boldsymbol{g} + (\boldsymbol{g} \cdot \nabla) \boldsymbol{f} + \boldsymbol{f} \times (\nabla \times \boldsymbol{g}) + \boldsymbol{g} \times (\nabla \times \boldsymbol{f})$, where $\boldsymbol{f}$ and $\boldsymbol{g}$ are vectors in this formula, the first term of Equation $\eqref{eq:mag-moment-sp}$ can be expanded as follow,
\begin{equation}
\nabla (\boldsymbol{M}_{\mathrm{r} } \cdot \boldsymbol{B}_{\mathrm{sp} }) = (\boldsymbol{M}_{\mathrm{r} } \cdot \nabla) \boldsymbol{B}_{\mathrm{sp} } + (\boldsymbol{B}_{\mathrm{sp} } \cdot \nabla) \boldsymbol{M}_{\mathrm{r} } 
+ \boldsymbol{M}_{\mathrm{r} } \times (\nabla \times \boldsymbol{B}_{\mathrm{sp} }) + \boldsymbol{B}_{\mathrm{sp} } \times (\nabla \times \boldsymbol{M}_{\mathrm{r} })
\label{eq:mag-moment1-0}
\end{equation}
% 根据Maxwell方程组，$\nabla \times \boldsymbol{B}$是由宏观电流$\boldsymbol{j}$和位移电流$\varepsilon_0 \mu_0 \partial \boldsymbol{E} / \partial t$组成。由于检验质量被层层包裹，宏观电流在检验质量上并不存在（充放电也会有宏观电流），可以被忽略，在此，我们仅保留位移电流项。则5式可以写为：
According to the Ampere-Maxwell law, $\nabla \times \boldsymbol{B} = \mu_0 \boldsymbol{j} +\varepsilon_0 \mu_0 \partial \boldsymbol{E} / \partial t$, which contains conduction current and displacement current.
% 由于TM是无拖拽的悬浮在卫星中，且被卫星层层包裹，应该是没有电子电流的。且现有的实验结果表明，TM的电子电流是可以忽略的。
Since the TM is drag-free and floating in the satellite, has no direct contact with the satellite, and it is wrapped by layers of the satellite, there should be no conduction current.
% And the experiments results show that the conduction current of TM is negligible. 
Thus, Equation $\eqref{eq:mag-moment1}$ can be written as, 
\begin{equation}
\nabla (\boldsymbol{M}_{\mathrm{r} } \cdot \boldsymbol{B}_{\mathrm{sp} }) = (\boldsymbol{M}_{\mathrm{r} } \cdot \nabla) \boldsymbol{B}_{\mathrm{sp} } + (\boldsymbol{B}_{\mathrm{sp} } \cdot \nabla) \boldsymbol{M}_{\mathrm{r} }
      + \boldsymbol{M}_{\mathrm{r} } \times (\frac{\varepsilon_0 \mu_0 \partial \boldsymbol{E}_{\mathrm{sp} }}{\partial t}) + \boldsymbol{B}_{\mathrm{sp} } \times (\nabla \times \boldsymbol{M}_{\mathrm{r} })
\label{eq:mag-moment1}
\end{equation}

% 类似的，Eq. $\eqref{eq:eq:mag-moment-sp}$式中的第2可以写为：
Similarly, the second term in Equation $\eqref{eq:mag-moment-sp}$ can be written as,
\begin{equation}
\begin{aligned}
\nabla (\boldsymbol{B}_{\mathrm{sp} } \cdot \boldsymbol{B}_{\mathrm{sc} })
&= (\boldsymbol{B}_{\mathrm{sp} } \cdot \nabla) \boldsymbol{B}_{\mathrm{sc} } + (\boldsymbol{B}_{\mathrm{sc} } \cdot \nabla) \boldsymbol{B}_{\mathrm{sp} }
+ \boldsymbol{B}_{\mathrm{sp} } \times (\frac{\varepsilon_0 \mu_0 \partial \boldsymbol{E}_{\mathrm{sc} }}{\partial t} ) 
+ \boldsymbol{B}_{\mathrm{sc} } \times (\frac{\varepsilon_0 \mu_0 \partial \boldsymbol{E}_{\mathrm{sp} }}{\partial t} )
\end{aligned}
\label{eq:mag-moment2}
\end{equation}

% 根据矢量运算规则$\nabla (uv) = u \nabla v + v \nabla u$，则Eq. $\eqref{eq:mag-moment-sp}$式中的第3项可以写为：
According to the vector operation rule $\nabla (uv) = u \nabla v + v \nabla u$, where $u$ and $v$ are scalars in this formula, the third term of Equation $\eqref{eq:mag-moment-sp}$ can be written as,
\begin{equation}
\nabla (B_{\mathrm{sp} }^2) = 2 B_{\mathrm{sp} } \nabla B_{\mathrm{sp} }
\label{eq:mag-moment3}
\end{equation}

% 把Eqs. 代入Eq. $\eqref{eq:mag-moment-sp}$式得：
Take Equations $\eqref{eq:mag-moment1}$, $\eqref{eq:mag-moment2}$, and $\eqref{eq:mag-moment3}$ into Equation $\eqref{eq:mag-moment-sp}$, the magnetic moment force $\boldsymbol{F}_{\rm M}$ is as follow,
\begin{equation}
\begin{aligned}
\boldsymbol{F}_{\rm M} &= 2 (\boldsymbol{M}_{\mathrm{isp} } \cdot \nabla) \boldsymbol{B}_{\mathrm{sc} } + [(\boldsymbol{M}_{\mathrm{r} } + 2\boldsymbol{M}_{\mathrm{isc} } \cdot) \nabla] \boldsymbol{B}_{\mathrm{sp} } \\
&+ \boldsymbol{M}_{\mathrm{isp} } \times \frac{\varepsilon_0 \mu_0 \partial \boldsymbol{E}_{\mathrm{sc} }}{\partial t} + \boldsymbol{M}_{\mathrm{isc} } \times \frac{\varepsilon_0 \mu_0 \partial \boldsymbol{E}_{\mathrm{sp} }}{\partial t} + 2 B_{\mathrm{sp} } \nabla B_{\mathrm{sp} }
\end{aligned}
\label{eq:mag-moment-sp-expand}
\end{equation}
% 将上式重新组织为如下形式，
Reorganize the Equation $\eqref{eq:mag-moment-sp-expand}$, and the acceleration noise can be written as the following 5 terms,
\begin{equation}
\left\{
\begin{aligned}
\boldsymbol{a}_{\rm M1} &= \frac{2}{m } (\boldsymbol{M}_{\rm isp} \cdot \nabla) \boldsymbol{B}_{\rm sc}    \\
\boldsymbol{a}_{\rm M2} &= \frac{1}{m } [(\boldsymbol{M}_{\rm r} + 2 \boldsymbol{M}_{\rm isc}) \cdot \nabla] \boldsymbol{B}_{\rm sp}    \\
\boldsymbol{a}_{\rm M3} &= \frac{1}{m }(\boldsymbol{M}_{\rm r} + 2 \boldsymbol{M}_{\rm isc}) \times \frac{\varepsilon_0 \mu_0 \partial \boldsymbol{E}_{\rm sp}}{\partial t}   \\
\boldsymbol{a}_{\rm M4} &= \frac{2}{m }(\boldsymbol{M}_{\rm isp}) \times \frac{\varepsilon_0 \mu_0 \partial \boldsymbol{E}_{\rm sc}}{\partial t}   \\
\boldsymbol{a}_{\rm M5} &= \frac{2}{m } M_{\rm isp} \nabla B_{\rm sp}   \\
\end{aligned}
\right.
\label{eq:mag-moment-sp12345}
\end{equation}

% 对于空间引力波探测，取

% 可以得到xx的量级为。

In order to further reduce the acceleration noise caused by magnetic field, magnetic shielding has been proposed, which is commonly represented by the magnetic shielding factor $\xi_{\rm m}$. The acceleration noise after considering the magnetic shielding is represented as $a_{\mathrm{M} } /\xi_{\rm m}$.

The TM with residual charge in the magnetic field is subject to the Lorentz force. In space, there are also energetic particles, e.g., galactic cosmic ray (GCR) \citep{Parker1965,Potgieter2013LRSP} and solar energetic particle (SEP) \citep{Reames2013,LiC2013}. The energetic particles can penetrate the protection of the TM, bombard the TM, and cause it to become charged \citep{Schumaker2003,Sumner2020}. The Lorentz force of charged TM in the space magnetic field is as following,
\begin{equation}
    a_{\mathrm{L} } = \frac{1}{m}q \boldsymbol{v} \times \boldsymbol{B}_{\mathrm{sp} }
    \label{eq:mag-Lorentz}
\end{equation}
where, $q$ is the charge of the TM, $\boldsymbol{v}$ is the velocity of the TM.

\subsection{Space environment data and models}

\subsubsection{In-situ observation of space plasma and magnetic field}

The heliocentric space-borne GW detectors (LISA and TJ) rotate around the Sun at a distance of about 1 AU from the Sun, and they consist of three satellites forming an equilateral triangular laser interferometric link with arms of millions of kilometers in length. Of these, LISA is planned to be deployed about 50 million kilometers behind the Earth, with an interferometric arm length of about 2.5 million kilometers; And TJ about 50 million kilometers in front of the Earth, with an arm length of about 3 million kilometers. 
The orbits of both LISA and TJ are completely dipped in the solar wind plasma \citep{LISA2017,Qiao2023Rev,Qiao2023-Helio}. Although there is no in-situ observation of space plasma and magnetic field near the orbits of LISA and TJ at present, since the symmetry of the solar wind parameters in the ecliptic plane, the observation at the 1st Sun-Earth Lagrange (L1) point, can be used to analyse the space environment problems for LISA.

\begin{figure}[h]
\centering
\includegraphics[width=0.86\textwidth, trim = 0 0 0 0]{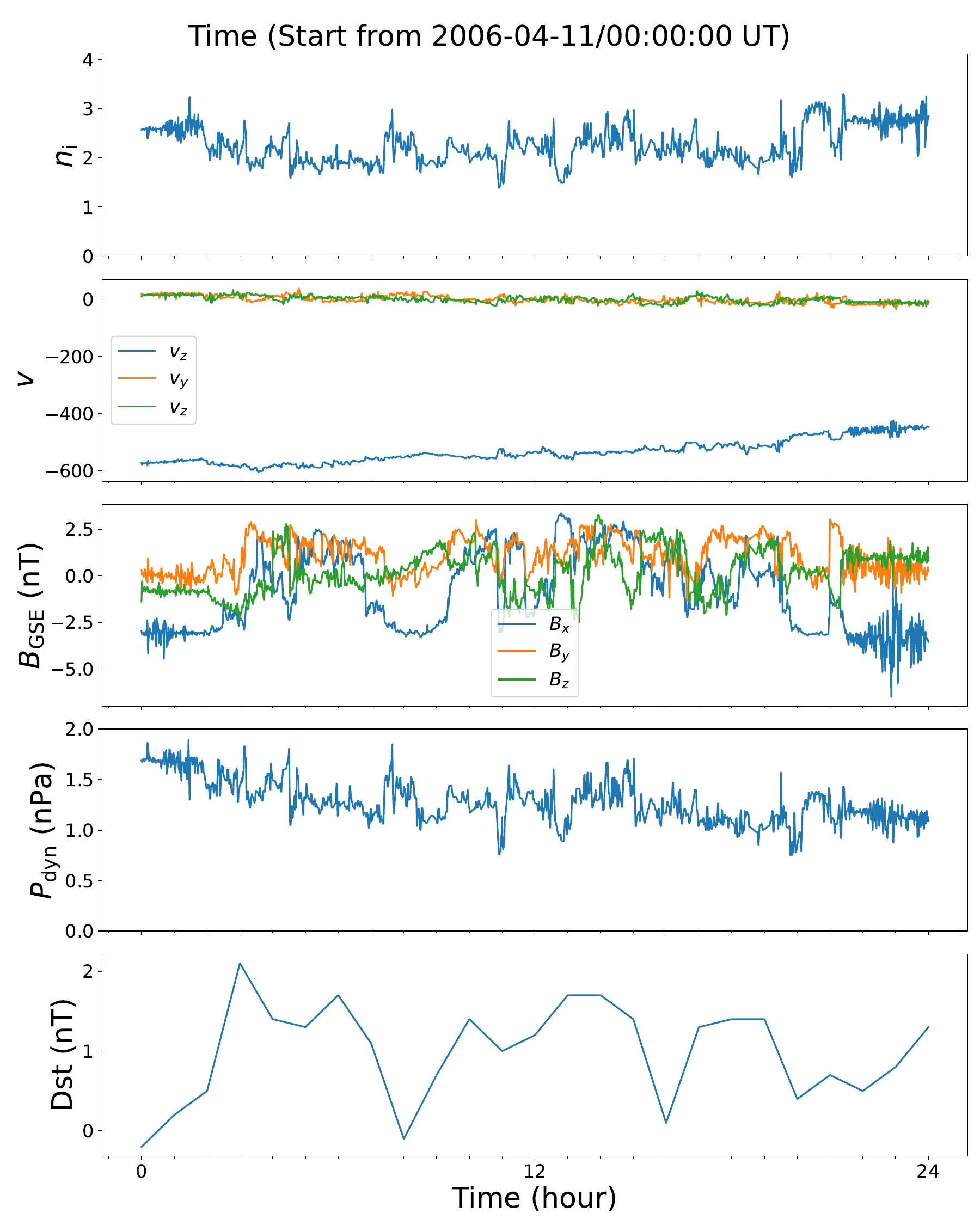}
\caption{Space plasma and magnetic field data from OMNI. $n_{\rm i}$, $v$, $B$ in GSE coordinate, $P_{\rm dyn}$, and Dst index in 2006-04-11. }
\label{fig-input}
\end{figure}

There are in-situ satellites to get the space plasma and magnetic field at L1 point, e.g. Wind \citep{Ogilvie1995}, ACE \citep{Stone1998}, and so on. They can detect the space plasma number density $n_{\rm i}$ and $n_{\rm e}$, velocity $v$, temperature $T$, space magnetic field $B_{\rm x}, B_{\rm y},B_{\rm z}$, and other parameters at L1 point. Based on the in-situ observations in the past decades, the solar wind dataset OMNI has been established \citep{King2005}, and the in-situ observation data can be used to study the effects of space environment on the space-borne GW detections.
In particular, since acceleration noise due to space magnetic field is localised, the in-situ observation data can be used directly to calculate space magnetic acceleration noise \citep{Su2020} (Figure \ref{fig-input}).

The geocentric GW detector, TQ, is at an altitude of about 100000 km from the center of the Earth, also forms an equilateral triangular laser interferometric links to the detect the GWs, with the arm length of about 170000 km \citep{Zhang2021PRD,Ye2021,Jia2023}. 
The Earth's magnetosphere is the result of the interaction of the solar wind with its intrinsic magnetic field, and the solar wind blowing onto the Earth's magnetosphere causes the magnetopause to be compressed, and the magnetotail to be stretched \citep{Parker1958PoF}. The magnetopause and bow shock are often considered to be the boundary between the solar wind and the Earth's magnetosphere \citep{Liu2015,Lv2019}.
The subsolar point of the magnetosphere is often at a distance of about 10 $\rm R_E$ from the Earth, which can be compressed to $<7~\rm R_E$ when the solar wind dynamic pressure $P_{\rm_{dyn} }$ is strong, and it can be $>7~\rm R_E$ when $P_{\rm_{dyn} }$ is weak \citep{Lv2015}; the bow shock nose region is often at a distance of about 12 $\rm R_E$ from the Earth, and the bow shock nose region can be compressed to $<9~\rm R_E$ when $P_{\rm_{dyn} }$ is strong, and it can be $>15~\rm R_E$ when $P_{\rm_{dyn} }$ is weak \citep{Wang2018}. 
The orbit of TQ is at a distance of about $15.6~\rm R_E$ from the Earth's center, and the TQ orbit is exactly in the region where the solar wind interacts with the Earth's magnetosphere.
Besides the magnetopause and bow shock, the structures such as cusp regions, magnetotail, and lobe regions are formed during the interaction between the solar wind and the Earth's magnetosphere. As solar wind conditions change, the geometry and parameters of these structures also change, solar wind observations can be used as inputs, for modelling the interaction between the Earth's magnetosphere and the solar wind by using MHD.

Around the orbit of geocentric GW detectors, there are in-situ observation of space plasma and magnetic fields by the satellites e.g., Clusters \citep{Cluster2001-MFI,Cluster2001-CIS}, THEMIS \citep{THEMIS2008}, and MMS \citep{Burch2016}. Although the orbits of these satellites are different from TQ, 
these satellites are similar to TQ in that they all pass through the signature regions of bow shock, magnetosheath, magnetopause, cusp region, magnetotail, lobe region, and so on.
The observation of these satellites can roughly reflect the approximate space magnetic field and plasma around the TQ orbit. Therefore, the in-situ observations of Earth's magnetosphere can be used as approximate alternative data to study the space magnetic field effect of TQ.

\subsubsection{Space magnetic field model -- Tsyganenko Model}

Tsyganenko space magnetic field model is a widely used empirical model that describes Earth's magnetospheric magnetic field \citep{Tsyganenko2022}. Developed by Tsyganenko and his collaborators over several decades \citep{Tsyganenko1989,Tsyganenko1995, Tsyganenko1996,Tsyganenko2002A,Tsyganenko2002B, Tsyganenko2005}, the model is based on a large dataset of spacecraft observations and provides a parameterized representation of the geomagnetic field under varying solar wind and geomagnetic conditions. It is a cornerstone in the field of magnetospheric physics, aiding in the study of space weather and Earth’s magnetic environment.

% The Tsyganenko model is built from large datasets of in-situ magnetic field measurements.
% These datasets include measurements of the magnetic field across various regions of the magnetosphere under different solar wind and geomagnetic conditions.

The Tsyganenko model fits the measured magnetic field data using parameterized mathematical functions that represent different current systems contributing to the magnetosphere magnetic field \citep{Tsyganenko1996,Tsyganenko2005}.
These current systems include,
Ring Current, encircling Earth and intensifying during geomagnetic storms;
Tail Current, extending into the magnetotail, shaped by solar wind pressure;
Magnetopause Current, currents on the boundary between the magnetosphere and solar wind;
Field-Aligned Currents, connecting the magnetosphere to the ionosphere \citep{Ganushkina2018,Yue2020}.

The Tsyganenko model adjusts the magnetic field based on real-time or average geophysical and solar wind conditions, using input parameters such as solar wind dynamic pressure ($P_{\rm dyn}$), which is the dominant term in the magnetic pressure balance of the Earth's magnetosphere, governs the compression of the magnetosphere \citep{Watermann2009,Jackman2014};
Dst Index, indicates the strength of the ring current and geomagnetic activity \citep{Daglis2006};
interplanetary magnetic field (IMF), which impacts the exchange of energy and plasma between the solar wind and Earth’s magnetosphere by reconnection processes at the magnetopause, and the asymmetry of the magnetosphere \citep{Tang2021};
Position in geocentric solar magnetic (GSM) coordinates, the model calculates the magnetic field at specific locations within the magnetosphere \citep{Tsyganenko1995,Tsyganenko1996,Tsyganenko2005}.

\begin{figure}[h]
\centering
\includegraphics[width=0.86\textwidth]{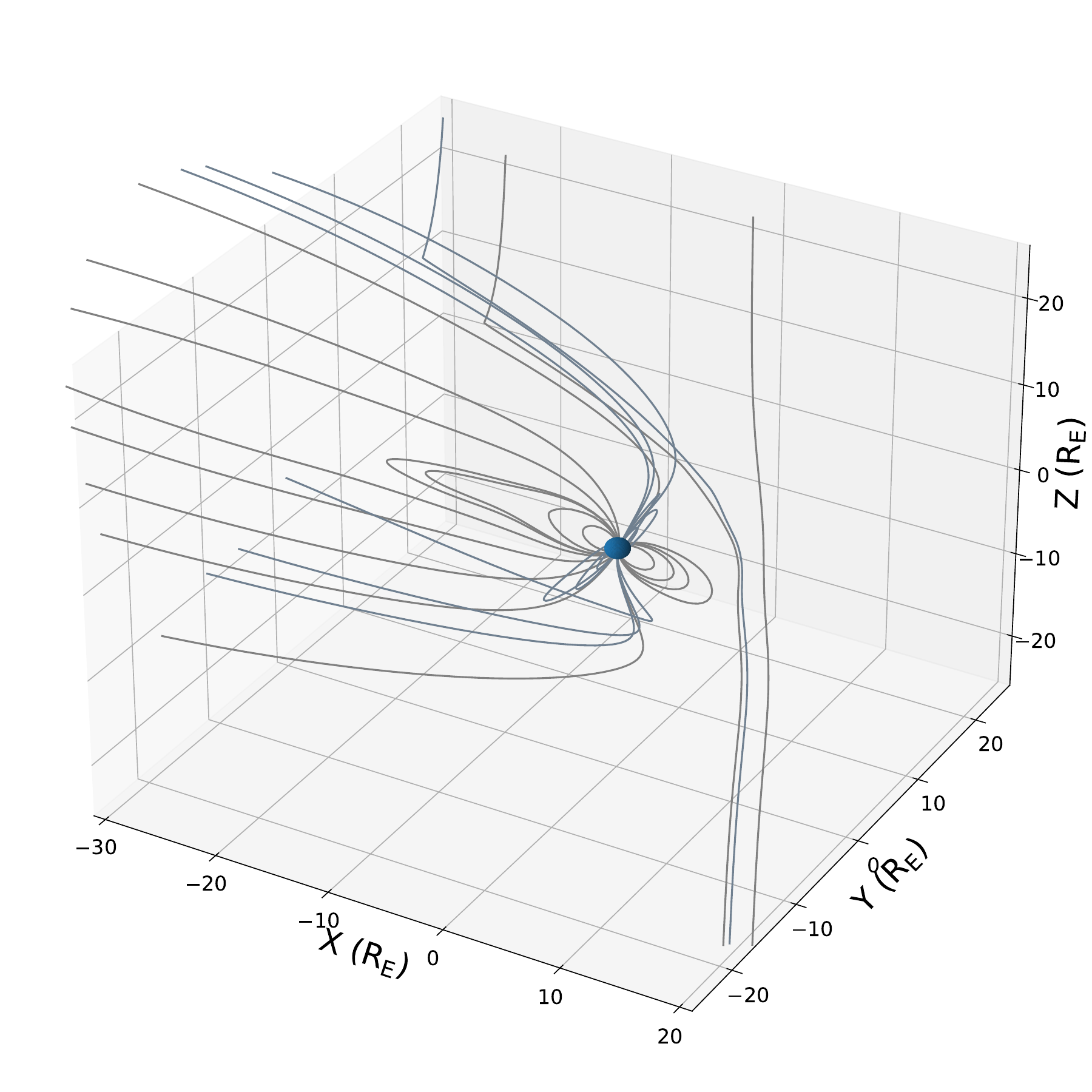}
\caption{The magnetic field around the geocentric GW detector that calculated by the Tsyganenko model \citep{Su2023}. }
\label{fig:B-TS3D}
\end{figure}

Based on the inputs, the Tsyganenko model uses a set of analytical functions to represent the contribution of each current system (including the ring current, magnetopause current, tail current, and field-aligned currents) to the overall magnetic field \citep{Tsyganenko1996,Tsyganenko2005}. 
And combining with the Earth’s dipole magnetic field, the Tsyganenko model calculates the magnetic field vector $\boldsymbol{B}$ with temporal and spatial variation.
The output $\boldsymbol{B}$ reflects the combined influence of all major magnetospheric current systems under the specified conditions (Figure \ref{fig:B-TS3D}).

% \begin{figure}[ht!]
% 	\begin{minipage}[b]{\textwidth}
% 		\centering
% 		\begin{minipage}{0.411\textwidth}
% 			\includegraphics[width = 5 cm]{sn-article-template/A-B-plt-3D-1.pdf}
% 		\end{minipage}
% 		\begin{minipage}{0.411\textwidth}
% 			\includegraphics[width = 5 cm]{sn-article-template/A-B-orbit-xz-arrow.pdf}
% 		\end{minipage}
% 	\end{minipage}
	
% 	\caption{The magnetic field lines calculated by the Tsyganenko model.}
	
% \label{fig:B-overview}
% \end{figure}

There are a wide applications for the Tsyganenko model in magnetosphere studies. 
It provides insights into the structure and dynamics of the Earth’s magnetosphere under different conditions, and helps to understand physical processes such as magnetic reconnection, particle trapping, and current systems.
The Tsyganenko model is also widely used in space weather research. 
It is used to study the effects of geomagnetic storms and substorms on the magnetosphere, and supports the modeling of radiation belts and plasma dynamics.
For satellite missions, the Tsyganenko model helps predict magnetic field conditions for spacecraft in Earth’s orbit, and supports the design and operation of space missions by estimating radiation and magnetic field exposures.
% Data Interpretation:
% Facilitates the interpretation of in-situ measurements from spacecraft by providing a reference magnetic field model.

Here, the Tsyganenko model is suitable for the study of space magnetic acceleration noise for geocentric GW detector, e.g. TQ.

\subsubsection{Space environment MHD simulation model: SWMF}

MHD simulation is used widely in the study of astrophysical \citep{Stone2020,Keppens2023}, laboratory \citep{Arber2015}, solar \citep{Rempel2017,Zhang2023} and space \citep{Toth2005} plasmas. 
Space Weather Modeling Framework (SWMF) \citep{Toth2005} is a well-known MHD simulation model in heliophysics, it has been widely used and validated \citep{Welling2010,Dimmock2013}.
The SWMF is a comprehensive computational framework designed to simulate and predict space weather phenomena. Developed and maintained by the University of Michigan’s Center for Space Environment Modeling (CSEM), it integrates multiple physics-based models to simulate various regions of the space environment, from the Sun to Earth and beyond.

SWMF connects different physics models as modules, allowing users to customize and couple models based on their research needs.
Examples of models include those for the solar corona, heliosphere, Earth's magnetosphere, ionosphere, thermosphere, and inner magnetosphere.
The SWMF integrates a suite of physics-based models representing different regions of the space environment, from the Sun to Earth and beyond. Each model specializes in a specific domain:
Solar Corona model (SC) and Inner Heliosphere model (IH) simulate the Sun's outer atmosphere and the propagation of the solar wind.
Global Magnetosphere model (GM) and Inner Magnetosphere model (IM) capture the dynamics of Earth's magnetic field and energetic particles.
Ionosphere Electrodynamics model (IE) and Thermosphere-Ionosphere model (TI) represent the upper atmosphere's electrodynamics and thermodynamics, and additional models, such as the SEP and Plasmasphere (PS) models, address specialized phenomena like particle acceleration and cold plasma behavior \citep{Toth2005}.
The SWMF’s modular design allows researchers to run these models independently or as coupled systems, facilitating comprehensive simulations of multi-region interactions.

\begin{figure}[h]
\centering
\includegraphics[width=1.0\textwidth]{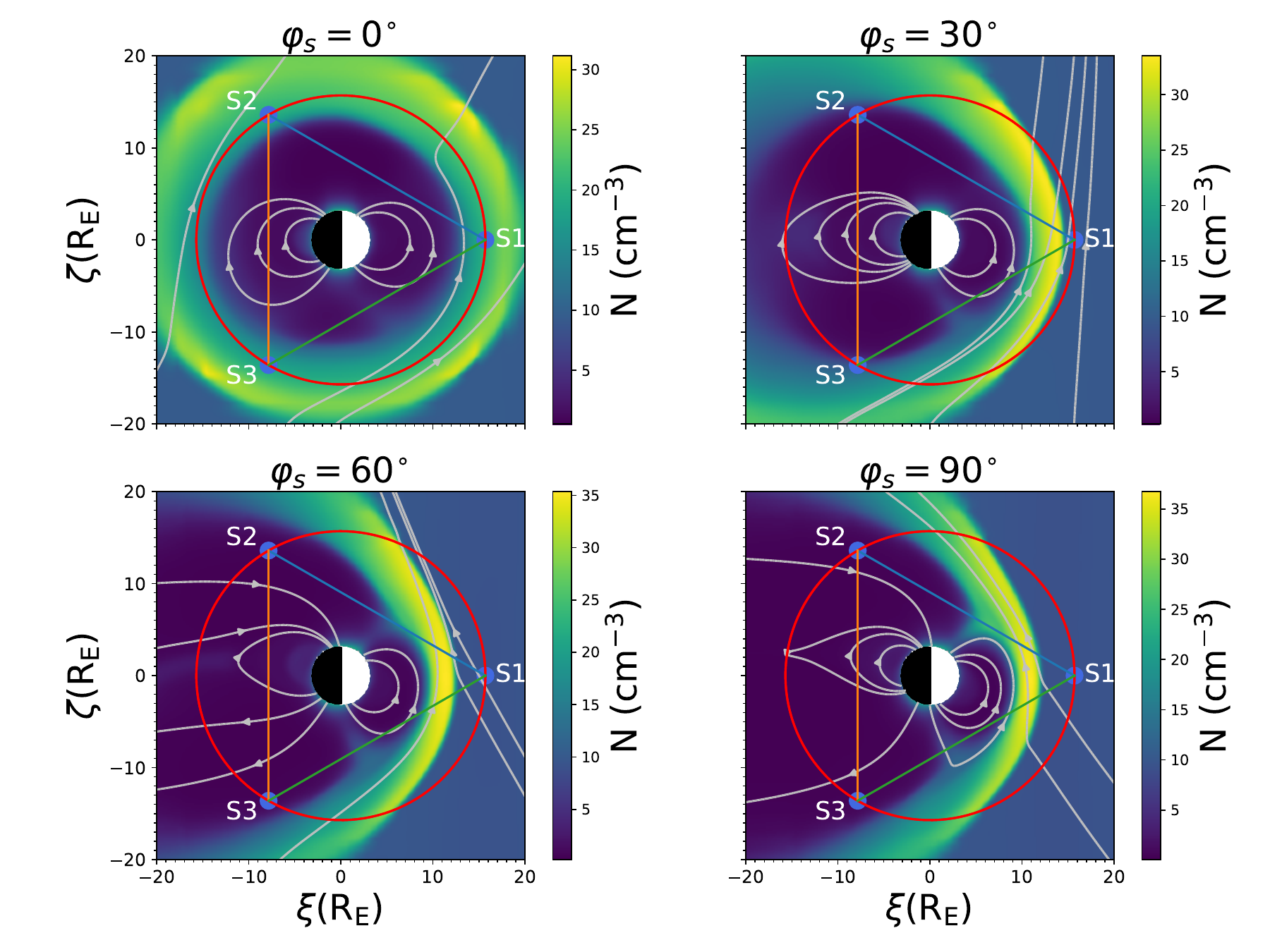}
\caption{The simulation results of the plasma and magnetic field around the orbital plane of GW detectors by the SWMF \citep{Su2020}. }
\label{fig-SWMF-nB}
\end{figure}

At the core, the SWMF is built upon the principles of physics that govern the behavior of plasma and magnetic fields across different regions of space. It employs MHD to describe the macroscopic behavior of plasmas in domains such as the solar corona and Earth’s magnetosphere \citep{Toth2005}. Kinetic and particle-based approaches are used in regions where particle dynamics dominate, such as the inner magnetosphere and radiation belts. Additionally, the SWMF integrates electrodynamics to model ionospheric currents and thermodynamics to describe the upper atmosphere's behavior. Each model within the SWMF represents a specific region of the space environment, and the framework facilitates their coupling to capture the complex interactions between these regions \citep{Toth2005}. The plasma and magnetic field around the orbital plane of GW detectors that obtained by the SWMF is shown in Figure \ref{fig-SWMF-nB}.

The SWMF has diverse applications in both scientific and practical contexts \citep{Takahashi2018}. It is a valuable tool for researchers studying the fundamental physics of the Sun-Earth system, enabling simulations of solar flares, CME, solar wind-magnetosphere coupling, geomagnetic storms, and ionospheric disturbances, and so on. The framework also supports space weather forecasting efforts, providing real-time predictions for agencies such as NASA and NOAA. These forecasts are crucial for protecting satellites, safeguarding astronauts during space missions, and mitigating the risks posed by space weather to critical infrastructure on Earth.

Beyond Earth, the SWMF is used to model the space environments of other planetary systems, such as Mars and Jupiter, aiding planetary science and exploration. Its adaptability and scalability also make it an essential tool for planning for future missions in an increasingly space-dependent world.

MHD simulation can get the spatial and temporal evolution of the plasma \citep{Feng2020,Zhou2021,Jiang2021}.
Thus, MHD simulations are suitable for the study of LPN in the detection of space-borne GW \citep{Su2020,Su2021}. 
It should be noted that plasma is quasi-neutral, so that the electrons and ions number densities are approximately equal. In the MHD simulation, the electrons and protons number densities are set to be equal, and the deviation from quasi-neutrality in the plasma along the laser link and in time is ignored. The proton number density is easier to obtain in the observation, and the deviation from quasi-neutrality in the plasma observation is also ignored, so that the observation of proton number density is set to be equal to electron number density in the study of LPN.

% For space-borne GW detectors, e.g. TQ, LISA, and TJ, the IH and GM models can be applied to the study of laser propagation noise.

% The MHD simulations are suitable to study the LPN for space-borne GW detections.

\section{Laser propagation effect in space plasma for space GW detection}
\label{sec3}

\subsection{Cases study of laser propagation noise}
\label{sec3-1}

The dispersion effect of laser propagation in space plasma leads to time delay, and generates OPD noise, which affects the space-borne GW detection. As shown in Equation \eqref{eq:dl}, the LPN is proportional to the integral of $n_{\rm e}$ along the laser link. At present, all the observations of space plasma are in-situ, and the in-situ observations can only obtain the space plasma data at one point along the laser link. Since the plasma parameters of the solar wind are variable rather than stable, the plasma distribution along the laser link is also variable. The MHD simulation can obtain the plasma distribution and evolution along the laser link, making it suitable for studying the LPN of space-borne GW detections.

\begin{figure}[h]
\centering
\includegraphics[width=0.86\textwidth]{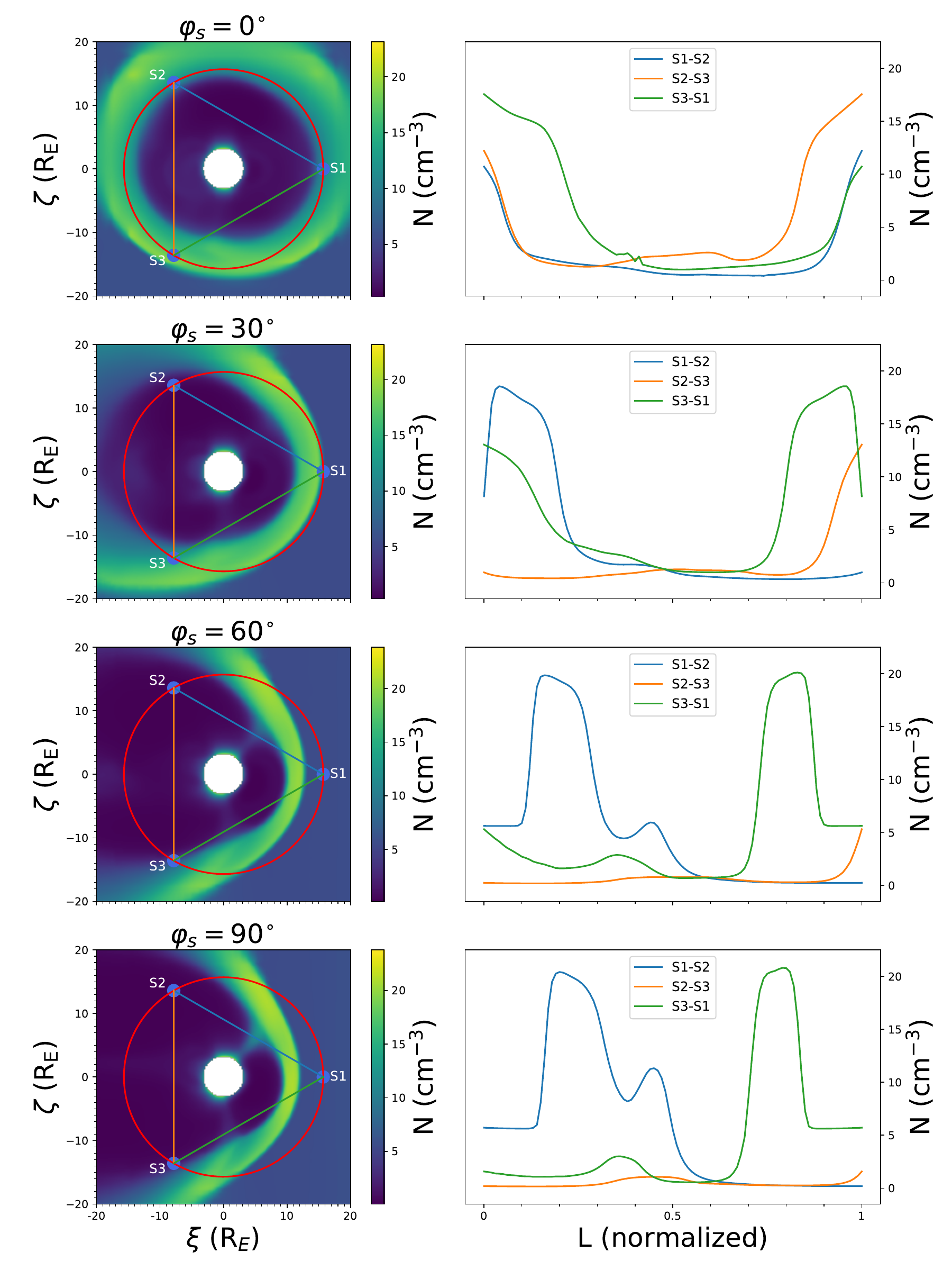}
\caption{The left panels are the number density distribution of electron on the orbital plane of the geocentric GW detector, the red circle is the TQ's orbit, the blue, orange, and green lines are the laser links; The right panels are the electron number density along the laser links \citep{Su2021}. }
\label{fig3-1-Overview}
\end{figure}

An MHD simulation model, SWMF \cite{Toth2005}, with the temporal resolution of 60 s and the finest spatial resolution of about 1/8 $\rm R_E$, is used to investigate the LPN for space-borne GW detection \cite{Lu2020,Su2021}.
Taking the solar wind parameters (e.g., plasma number density $n_{\rm e}$, bulk flow velocity of space plasma $v$, magnetic field $B$, etc.) as inputs, the SWMF simulation can yield the evolutionary characteristics of the plasma parameters in the vicinity of the orbital plane of the geocentric GW detectors, and the results of plasma parameters around TQ's orbit planes are shown in Figure \ref{fig3-1-Overview}.
The results of the SWMF simulation show the typical structures of the Earth's magnetosphere, such as the bow shock, the magnetopause, the magnetosheath, the cusp region, and so on. 
As shown in the Figure \ref{fig3-1-Overview}, the plasma number density $n_{\rm e}$ of the solar wind is higher than that in the magnetosphere. Since the bow shock, magnetosheath and magnetopause are formed by the compression of the Earth's magnetosphere by the solar wind blowing to the Earth, the magnetosheath, which is the downstream of the bow shock, is a compressed region, and its density is significantly higher than that of the other regions. 
The three satellites of TQ are deployed in geocentric orbit with an orbital altitude of about 100000 km, the three satellites form an equilateral triangular, and they are connected by the laser interferometric links. 
The normal of TQ's orbital plane is aligned with a white dwarf binary systems, J0806, and the angle between the orbital plane normal to the ecliptic plane is 4.7 degrees, it means that the orbital plane is almost perpendicular to the ecliptic plane \citep{Zhou2021PRD,Jia2023}. The angle between the normal of TQ orbital plane and the direction of the Sun-Earth line is defined as $\varphi_s$, which has a periodicity of 1 year. As shown in Figure \ref{fig3-1-Overview}, the laser links of TQ pass through various regions of the Earth's magnetosphere and solar wind, where $n_{\rm e}$ along the laser link increases as it passes through the magnetosheath.

\begin{figure}[h]
\centering
\includegraphics[width=0.9\textwidth]{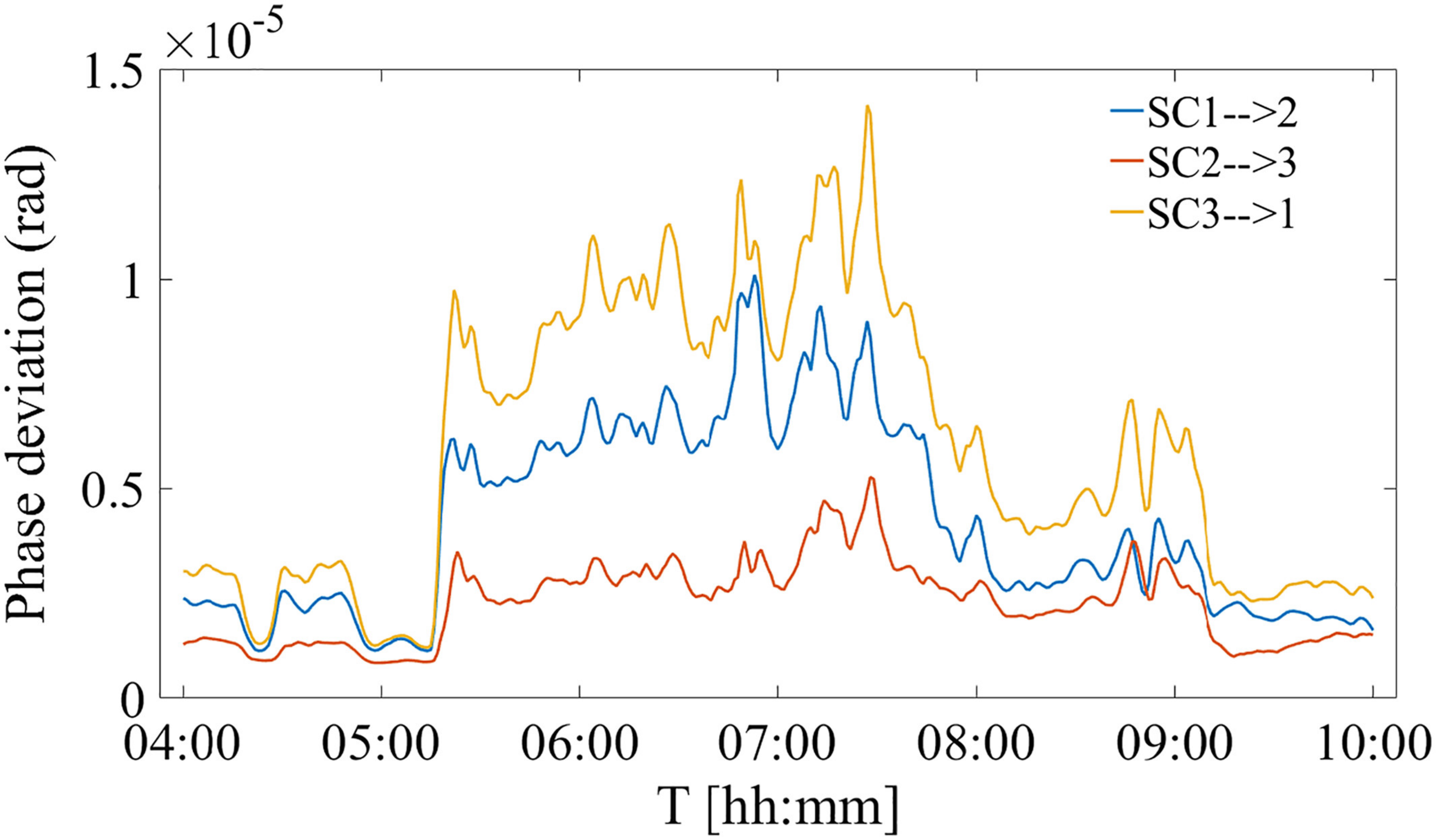}
\caption{The LPN time series of the single links for geocentric GW detector TQ \citep{Lu2020}. }
\label{fig3-Lu2020}
\end{figure}

The SWMF simulation is used to obtain plasma parameters evolution with a time length of 6 hours, and based on simulation results, the LPN of the TQ is calculated \citep{Lu2020}, and the time series of the LPN is shown in Figure \ref{fig3-Lu2020}. Note that the result is shown in phase deviation, the corresponding LPN needs to be multiplied by the laser wavelength $\lambda$, and the LPN is on the order of 1 pm. Based on the simulation result, the LPN of single links are calculated, and it showed that the LPN at 6 mHz is about 20\% of the TQ's displacement requirement \citep{Lu2020}. 
Changing the orbital altitude of TQ satellites from $10^5$ to $0.7\times 10^5$ and $1.5\times 10^5$ km, so that the arm length of the laser links are $0.7\sqrt{3}\times 10^5$ and $1.5\sqrt{3}\times 10^5$ km. The LPN in the cases of arm lengths $0.7\sqrt{3}\times 10^5$ and $1.5\sqrt{3}\times 10^5$ km are also investigated.
It is found that the LPN are about 14\% and 41\% of the TQ's displacement requirement for the arm lengths of $0.7\sqrt{3}\times 10^5$ and $1.5\sqrt{3}\times 10^5$, respectively \citep{Lu2020}.
This is because when the arm length increases, the proportion of the laser link in the solar wind and magnetosheath is higher than when the laser link is short. And considering that $n_{\rm e}$ in the solar wind and magnetosheath are about several times and tens times higher than in the magnetosphere, it leads to an increase in the ratio of LPN to TQ's displacement noise requirement as the arm length increases.

\begin{figure}[h]
\centering
\includegraphics[width=0.96\textwidth]{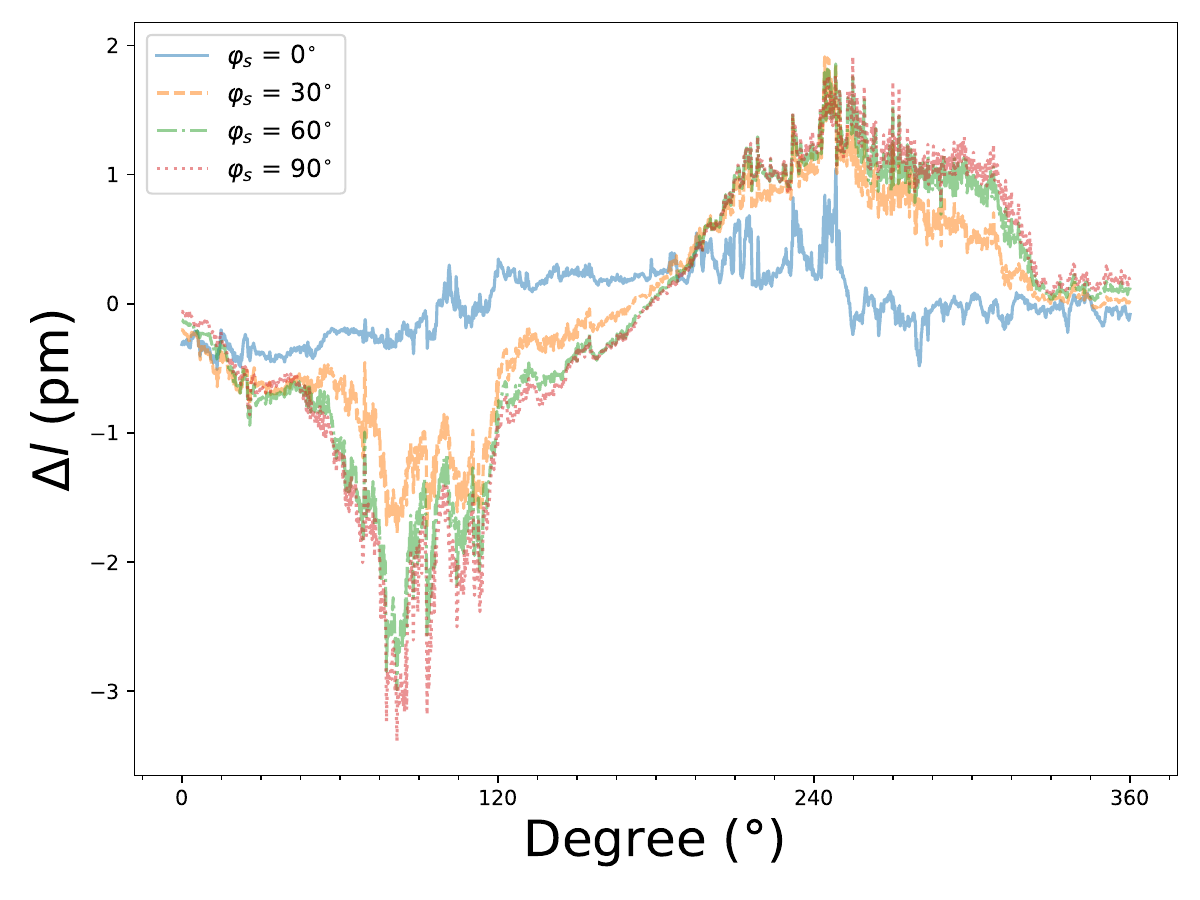}
\caption{The LPN time series of the Michelson combination for geocentric GW detector TQ \citep{Su2021}. }
\label{fig3-2}
\end{figure}

\begin{figure}
\centering
\includegraphics[width=0.9\textwidth]{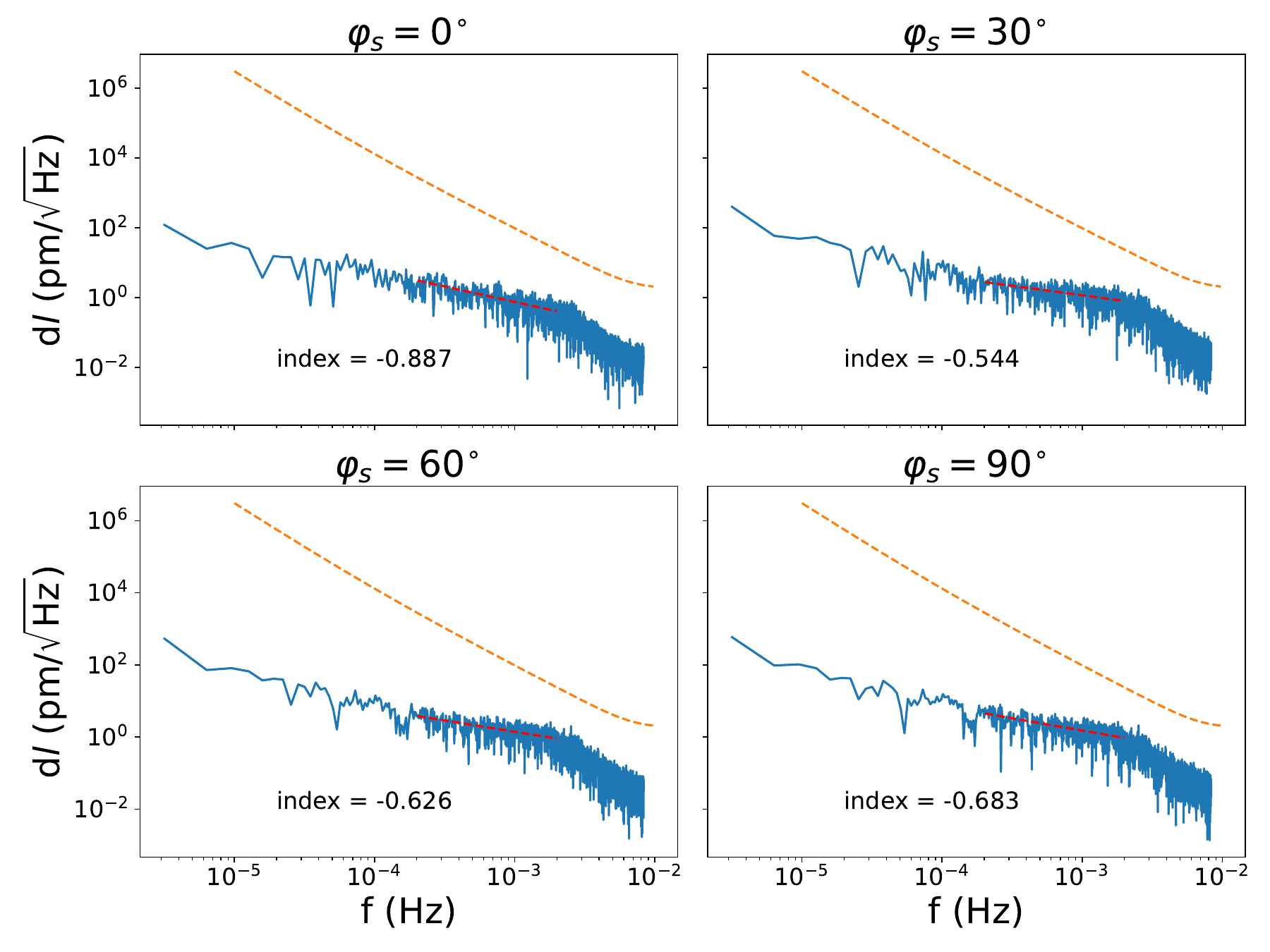}
\caption{The LPN's ASDs of the Michelson combination for geocentric GW detector TQ \citep{Su2021}. }
\label{fig3-3}
\end{figure}

In order to study the LPN over an entire TQ's cycle (3.65 days), the spatial and temporal distribution of the plasma during more than a cycle of TQ is obtained by using SWMF simulations. Based on the MHD simulation results, the LPN is investigated for $\varphi_s=0^{\circ}$, $30^{\circ}$, $60^{\circ}$, and $90^{\circ}$ cases, where the acute angle of $\phi_s$ is taken as $\varphi_s$ \citep{Su2021}.
Although the LPN noise of single links are important for analysing the laser propagation effect, the laser interferometric method is used in practical space-borne GW detection. Michelson interferometric combination is a typical laser interferometric combination.
The LPN in time domain for the single-link and Michelson combination cases in a TQ orbital cycle are shown in Figure \ref{fig3-2} \citep{Su2021}.
It shows that the LPN is more consistent in the time domain at $\varphi_s=30^{\circ}$, $60^{\circ}$, and $90^{\circ}$, whereas the LPN of $\varphi_s=0^{\circ}$ is smaller than those of $\varphi_s=30^{\circ}$, $60^{\circ}$, and $90^{\circ}$. 
As shown in Figure \ref{fig3-2}, the LPN is about 1 pm in the time domain for a single link and about $\pm 3$ pm for the Michelson combination. These results show that the LPN is on the order of magnitude of TQ's displacement requirement in the time domain, and suggest that the LPN is a noise source that deserves to be taken seriously. In addition, the ASD of the LPN for Michelson combination case is calculated and displayed in Figure \ref{fig3-3} \cite{Su2021}.
It shows that the LPN of Michelson combination is below the requirement of TQ's displacement noise over the full frequency band, and the ratio of the LPN to the requirement of TQ's displacement noise is largest at about 10 mHz with the value of about 30\%. It suggests that although LPN does not exceed the TQ's requirement in frequency domain, it is still a noise source that deserves attention.

In addition, wavefront distortion can be generated during laser propagation in space plasma. Based on the SWMF, the wavefront distortion is studied for TQ, and the result show that the wavefront distortion is on the order of 10$^{-9}$ rad, which is 3 orders of magnitude lower than the budget (10$^{-6}$ rad) of TQ \citep{Lu2020PlST}. It indicates that the wavefront distortion during laser propagation is neglected.

% \subsection{The laser propagation noise under different solar wind conditions}
% \label{sec3-2}

Various solar activities can cause variations of the plasma number density $n_{\rm e}$. For example, on long timescales, $n_{\rm e}$ of the solar wind has a periodicity of about 11 years \citep{Hao2015}, and on short timescales, when there is a solar eruption encounter the GW detectors, it can lead to dramatic changes of $n_{\rm e}$ along the laser link. The value of $n_e$ can differ by hundreds of times in different solar wind scenarios, leading to large differences in the LPN. Thus, it is therefore necessary to evaluate the LPN in different solar wind scenarios. Kp index is an significant index that reflects geomagnetic activity, the MHD simulation is used to study the LPN of TQ when Kp index is large, and find that when the Kp index is large, the LPN also increases \citep{Jing2022}.

For heliocentric GW detectors, such as LISA, \cite{Smetana2020} studied the LPN of LISA for 20 different solar wind conditions during 1997-1998. The time series data of $n_{\rm e}$ is got from the observations by the WIND/SWE instrument at the L1 point, and the duration of $n_{\rm e}$ for the 20 cases are at least 24 hours. Since the observations of Wind/SWE is in-situ, the distribution of $n_{\rm e}$ along the laser link of LISA cannot be obtained. Due to the lack of the spatial distribution of $n_e$, \cite{Smetana2020} assumed that $n_e$ along the LISA laser link arm is uniform with only temporal variations. The minimum value $n_{\rm e}$ for the 20 events they is on the order of 0.1 $\rm cm^{-3}$, and the maximum value reaches the order of 10 $\rm cm^{-3}$; 6 of the total 20 events are eruptions events such as CMEs and magnetic storms. In addition, the ASDs of the LPN is calculated, and compare the ASDs with the displacement measurement requirement of LISA, it shows that the LPN of LISA is about several times higher than the displacement noise measurement of LISA \citep{Smetana2020}.

\begin{figure}[h]
\centering
\includegraphics[width=0.80\textwidth]{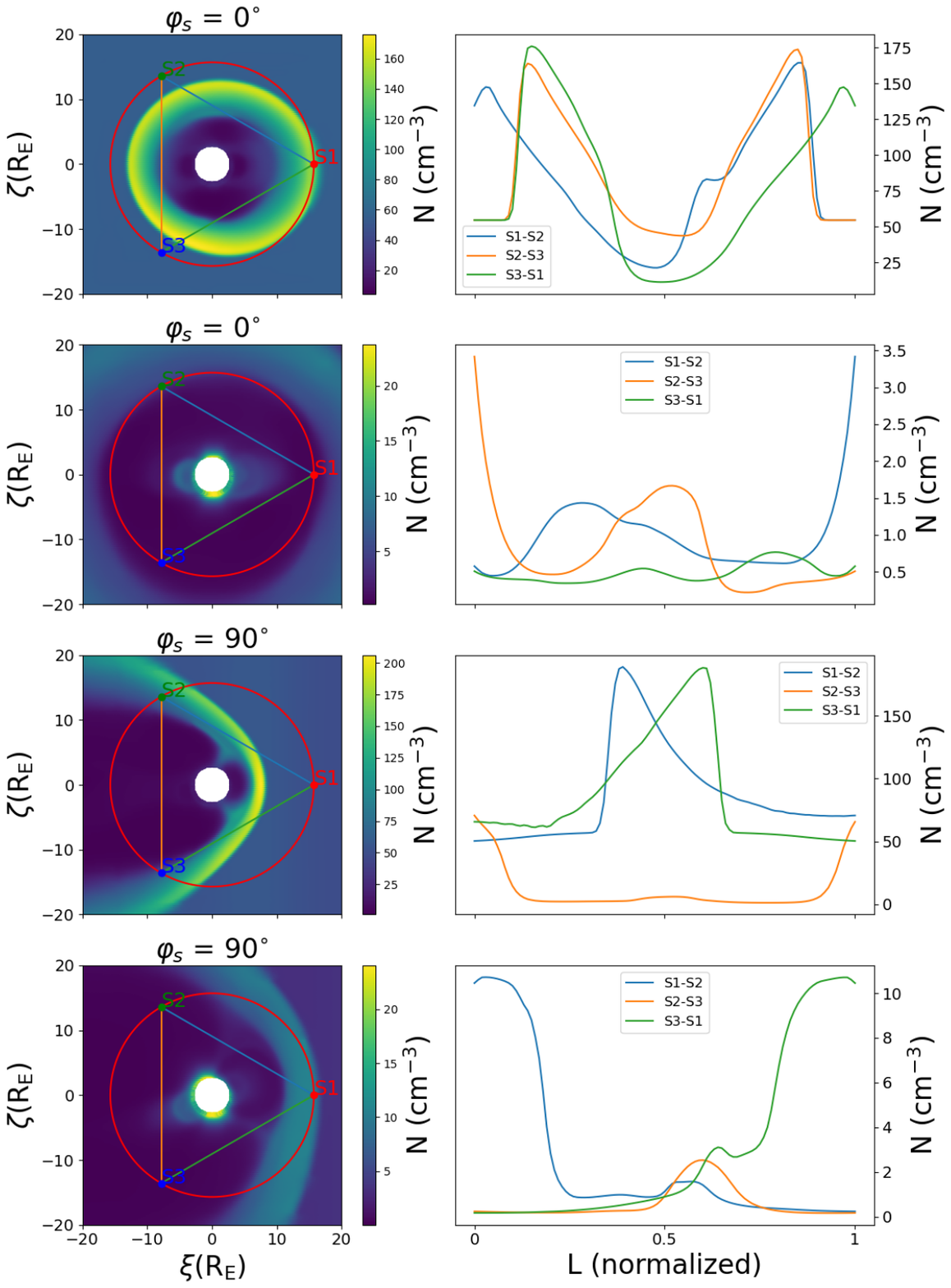}
\caption{Similar to Figure \ref{fig3-1-Overview}. The first and third rows are the number density distribution on the orbital plane and along the laser links for $\overline{P}_{\rm dyn}$ is large, The second and forth rows are the number density distribution on the orbital plane and along the laser links for $\overline{P}_{\rm dyn}$ is small \citep{Liu2024}. }
\label{fig3-4}
\end{figure}

For geocentric GW detectors, such as TQ, \cite{Liu2024} investigates the LPN based on SWMF for solar wind dynamic pressure are small ($\overline{P}_{\rm dyn} \approx 0.79$ nPa) and large ($\overline{P}_{\rm dyn} \approx 5.05$ nPa). The MHD simulation results are shown in the left panels of Figure \ref{fig3-4}, which illustrate the cases of large $P_{\rm dyn}$ (rows 1 and 3) and small $P_{\rm dyn}$ (rows 2 and 4) for $\varphi_{\rm s} = 0^{\circ}$ (upper panels) and $\varphi_{\rm s} = 90^{\circ}$ (lower panels), respectively. 
For strong $P_{\rm dyn}$, on $\varphi_{\rm s} = 0^{\circ}$ orbital plane, the Earth's magnetosphere is compressed to within about 16 $\rm R_E$, and TQ's orbit is almost outside the bow shock; on the $\varphi_{\rm s} = 90^{\circ}$ orbital plane, the subsolar point is compressed to within about 10 $\rm R_E$. For weak $P_{\rm dyn}$, on $\varphi_{\rm s} = 0^{\circ}$ orbital plane, the bow shock and magnetopause are expanded beyond about 16 and 20 $\rm R_E$, respectively, and TQ's orbit is almost inside magnetopause; on the $\varphi_{\rm s} = 90^{\circ}$ orbital plane, the subsolar point is expanded beyond about 16 $\rm R_E$, and TQ's orbit is almost inside the bow shock. For strong $P_{\rm dyn}$, $n_{\rm e}$ along the laser link increases and the value can exceed 100 $\rm cm^{-3}$; for $P_{\rm dyn}$, $n_{\rm e}$ along the laser link decreases to the order of 1 $\rm cm^{-3}$. 
% The difference of the maximum of $n_{\rm e}$ for large $P_{\rm dyn}$ and small $P_{\rm dyn}$ is more than one order of magnitude.
The difference between the maximum of $n_{\rm e}$ for the large $P_{\rm dyn}$ and for the small $P_{\rm dyn}$ is more than an order of magnitude.

% \begin{figure}[h]
% \centering
% \includegraphics[width=0.9\textwidth]{O-t-single-Michelson-cases.pdf}
% \caption{O-asdf}
% \label{fig3-5}
% \end{figure}

Combining the simulation result of $n_{\rm e}$ and Equation \eqref{eq:dl}, the time series of LPN for strong and weak $P_{\rm dyn}$ can be obtained. 
% As shown in Figure \ref{fig3-5}, 
The LPN can reach 10 pm when $P_{\rm dyn}$ is strong, while the LPN is about 1 pm when $P_{\rm dyn}$ is weak. The LPN under strong $P_{\rm dyn}$ is about 10 times that under weak $P_{\rm dyn}$. In addition, the ASDs of LPN can be obtained.
% the ASDs of LPN are shown in Figure \ref{fig3-6}. 
For $P_{\rm dyn}$ is weak, the ratio of the LPN to requirement of TQ’s displacement noise is largest at about 10 mHz with the value of about 10\%. Whereas, for $P_{\rm dyn}$ is strong, the ratio of the LPN to requirement of TQ’s displacement noise is largest at about 10 mHz with the value of nearly 100\%. 
The strong $P_{\rm dyn}$ event corresponds to a strong solar eruption event, the well known Halloween event, and in general, $P_{\rm dyn}$ is strong when the solar eruption is strong, which indicates that the effect of the LPN noise is crucial when a strong TQ encounters a strong solar eruption.
% it suggests that 

% \begin{figure}[h]
% \centering
% \includegraphics[width=0.9\textwidth]{O-ASD-Michelson-cases.pdf}
% \caption{asdf}
% \label{fig3-6}
% \end{figure}

\subsection{Statistical study of laser propagation noise}
\label{sec3-3}

The planned run time of the space-borne GW detection is more than 4 years and, if everything goes well, the overall run time can be up to more than 10 years \citep{LISA2017}, which is about 1 solar cycle. Statistical studies of the LPN of space-borne GW detections are needed to assess the LPN noise for various solar activity scenarios over a long time scale.

For heliocentric GW detector, LISA, based on the data of $n_{\rm e}$ for more than 600 days that obtained by Wind/SWE, the evaluation of the LPN noise is studied \citep{Jennrich2021}. By assuming that the spectrum index of $n_{\rm e}$ in the solar wind is consistent with the spectra indices of the space magnetic field and plasma velocity, both of which roughly obey the Kolmogorov spectrum, and under the conditions of the Taylor hypothesis \citep{Perez2021}, 
the relationship between the spectrum of the local plasma number density ($S_{\mathrm{Ne}} (f)$) and the spectrum of the plasma number density on the laser link ($S_{L}(f)$) was derived \cite{Jennrich2021},
\begin{equation}
    S_{L}(f) \approx(L \chi)^{2}\left(\frac{25}{9}\right) \beta^{5 / 3}\left(\frac{V}{2 \pi L f}\right) S_{\mathrm{Ne}}(f)
    \label{eq:dl-PSD-sw}
\end{equation}
where, $L$ is the arm length of heliocentric GW detector, $V$ is the bulk flow velocity of space plasma, $\beta$ is geometrical factor, $f$ is frequency, $\chi = \frac{\lambda^{2} e^{2}}{2 \pi m_{\mathrm{e} } c^{2}}$. Here, $\lambda$ is laser wavelength, $c$ is the light speed, $m_{\mathrm{e} }$ is the mass of electron.

\begin{figure}[h]
\centering
\includegraphics[width=0.96\textwidth]{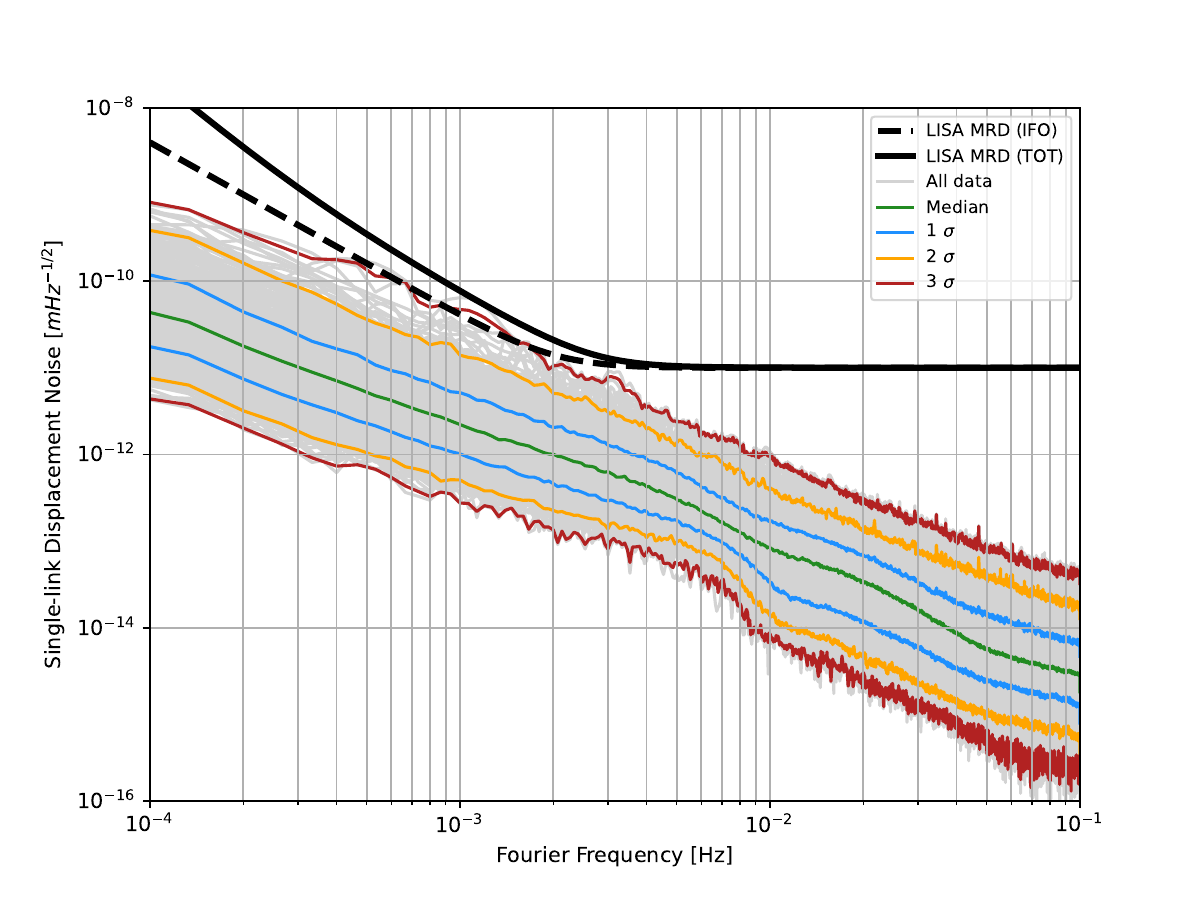}
\caption{The LPN's ASDs of single link for LISA based on the in-situ observation of Wind \citep{Jennrich2021}. The gray curves are the daily ASDs of the LPN. The black curves are the single link sensitivity requirements of LISA \citep{Jennrich2021}. }
\label{fig3-LISA}
\end{figure}

Based on Equation \eqref{eq:dl-PSD-sw} and the observations of Wind/SWE, about 600 daily ASDs of LPN for the single link between 1997 and 1998 is obtained \cite{Jennrich2021}. The statistical results of LPN for LISA are shown in Figure \ref{fig3-LISA}, the median ASD of LPN is about 1 order lower than the displacement noise requirement of LISA \cite{Jennrich2021}. However, beyond the 3-sigma interval, the LPN noise is close to the displacement noise requirement of LISA, which indicates that the LPN noise has a serious impact on LISA under some special solar wind conditions, although the incidence of such events is not high.

\begin{figure}[h]
\centering
\includegraphics[width=0.9\textwidth]{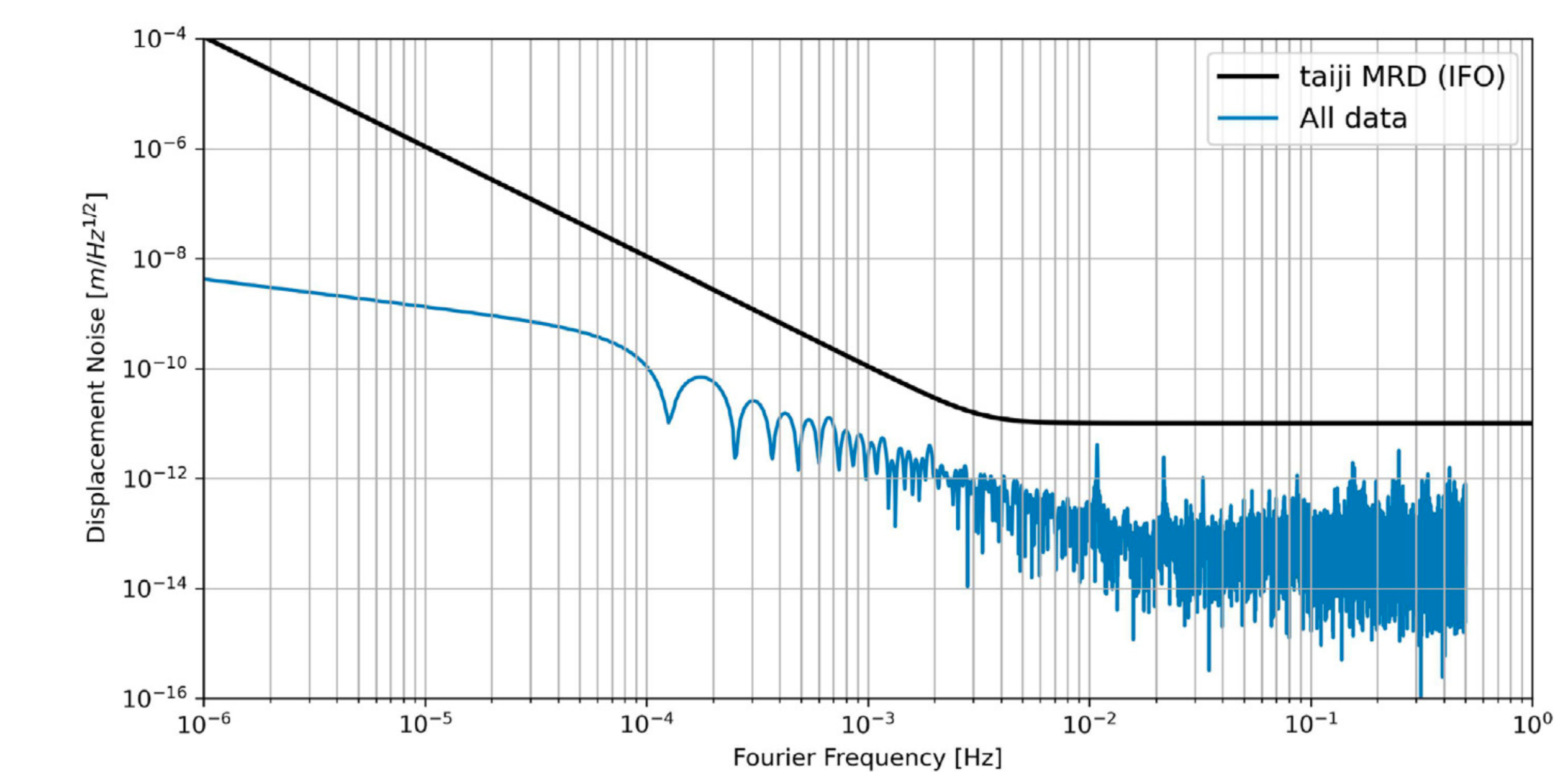}
\caption{The LPN's ASDs of single link over the entire period for TJ based on the in-situ observation of Wind is represented as blue curves. The black curve is the single link sensitivity requirements of TJ. \citep{Xie2024}
}
\label{fig3-TJ}
\end{figure}

Based on Lomb–Scargle spectral analysis method, and 6.5-year data of $n_{\rm e}$, the LPN of TJ is studied, and the ASDs of the LPN is shown in Figure \ref{fig3-TJ} \citep{Xie2024}. Due to solar wind conditions of TJ is similar to LISA, the LPN of TJ \citep{Xie2024} is similar to that of LISA \citep{Jennrich2021}.

\begin{figure}[h]
\centering
\includegraphics[width=0.75\textwidth]{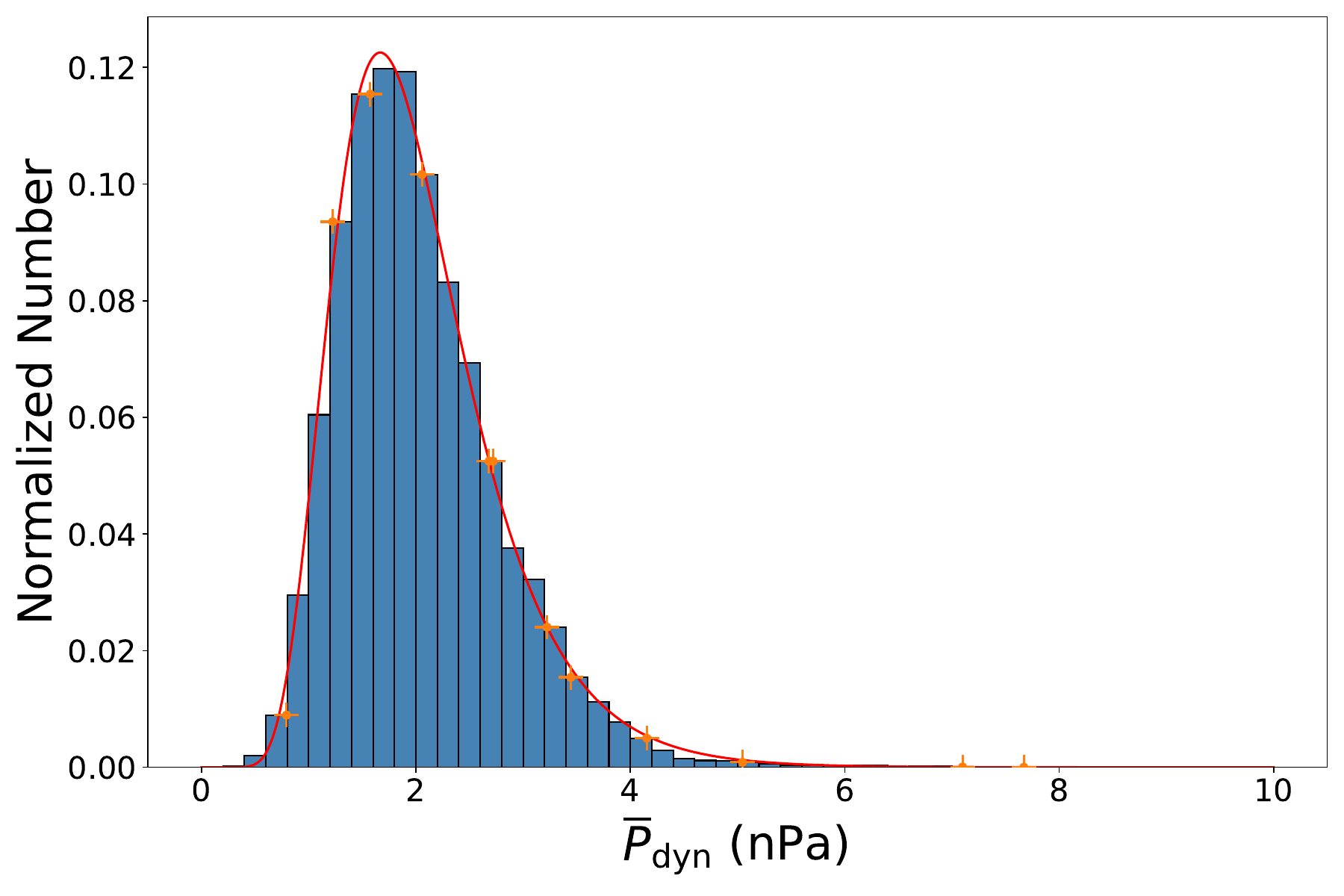}
\caption{Histogram of the solar wind $\overline{P}_{\rm dyn}$ from 1999 to 2021 \citep{Liu2024}. The cases studied in \citep{Liu2024} are marked as 12 orange pluses.}
\label{fig3-histPdyn}
\end{figure}

For geocentric GW detector TQ, due to the multiple structures of the Earth's magnetosphere, which are more complex than the solar wind, it is not possible to use Equation \eqref{eq:dl-PSD-sw} for analytical calculations, but instead requires MHD simulations.
In order to assess the LPN of TQ throughout a solar cycle, the SWMF is used to simulate the plasma parameters in the vicinity of TQ's orbit for 12 different scenarios of solar wind conditions \citep{Liu2024}. The space weather parameters (${P}_{\rm dyn}$, ${\rm Kp}$, ${\rm AE}$, ${\rm Sym\text{-}H}$, ${B}_{\rm y}$, ${B}_{\rm z}$) are different for each of these 12 events. Figure \ref{fig3-histPdyn} shows the mean solar wind dynamic pressure $\overline{P}_{\rm dyn}$ over 23 years (from 1999 to 2021), $\overline{P}_{\rm dyn}$ of the 12 events are marked with a plus sign in Figure \ref{fig3-histPdyn}, which cover almost the entire distribution of $\overline{P}_{\rm dyn}$ in more than two solar cycles. 

\begin{figure}[h]
\centering
\includegraphics[width=0.96\textwidth]{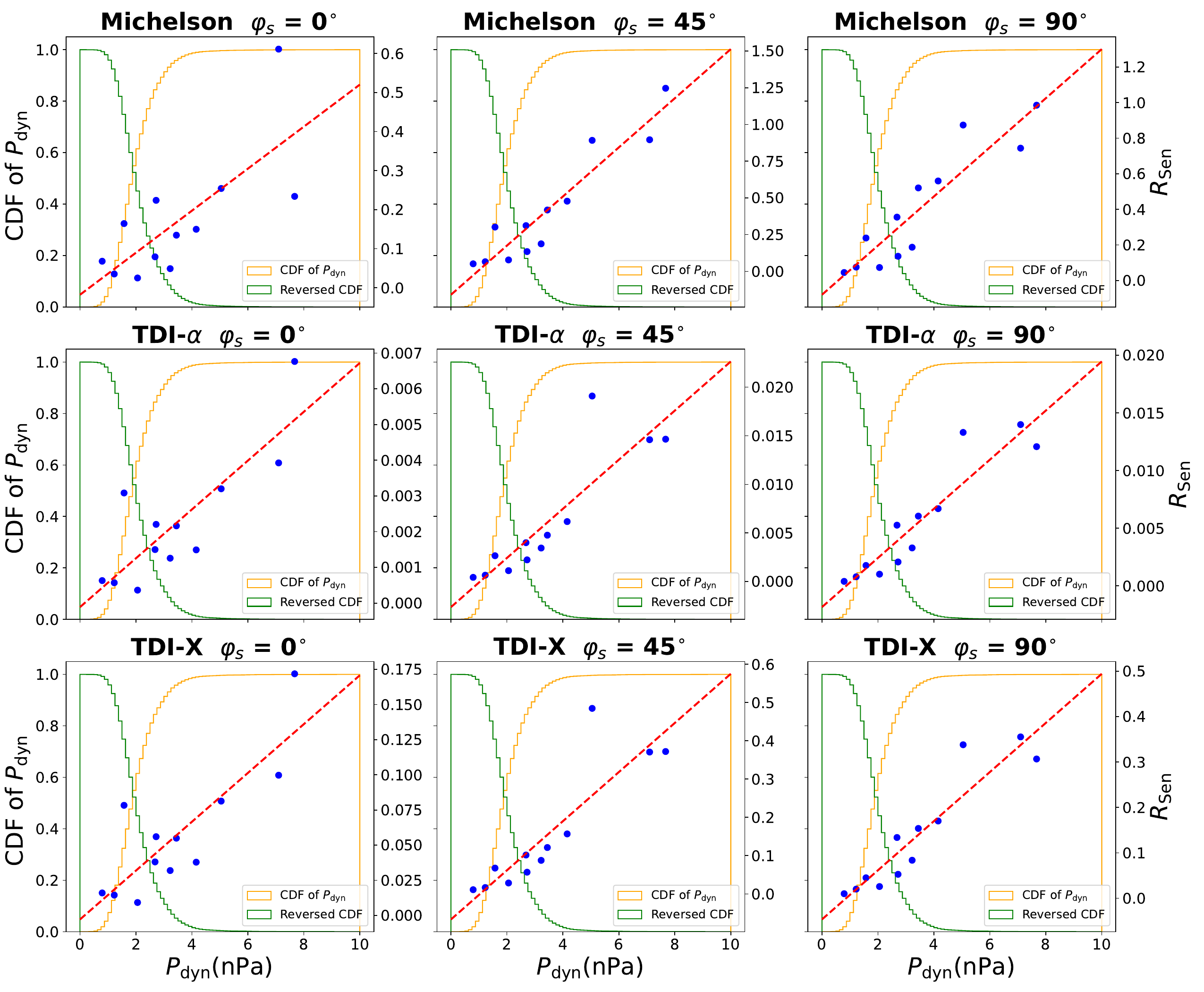}
\caption{The CDF and reversed CDF of $\overline{P}_{\rm dyn}$, and the relationship between the amplitude of LPN's ASDs and $\overline{P}_{\rm dyn}$ \citep{Liu2024}. }
\label{fig3-cc-lin}
\end{figure}

In addition, the ASD of the Michelson combination of LPN for these events is calculated, and fits the ASDs of these events by using $\mathrm{ASD} =a f ^ b $ to obtain the amplitude $a$ and spectral index $b$ \citep{Liu2024}. The correlation coefficients of $a$ and $b$ of these events with solar wind parameters (${P}_{\rm dyn}$, ${\rm Kp}$, ${\rm AE}$, ${\rm Sym\text{-}H}$, ${B}_{\rm y}$, ${B}_{\rm z}$) are calculated. The linear correlation coefficient between $b$ and all solar wind parameters is found to be less than 0.4, and none of the correlations are apparent. The linear correlation coefficients of ${B}_{\rm y}$ and ${B}_{\rm z}$ with $a$ are also less than 0.4. While the linear correlation coefficient of other parameters (${P}_{\rm dyn}$, ${\rm Kp}$, ${\rm AE}$, ${\rm Sym\text{-}H}$) with $a$ are $\gtrsim$ 0.7, among which, the linear correlation of ${P}_{\rm dyn}$ with $a$ is the highest, reaching about 0.9. 
Due to the maximum of the ratio of LPN to the requirement of TQ's displacement noise in the frequency domain is around 10 mHz, the ratio of LPN to requirement of TQ's displacement noise at 10 mHz is marked as $R_\mathrm{O10mHz}$. As shown in Figure \ref{fig3-cc-lin}, the relationship between ${P}_{\rm dyn}$ and $R_\mathrm{O10mHz}$ is illustrated. Based on the linear correlation between ${P}_{\rm dyn}$ and $R_\mathrm{O10mHz}$ and the distribution of ${P}_{\rm dyn}$ over more than 2 solar cycles, the impact of LPN on TQ throughout the entire solar cycle is evaluated \citep{Liu2024}. The occurrence rates of $R_\mathrm{O10mHz}$ exceeding 20\% and 30\% are approximately 40\% and 15\%, respectively. Overall, the impact of LPN of Michelson combination on TQ is roughly within acceptable limits, but still requires attention.

\subsection{Suppression of laser propagation noise}
\label{sec3-4}

As shown in Equation \eqref{eq:dt}, the dispersion effect of EM waves propagating in space plasma can cause time delay $\Delta \tau$, which is proportional to the integration of $n_{\rm e}$ along the laser link ($\int_L n_{\mathrm{e} } \mathrm{d} s$). To deduct LPN noise, the integral of $n_{\rm e}$ along the laser link needs to be obtained.

Multiple measurements of the time $\tau$ or phase $\phi$ signals of multiple frequencies EM waves can obtain $\int_L n_{\mathrm{e} } \mathrm{d} s$. From Equation \eqref{eq:dt}, the time moments ($t_1$, $t_2$) required for EM waves with frequencies ($f_1$, $f_2$, and $f_1 \ne f_2$) to propagate in the same plasma background are as follows,
\begin{equation}
\left\{
\begin{aligned}
\tau_1 &= \frac{L}{c} + \frac{K}{2 c f_1^2} N_{\rm T}   \\
\tau_2 &= \frac{L}{c} + \frac{K}{2 c f_2^2} N_{\rm T}    \\
\end{aligned}
\right.
\label{eq:dt-f12}
\end{equation}
here, $N_{\rm T} = \int_{L} n_{\mathrm{e}} \mathrm{d} s$, is the total electron content (TEC). $\tau_1$ and $\tau_2$ can be measured, thus the difference between $\tau_1$ and $\tau_2$ can be derived. Therefore, $N_{\rm T}$ can also be obtained as follow,
\begin{equation}
    N_{\mathrm T} = \frac{2c}{K} \frac{f_1^2 f_2^2}{f_2^2 - f_1^2} (\tau_1 - \tau_2)
    \label{eq:TEC-DGD}
\end{equation}
Thus, we can deduct the dispersion effect by differential group delay (DGD).

We can also obtain TEC from the perspective of phase. The phase variations $\phi$ of EM waves with frequency $f$ after propagating a distance $L$ is,
\begin{equation}
    \phi = \frac{2 \pi f L}{c/N} = \frac{2 \pi f L}{c} \int N \mathrm{d} l
\end{equation}
Considering Equation \eqref{eq:A-H-f_p}, the phase variations $\phi_1$ and $\phi_2$ of EM waves with frequencies $f_1$ and $f_2$ after propagating a distance $L$ in the same plasma background are as follows,
\begin{equation}
\left\{
\begin{aligned}
\phi_1 &= \frac{2 \pi f_1 L}{c} + \frac{\pi K}{c f_1} N_\mathrm{T}  \\
\phi_2 &= \frac{2 \pi f_2 L}{c} + \frac{\pi K}{c f_2} N_\mathrm{T}  \\
\end{aligned}
\right.
\label{eq:phi-f12}
\end{equation}
Ignoring the integer ambiguity issue, $N_\mathrm{T}$ is as follows,
\begin{equation}
    N_\mathrm{T} = \frac{c}{\pi K}\frac{f_1^2 f_2^2}{f_1^2-f_2^2}\left(\frac{\phi_1}{f_1} - \frac{\phi_2}{f_2} \right)
    \label{eq:TEC-DCP}
\end{equation}
Since we can measure $\phi_1$ and $\phi_2$, thus, we can get $N_\mathrm{T}$ along the laser link, and deduct the dispersion effect by differential carrier phase (DCP).

DGD can measure the absolute value of TEC, and DCP is suitable for measuring the relative variation of TEC. The methods of multiple measurements of the time (DGD) or phase (DCP) signals of EM waves of different frequencies to get $N_{\rm T}$ along the laser link are dual-frequency scheme of the dissipation for dispersion. The dual-frequency scheme and similar methods are used in fields such as ionosphere, global navigation satellite system (GNSS) and radio astronomy \citep{Tapley2004,Yang2011,Petroff2019,Landerer2020}.

In the space-borne GW detectors, the laser frequency $f \approx 2.82 \times 10^{14}$ Hz (wavelength $\lambda \approx 1064 $ nm). The difference between the sidebands of the laser and the carrier frequency are in the GHz range, thus, $\Delta f/f \sim 10^{-5}$. 
Taking the typical $n_{\rm e} \sim \rm 10~ cm^{-3}$, the typical arm length $L \sim 10^5$ and $10^6$ km for TQ and LISA, respectively. 
Thus, $N_{\rm T} \sim 10^{15}$ to $10^{16}$ m$^{-2}$.

Taking the carrier frequency as $f_1$, and sidebands frequency as $f_2$. Thus, $\Delta f/f = (f_1-f_2)/f \sim 10^{-5}$, and $f_1 \approx f_2$.
According to Equation \eqref{eq:TEC-DGD}, the time measurement accuracy requirement for eliminating TEC by using DGD method can be estimated as follow,
\begin{equation}
    N_\mathrm{T} \approx \frac{c}{K} \frac{f}{\Delta f}f^2 \Delta t
    \label{eq:TEC-DGD-dt}
\end{equation}
here, $\Delta t = \tau_1 - \tau_2$. According to Equation \eqref{eq:TEC-DGD-dt} and the parameters above, $\Delta t $ can be estimated with the value of about $10^{-28}$ to $10^{-27}$ s. 
Similarly, according to Equation \eqref{eq:TEC-DCP}, the phase measurement accuracy requirements for eliminating TEC by using DCP method can be estimated as follow,
\begin{equation}
    N_\mathrm{T} \approx \frac{c}{\pi K} \frac{f}{\Delta f} f \Delta \phi
    \label{eq:TEC-DCP-dphi}
\end{equation}
here, $\Delta \phi = \phi_1 - \phi_2$. According to Equation \eqref{eq:TEC-DCP-dphi} and the parameters above, $\Delta \phi $ can be estimated with the value of about $10^{-11}$ to $10^{-10}$ rad. 

With current human technology, time measurement accuracy ability is far from reaching $\Delta t \approx 10^{-28}$ to $10^{-27}$ s, and phase measurement accuracy is far from reaching $\Delta \phi \approx 1.6\times10^{-10}$ to $1.6\times10^{-11}$ rad. 
It indicates that using laser carrier and sideband frequencies for the DGD and DCP methods cannot eliminate the LPN noise in space-borne GW detection.

\begin{figure}[h]
\centering
\includegraphics[width=0.96\textwidth]{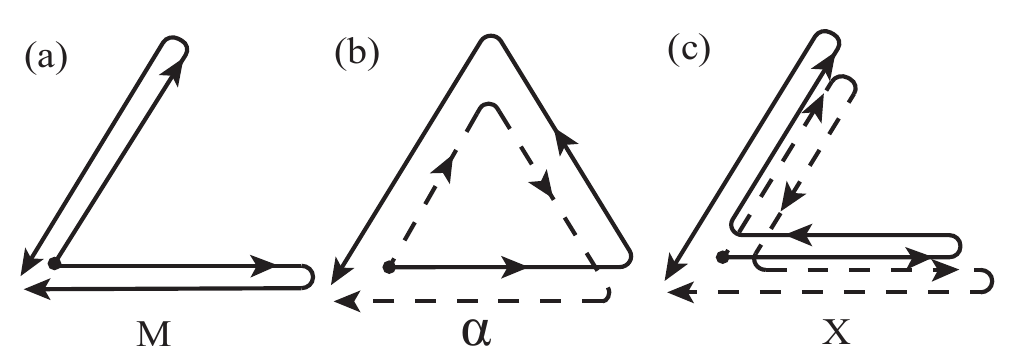}
\caption{The schematic of Michelson, TDI--$\alpha$, and TDI--$X$ combinations \citep{Su2021}. }
\label{fig3-TDIpath}
\end{figure}

Besides dual-frequency scheme, \citep{Su2021} have revealed that time-delay interferometry (TDI) can suppress the OPD noise caused by laser propagation in space plasma.
TDI was proposed to address the challenges of space-based gravitational wave detection, specifically the overwhelming laser frequency noise in missions like LISA \citep{Armstrong1999,Estabrook2000,Tinto1999}. 
By constructing delayed interferometric combinations of signals, TDI can effectively cancel laser frequency noise and compensating for arm length variations \citep{Tinto2005,Tinto2023-2}.
TDI can also help mitigate other noise sources, such as clock noise, in GW detections \citep{Tinto2014,Tinto2023-n}. 
The schematic diagrams of interferometry combination, Michelson, TDI--$\alpha$, and TDI--$X$ are shown in Figure \ref{fig3-TDIpath} \citep{Su2021}.

Here, denoting the phase noise cause by the laser propagation in the space plasma as,
\begin{equation}
    s_{ij}(t) = \frac{2 \pi \Delta l_{ij}(t) }{\lambda}
\end{equation}
where, footnotes $i$ and $j$ denote satellites $i$ and $j$, $\Delta l_{ij}(t) $ is the LPN for a single link between satellites $i$ and $j$.

The distance between satellites $i$ and $j$ is set as $L_{ij}$, and set the light speed $c = 1$.
For TQ, $L_{ij} = L \approx 0.6$ s, and the temporal resolution of the MHD simulation $\Delta t$ = 60 s \citep{Su2021}, $\Delta t \gg L$. It suggest that MHD simulations are unable to resolve space plasma disturbances during laser propagation here, so reasonable to approximate $\int^{i}_{j}n_{\mathrm{e}} \mathrm{d}s = \int^{j}_{i}n_{\mathrm{e}}\mathrm{d}s$. After which, for TDI–$\alpha$ combination, $s_{\alpha}$ is as follow \citep{Su2021}, 
\begin{equation}
\label{eq:TDI-alpha1}
\begin{aligned}
s_{\alpha} = s_{12}(t - 2L)+s_{31}(t) - s_{12}(t)-s_{31}(t - 2L)
\end{aligned}
\end{equation}
And for TDI–$\alpha$ combination, $s_{X}$ is as follow \citep{Su2021},
\begin{equation}
\label{eq:TDI-X1}
\begin{aligned}
s_{X} = s_{12}(t-3L)+s_{12}(t-2L)+s_{31}(t-L)+s_{31}(t) \\
-s_{12}(t-L)-s_{12}(t)-s_{31}(t-3L)-s_{31}(t-2L)  \,.
\end{aligned}
\end{equation}

Since $\Delta t \gg L \approx 0.6$ s, the time delay of $s_{ij}$ is got by linear interpolation, $s_{ij}(t -  \delta t) =  s_{ij}(t) + (s_{ij}(t -  \Delta t)- s_{ij}(t))(\delta t/\Delta t)$. Thus, $s_{\alpha}$ and $s_{X}$ is reduced as,
\begin{equation}
\label{eq:TDI-alhpa2}
\begin{aligned}
s_{\alpha} = (s_{12}(t - \Delta t)- s_{12}(t) )\frac{2L}{\Delta t} - (s_{31}(t - \Delta t) - s_{31}(t))\frac{2L}{\Delta t} \,,
\end{aligned}
\end{equation}
\begin{equation}
\label{eq:TDI-X2}
\begin{aligned}
s_{X} =  2s_{\alpha}
\end{aligned}
\end{equation}

\begin{figure}[ht]
\centering
\begin{minipage}[b]{\textwidth}
\centering
    \begin{minipage}[b]{0.411\textwidth}
    \includegraphics[width = 5.21 cm]{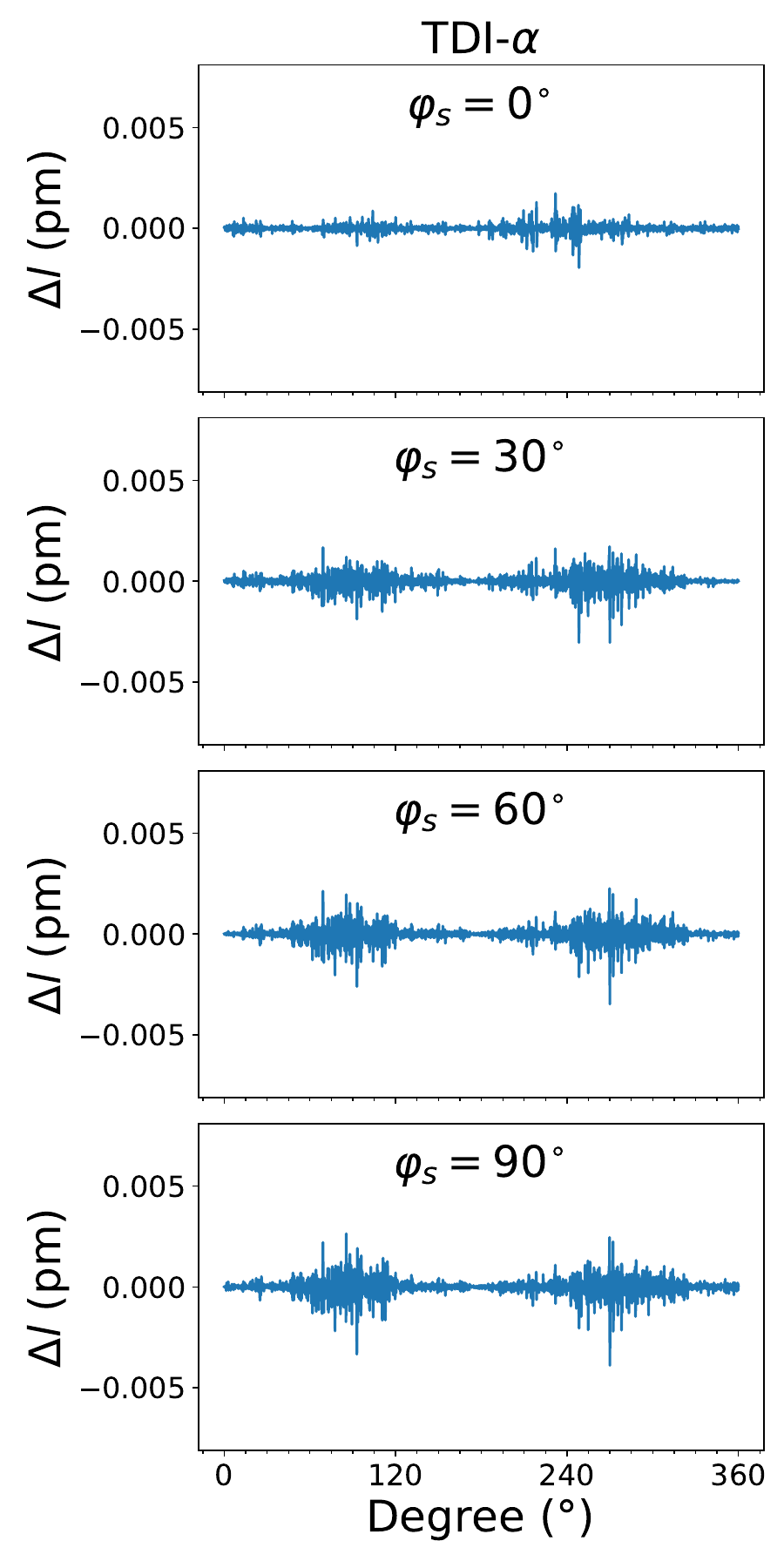}
    \end{minipage}
    \begin{minipage}[b]{0.411\textwidth}
    \includegraphics[width = 5.21 cm]{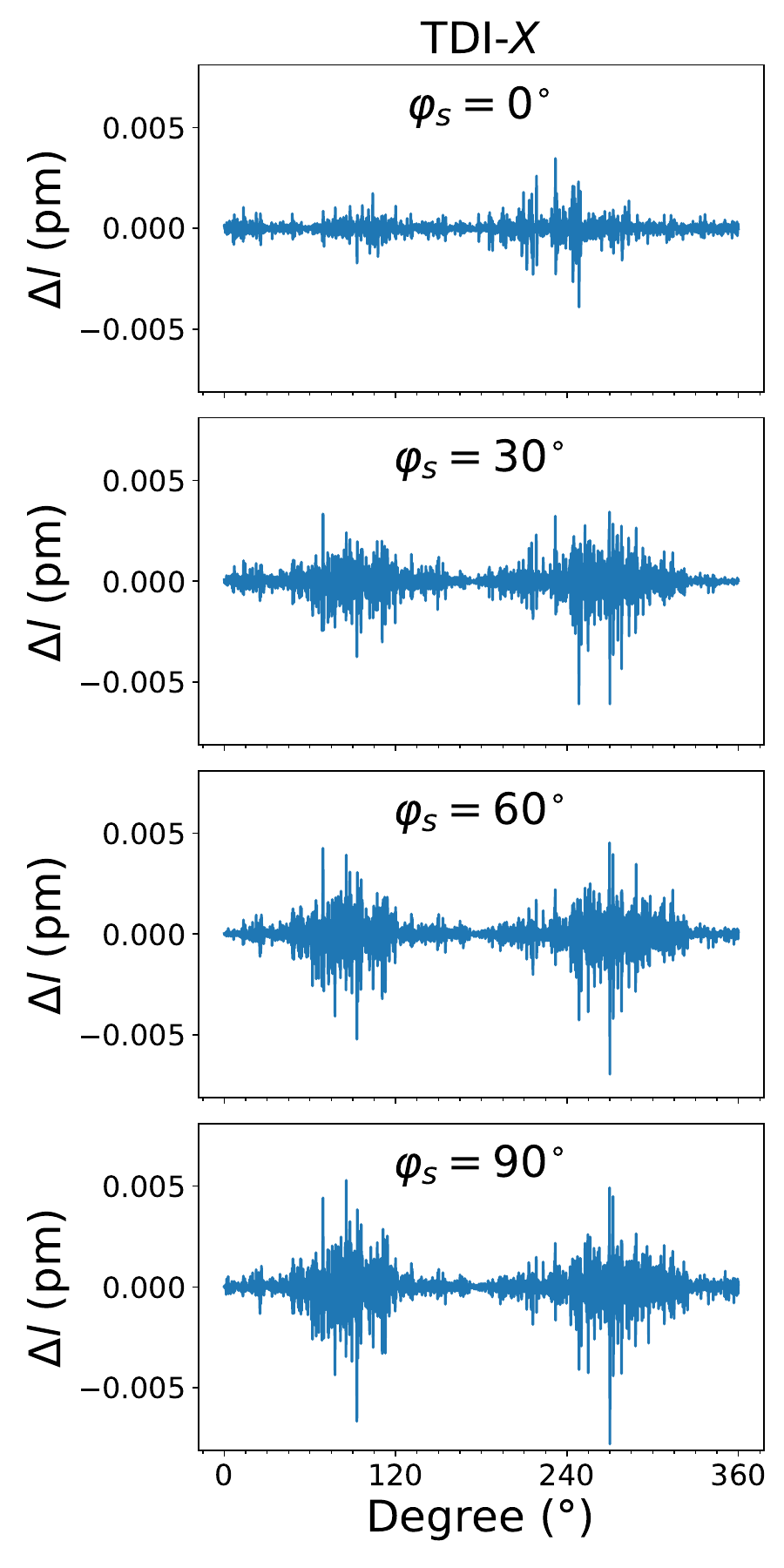}
    \end{minipage}
\end{minipage}
\caption{The LPN's time series of TDI--$\alpha$ (left panels) and TDI--$X$ (right panels) combinations for geocentric GW detectors TQ \citep{Su2021}. }
\label{fig3-TDI-t}
\end{figure}

According to Equations $\eqref{eq:TDI-alhpa2}$ and $\eqref{eq:TDI-X2}$, the LPN of TDI--$\alpha$ and --$X$ combinations can be calculated.
Figure \ref{fig3-TDI-t} shows the time series results of the LPN noise for the TDI combinations of TQ. The left and right columns are for the TDI--$\alpha$ and TDI--$X$ combinations, respectively. And rows 1, 2, 3, and 4 show the results of $\varphi_s$=0$^{\circ}$, 30$^{\circ}$, 60$^{\circ}$, and 90$^{\circ}$, respectively. Comparing Figures \ref{fig3-2} and \ref{fig3-TDI-t}, it can be seen that the LPN of the TDI combinations are about 0.001 to 0.01 pm, which is 2 to 3 orders of magnitude lower than that of the Michelson combination, suggesting that TDI can significantly suppress the LPN in the time domain.

\begin{figure}[ht]
%	\figurenum{7}
\centering
\begin{minipage}[b]{\textwidth}
\centering
    \begin{minipage}[b]{0.411\textwidth}
    \includegraphics[width = 5.21 cm]{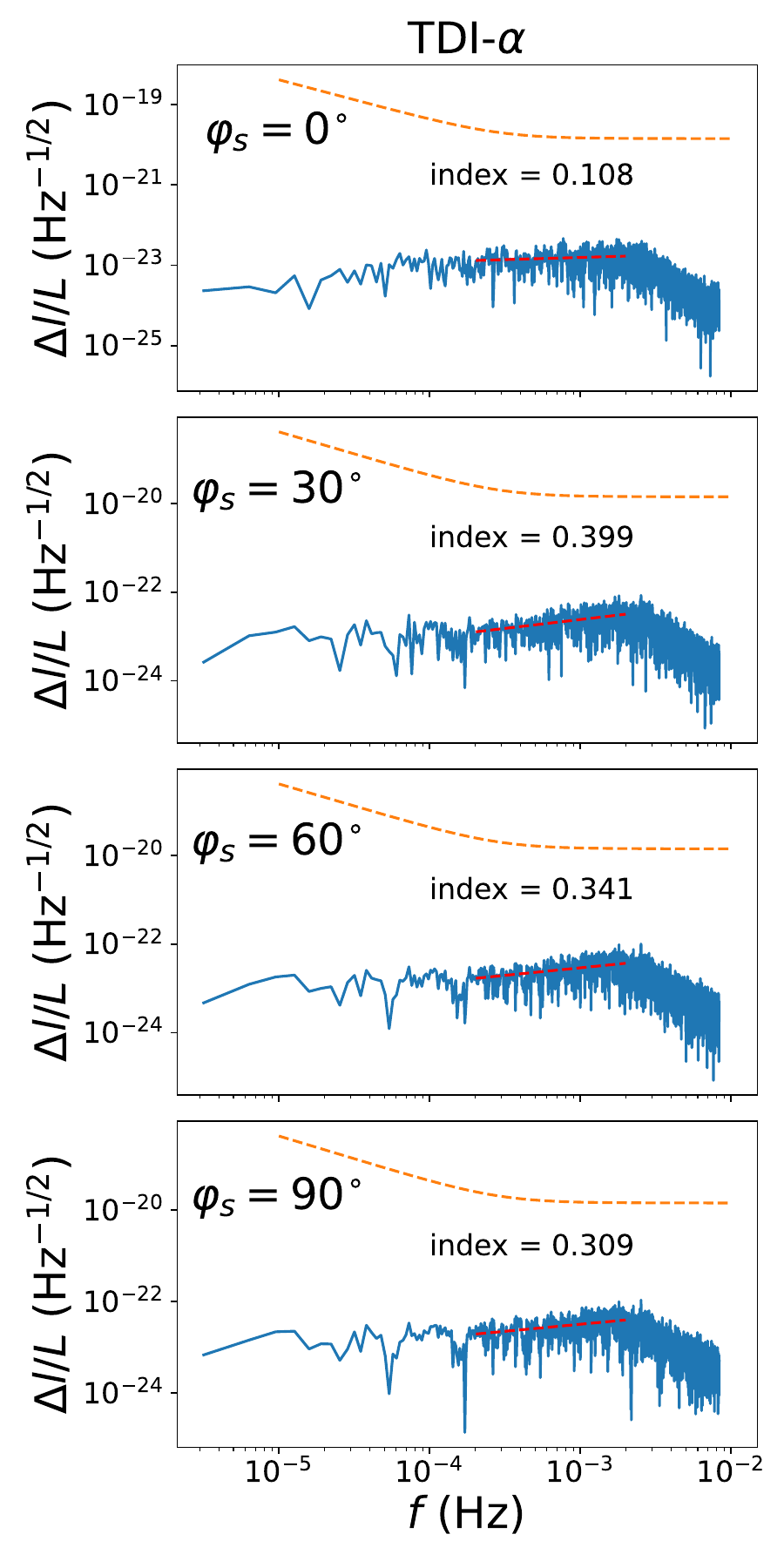}
    \end{minipage}
    \begin{minipage}[b]{0.411\textwidth}
    \includegraphics[width = 5.21 cm]{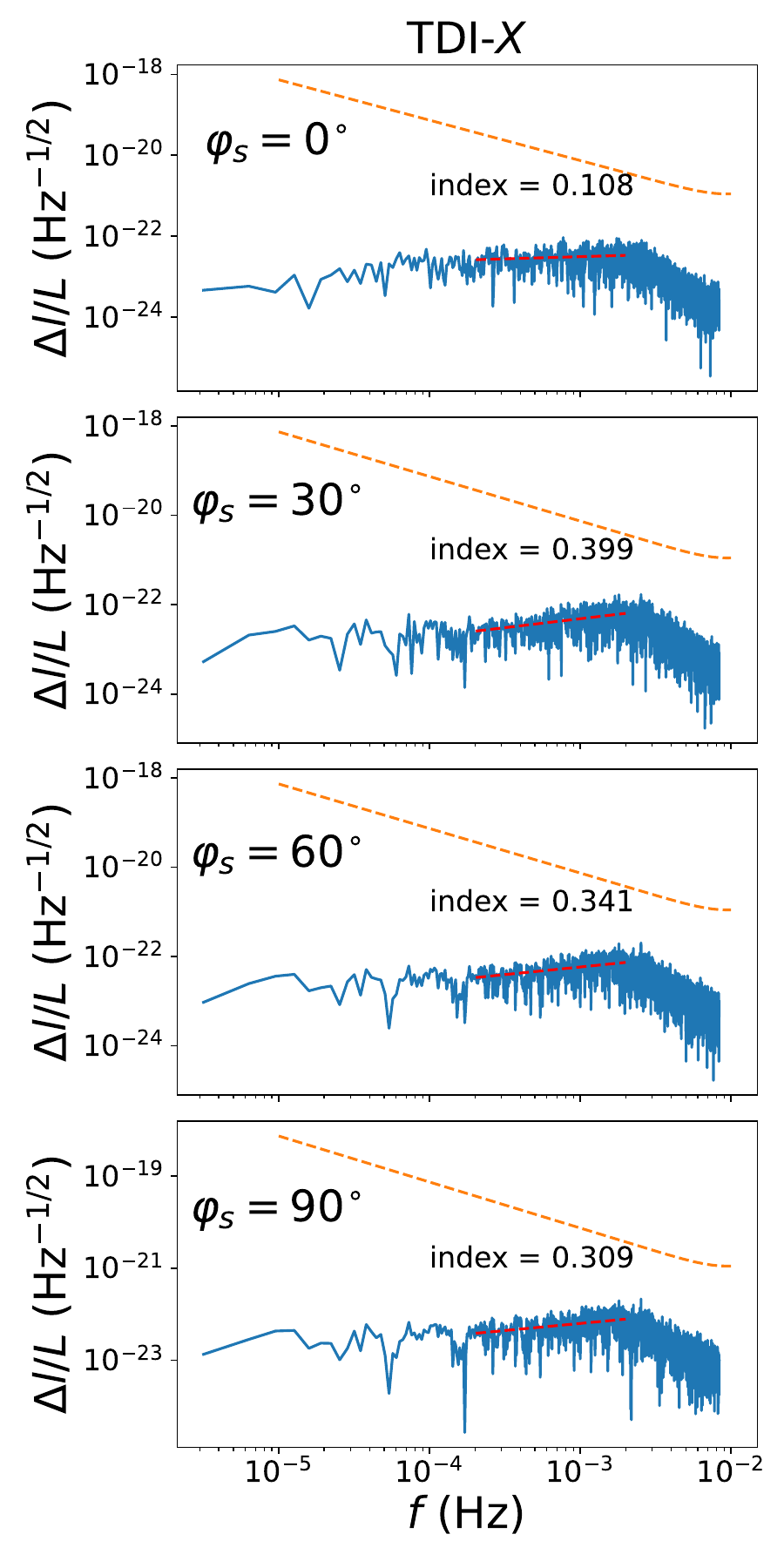}
    \end{minipage}
\end{minipage}
\caption{The LPN's ASDs of TDI--$\alpha$ (left panels) and TDI--$X$ (right panels) combinations for geocentric GW detectors TQ are represented as blue curves, and the displacement requirement of TQ is represented as orange curve \citep{Su2021}. }
\label{fig3-TDI-ASD}
\end{figure}

In addition, the ASD of LPN under TDI combination was calculated.  Figure \ref{fig3-TDI-ASD} represents the ASDs of LPN noise of the TDI combinations $\alpha$ and $X$ for TQ; the left and right columns of Figure \ref{fig3-TDI-ASD} are the ASDs of TDI--$\alpha$ and TDI--$X$ combinations, respectively; and rows 1, 2, 3, and 4 of Figure \ref{fig3-TDI-ASD} are the ASDs of $\varphi_s$=0$^{\circ}$, 30$^{\circ}$, 60$^{\circ}$, and 90$^{\circ}$, respectively. Comparing the ASD of LPN for Michelson combination (Figure \ref{fig3-3}) and TDI combinations (Figure \ref{fig3-TDI-ASD}), it can be found that the ASDs of the TDI combinations are significantly lower than those of the Michelson combination in the low-frequency band ($\lesssim 2$ mHz); unlike the LPN of the Michelson combination, which is a colored noise with spectra indices of the ASDs about $-1$ to $-0.5$, the ASDs of LPN for the TDI combinations are presented as approximate white noises in the low-frequency band ($\lesssim 2$ mHz) with spectra indices $\approx 0$.

For Michelson combination, the ASD of the equivalent strain noise ($\sqrt{S_n^{M}}$) for TQ is written as \citep{Hu2018},
\begin{equation}
    S_n^{M} = S_n^{x} + S_n^{a} (1 + \frac{10^4~\mathrm{Hz}}{f} )
\end{equation}
here, $f$ is frequency, $S_n^{x}$ is the equivalent strain noise of the displacement measurement, $S_n^{a}$ is the equivalent strain noise caused by acceleration noise.
The ASD of equivalent strain noises for TDI--$\alpha$ ($\sqrt{S_n^{\alpha}}$) combinations is as follow \citep{Armstrong1999,Estabrook2000},
\begin{equation}
\begin{aligned}
S_n^{\alpha} = [4\mathrm{sin}^2(3 \pi f L/c) + 24\mathrm{sin}^2(\pi f L/c)]S_n^a + 6S_n^x
\end{aligned}
\end{equation}
And for TDI--$X$, the ASD of equivalent strain noises is as follow \citep{Armstrong1999,Estabrook2000},
\begin{equation}
\begin{aligned}
S_n^{X} = [4\mathrm{sin}^2(4 \pi f L/c) + 32\mathrm{sin}^2(2 \pi f L/c)]S_n^a + 16\mathrm{sin}^2(2\pi f L/c) S_n^x
\end{aligned}
\end{equation}

\begin{figure}[h]
\centering
\includegraphics[width=0.95\textwidth]{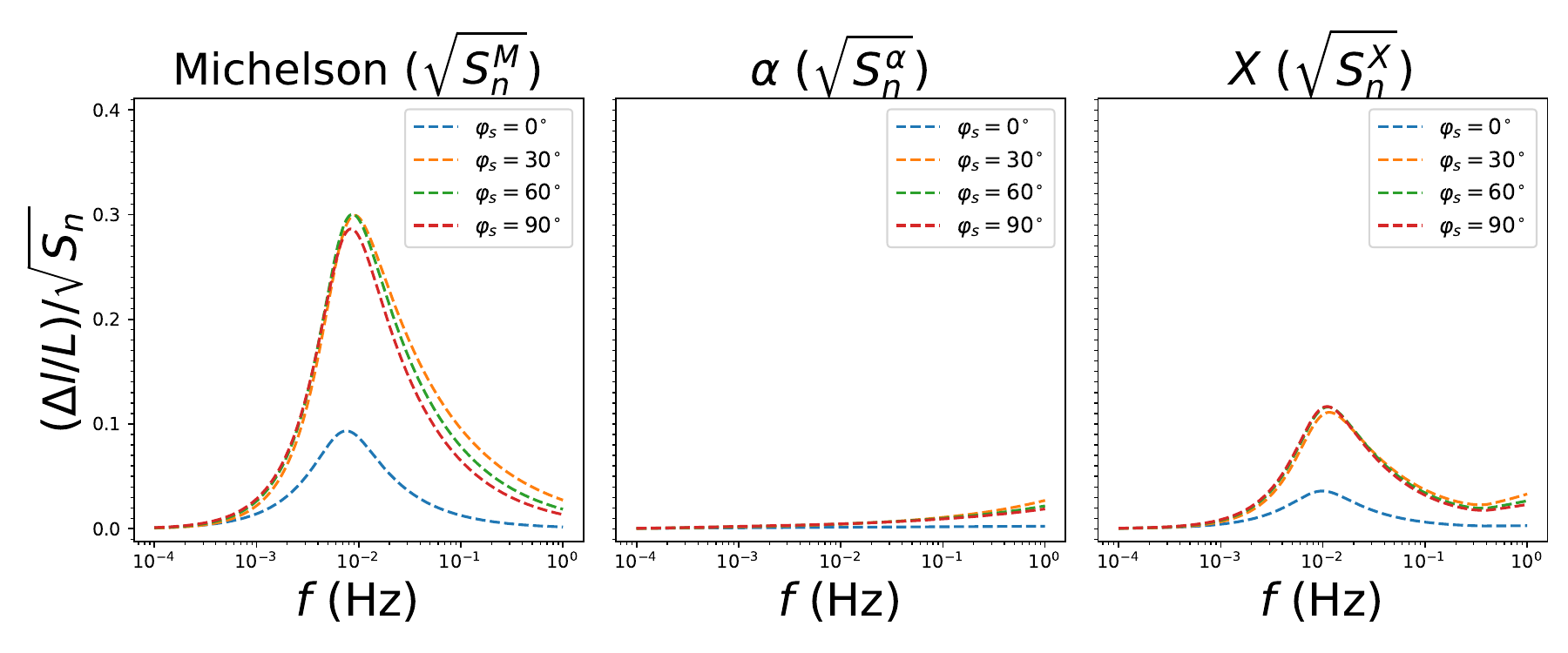}
\caption{The ratios of the normalization of LPN to the equivalent strain of TQ ($(\Delta l/L)/\sqrt{S_n}$), the left, middle, and right panels are $(\Delta l/L)/\sqrt{S_n}$ of Michelson, TDI--$\alpha$ and TDI--$X$ combinations, respectively \citep{Su2021}. }
\label{fig3-TDI-ratio}
\end{figure}

In order to evaluate the LPN for Michelson, TDI--$\alpha$, and TDI--$X$ combinations, the best-fit spectra of the LPN are used to get the ratio between the equivalent strains of LPN $(\Delta l/L)$ and $\sqrt{S_n^M}$, $\sqrt{S_n^{\alpha}}$, $\sqrt{S_n^X}$ \citep{Su2021}. As shown in Figure \ref{fig3-TDI-ratio}, both $(\Delta l/L)/\sqrt{S_n^{\alpha}}$ and $(\Delta l/L)/\sqrt{S_n^X}$ are lower than $(\Delta l/L)/\sqrt{S_n^M}$, it suggests that TDI--$\alpha$ and --$X$ combinations can suppress the LPN noise. The maximum of $(\Delta l/L)/\sqrt{S_n^M}$ and $(\Delta l/L)/\sqrt{S_n^X}$ are at around 10 mHz with the value of about 0.1 and 0.3, respectively \citep{Su2021}. 

Since TDI is revealed that can suppress the LPN, this result has been repeatedly confirmed \citep{Jing2022,Zhao_2024,Xie2024}. Based on the high temporal and spatial resolution MHD simulations, and combining with the assumption that the perturbations of $n_{\mathrm{e} }$ follow Kolmogorov spectra, the TDI is confirmed that can suppress LPN \citep{Jing2022}. Geometric method is applied to search TDI combinations, and 45 second-generation TDI is indentified \citep{Zhao_2024}; Based on the assumption of Kolmogorov spectrum, the LPN of 45 TDI combinations is evaluated by an analytical model, and find that after using TDI, if the LPN noise of LISA is required to be lower than the noise requirement, $n_e$ needs to be $\leq$ 100 cm$^{-3}$ at 10 mHz \citep{Zhao_2024}. The TDI combinations is also find that can suppress the LPN for TJ \citep{Xie2024}.

The LPNs of TDI combinations throughout more than a solar cycle is also studied \citep{Liu2024}. \cite{Liu2024} calculate the ASDs of the LPN of the TDI--$\alpha$ and --$X$ combinations for 12 cases with different $\overline{P}_{\rm dyn}$. Based on the fitting results of the ASDs, the values of $(\Delta l/L)/\sqrt{S_n^{\alpha}}$ and $(\Delta l/L)/\sqrt{S_n^X}$ at 10 mHz can be obtained, and the results are shown in the middle and bottom panels of Figure \ref{fig3-cc-lin}. The result shown that for events with $\overline{P}_{\rm dyn}>5$~nPa, $(\Delta l/L)/\sqrt{S_n^X}$ can reach $>0.3$. Fortunately, according to CDF of $\overline{P}_{\rm dyn}$ during more than 2 solar cycles, the occurrence probability of $(\Delta l/L)/\sqrt{S_n^X}>0.3$ is less than 1\% \citep{Liu2024}. It further demonstrates the practicality of TDI throughout the solar cycles.

% The reason that TDI can suppress LPN noise is as follows. The plasma along different laser interferometry arms is different, i.e., the plasma is spatially different along interferometry links for Michelson conbination. Whereas the interferometry links of TDI are spatially symmetric or identical, and the spatial difference of LPN between different laser links can be eliminated by spatially identical interferometry links constructed TDI combinations, but the temporal difference due to TDI is preserved.

From the perspective of plasma turbulence theory, TDI's ability to suppress LPN can be explained as follows.
For a single arm, the distance over which plasma turbulence can propagate is the laser arm length $L$. The Michelson interferometric combination has two arms, and the plasma distributions along the two arms are different and can be considered uncorrelated. Thus, the typical length ($S_{\rm M}$) that plasma disturbance can propagate for Michelson interferometric combination is $L$, i.e., $S_{\rm M} \sim L$. The LPN distributions of single link and Michelson combination are both of the order of 1 pm, it verifies that $S_{\rm M} \sim L$. According to the cascade theory of plasma turbulence, the plasma turbulence propagation velocity is the Alfvén speed $v_{\rm A}$. Thus, the typical timescale $\tau_{\rm M}$ of plasma turbulence propagate along the laser link is $L/v_{\rm A}$. Taking the typical Alfvén velocity of the space plasma as about 300 km/s, $\tau_{\rm M}$ is about $10^3$ s for TQ and about $10^4$ s for LISA and TJ.

TDI combinations involve two laser beams passing through the same optical path in different orders to produce interference. Thus, the two laser beams of TDI are symmetric or identical in spatial, but asymmetry or differences in temporal. The temporal asymmetry can be manifested by the time delay factor $L/c$ in Equations \eqref{eq:TDI-alpha1} and \eqref{eq:TDI-X1}. It indicates that the TDI combination still retains plasma turbulence on the timescale $\tau_{\rm T} \approx L/c$. It means that TDI cannot suppress the LPN noise on the timescale less than $L/c$. For space-borne GW detections, $\tau_{\rm T}$ is about 1 s for TQ, and about 10 s for LISA and TJ.

The turbulence of space plasma follows the Kolmogorov theory, with the PSD spectrum index of $-5/3$, i.e., the ASD spectrum index of $-5/6$. Thus, the ratio of the plasma turbulence magnitudes at the timescales $\tau_{\rm M}$ and $\tau_{\rm T}$ is as follow,
\begin{equation}
    \frac{\delta n(\tau_{\rm M})}{\delta n(\tau_{\rm T})}=\left(\frac{1/\tau_{\rm M}}{1/\tau_{\rm T} }\right)^{-5/6} \approx {1000}^{5/6}
    \label{eq:TDI-plasma}
\end{equation}
Comparing LPN of Michelson and TDI combinations, the results are consistent with those in Equation \eqref{eq:TDI-plasma}.

Overall, the dual-frequency scheme can accurately solve the TEC along the laser link, deduct the LPN; however, the dual-frequency scheme needs to add a new frequency laser, which will increase the complexity of the system and cost of the whole space-borne GW detection. The TDI scheme can suppress the LPN without additional hardware, and is especially effective in suppressing the LPN in the low-frequency band. Considering the effectiveness of the TDI scheme throughout the solar cycles, TDI is a feasible solution.

\section{Space magnetic effect for space GW detection}\label{sec4}

\subsection{Space magnetic acceleration noise based on observations and MHD simulation}
\label{sec4-1}

The effects of acceleration noise due to magnetic fields on the space-borne GW detection have been studied more than 2 decades \citep{Schumaker2003,Hanson2003}. 
In this process, amount results on the magnetic field effects of GW detection have been accumulated, such as the measurement and manufacturing of the magnetic field properties of the TM \citep{Diaz-Aguil2010,Yin2021,Lou2023}, the magnetic field measurement system of the GW satellite \citep{Diaz-Aguil2012}, and the overall magnetic effects of various instruments on the GW detectors \citep{Diaz-Aguil2011}. However, these results mainly discuss the magnetic field generated by the instruments onboard the satellite \citep{Sun2023}.

\begin{figure}[h]
\centering
\includegraphics[width=0.96\textwidth]{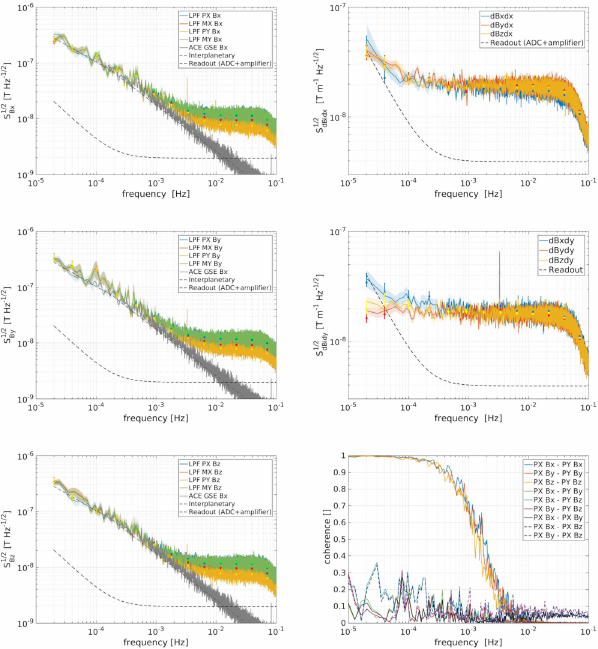}
\caption{The ASDs of magnetic field measured LPF and from the ACE magnetometer. The ASDs measured by 4 LPF magnetometers are represented as colored curves, and the ASDs measured by ACE is represented as grey curve \citep{LPF-M-mnras2020}. }
\label{fig4-LPF}
\end{figure}

For heliocentric GW detector, LISA, the acceleration noise due to space magnetic field has been measured by LISA PathFinder (LPF) \citep{LPF-M-mnras2020}.
LPF is a European Space Agency (ESA) mission designed to test key technologies required for the future space-based GW observatory, LISA \citep{LPF2012}. Since magnetic forces can influence the motion of the TM, LPF carries a magnetic diagnostics subsystem with the measurement sensitivity of 10 nT Hz$^{-1/2}$ \citep{LPF-M-mnras2020} to measure the magnetic field around the TMs. The magnetic field diagnostics system consists of four tri-axial fluxgate magnetometers and two coils which deployed around the TMs. The magnetic diagnostics have measured magnetic field around the TMs for hundreds of days, and the magnitude of the magnetic is on the order of 1000 nT \citep{LPF-M-mnras2020}. The ASD of the magnetic field is calculated based on the time series measurements. Note that the magnetic field measured by LPF could not distinguish between the space magnetic field and the satellite magnetic field. Both ACE and LPF are located around Sun--Earth L1 point, the magnetic field observed by the ACE is pure space magnetic field, thus, the comparison of the magnetic field observation data of the ACE and LPF satellites can separate the data of the LPF satellite into the space magnetic field and the satellite magnetic field \citep{LPF-M-mnras2020}. 
As shown in Figure \ref{fig4-LPF}, the results show that the magnetic field in the low-frequency band is consistent with the interplanetary magnetic field measurements of ACE. It suggests that for LPF, the magnetic field in low frequency bands ($\lesssim$ 1 mHz) is mainly caused by the interplanetary magnetic field, which is roughly presented as Kolmogorov spectrum. In the high-frequency band ($\gtrsim$ 1 mHz), the magnetic field appears as white noise with spectrum index $\approx 0$ for LPF, while the magnetic field observed by ACE still appears as Kolmogorov spectrum, it indicates that the magnetic noise onboard electronic systems are white noise \citep{LPF-M-mnras2020}. 
In addition, \cite{LPF-M-mnras2020} reveal that solar wind velocity plays a significant role in driving magnetic fluctuations, during high-speed solar wind events ($\sim 500$~km~s$^{-1}$), the amplitude of magnetic fluctuations increases significantly compared to low-speed events ($\sim 300$~km~s$^{-1}$). This work is crucial for improving the design of magnetically quiet for space-based GW detectors.
Recently, based on the magnetic measurements onboard the LPF, the interplanetary and spacecraft magnetic-induced acceleration noise is found to be about 0.25 and 1 $\rm fm~s^{-2}~Hz^{-1/2}$ at 1 and 0.1 mHz, respectively \citep{LPF2025PRL,LPF2025PRD}.

For geocentric GW detector, TQ, there are no corresponding in-situ measurements of the space magnetic field at the altitude of TQ's orbit, the acceleration noise due to space magnetic field is estimated based on MHD simulations \citep{Su2020} and semi-empirical models \citep{Su2023}, e.g., Tsyganenko model \citep{Tsyganenko1996}.

According to Equations \eqref{eq:mag-moment-sp12345} and \eqref{eq:mag-Lorentz}, the acceleration noise due to space magnetic field is related to the values of some TM's physical parameters, such as the mass of TM $m$, the side length of the TM $r$, magnetic susceptibility $\chi_{\rm m}$, magnetic shielding factor $\xi_{\rm m}$, remanent magnetic moment $\boldsymbol{M}_{\rm r}$, and residual charge $q$. In the past decades, technology has been advancing, the values of some of these parameters have been iteratively updated \citep{Qiao2023,Xu2024,Ma2025}, and the design, testing, and manufacturing of the TM has been rapidly improving in the last few years \citep{Yan2024,Yu2025}. 

\begin{table}[]
\begin{tabular}{ccc}
\hline
       & LISA & TQ \\  \hline
$m$ (kg) & 1.96 &  2.45  \\  \hline
$r$ (cm) & 4.6 & 5  \\  \hline
$\chi_{\rm m}$ & $2.5\times10^{-5}$ & $1\times10^{-5}$  \\  \hline
$\xi_{\rm m}$  & 10 & 10  \\  \hline
$M_{\rm r}$ (Am$^2$)  & $2\times10^{-8}$  & $2\times10^{-8}$  \\  \hline
$B_{\rm sc}$ (T)  & $1\times10^{-6}$  & $1.6\times10^{-6}$  \\  \hline
$q$ (C)  & $1.6\times10^{-12}$  & $1.6\times10^{-12}$  \\  \hline
\caption{The list of parameters for heliocentric (LISA) and geocentric (TQ) GW detectors that used in this paper, including the mass of TM $m$, the side-length of TM cube $r$, the magnetic susceptibility $\chi_{\rm m}$, the magnetic shielding factor $\xi_{\rm m}$, remanent magnetic moment $\boldsymbol{M}_{\rm r}$, the spacecraft magnetic field $\boldsymbol{B}_{\rm sc}$, and the residual charge of TM $q$. }
\label{tb1}
\end{tabular}
\end{table}

In this review, the values of the physical parameters for heliocentric and geocentric GW detectors are listed in Table \ref{tb1}. The TM is made of Au--Pt alloy, the TM mass and side length are 1.96 kg and 4.6 cm for LISA \citep{Diaz-Aguil2011}, and 2.45 kg and 5 cm for TQ \citep{Luo2016}, respectively. The experiments show that $\chi_{\rm m}$ can be less than $10^{-5}$ \citep{Lou2023}, and the theoretical calculations show that Au--Pt alloy with Fe or Bi impurity can even make $\chi_{\rm m}$ be less than $10^{-6}$ \citep{Ma2025}. Conservatively, $\chi_{ m}$ of LISA and TQ are taken as $2.5 \times 10^{-5}$ and $1 \times 10^{-5}$ in this review, respectively.
In the case of magnetic shielding, $\boldsymbol{a}_{\rm m}$ becomes $\boldsymbol{a}_{\rm m}/\xi_{\rm m}$, which suggests that $\xi_{\rm m} > 1$ can further suppress the acceleration noise due to magnetic field. $\xi_{\rm m}$ can be taken as 1 to 100 \citep{Cavalleri2009,Diaz-Aguil2011}, and in this review $\xi_{\rm m}$ of LISA and TQ are both taken as 10. For LISA, the satellite magnetic field around the TM is taken as $1.0\times 10^{-6}$ T\citep{Diaz-Aguil2011}; For TQ, take equivalent magnetic moment requirement of TQ satellite as 1 A m$^2$, and the equivalent distance from the TM as 0.5 m, the satellite magnetic field around the TM is about $1.6\times 10^{-6}$ T \citep{Luo2016,Su2016}.

Based on the GCR observation of LPF \citep{LPF-GCR-2018a,LPF-GCR-2018b,LPF-GCR-2019,Cesarini2022,Villani2024-GCR}, and modelling of GCR \citep{Heber2006,Chu2016,Shen2021GCR} for heliocentric and geocentric GW detectors, the simulation of charging of TM can be carried out, and the results show that the charging effects are comparable for heliocentric and geocentric GW detectors \citep{LPF-charge-2017,LPF-charge-2018,Han2024,Lei2024}. Besides GCR, occasional SEP is another important reason for the charging of the TM \citep{LiG2021,Ding2022,Wang2023SEP,Han2023}. The charging of TM due to GCR and SEP makes the charge management important \citep{Apple2022,Buchman2023,Gu2024}. In this review the residual charge of LISA and TQ are taken to be equal with the value of $1.6\times10^{-12}$ C.
When the charged TM passes through the space magnetic field, the Lorentz force can be partially compensated by the electric field force induced by the Hall effect \citep{Sumner2020}. Here, introduce the effective shielding coefficient $\eta$, the Lorentz force $\boldsymbol{a}_{\rm L}$ becomes $\eta \boldsymbol{a}_{\rm L}$. In this review, $\eta$ is taken as 0.03 \citep{Sumner2020}.

\begin{figure}[h]
\centering
\includegraphics[width=1.0\textwidth]{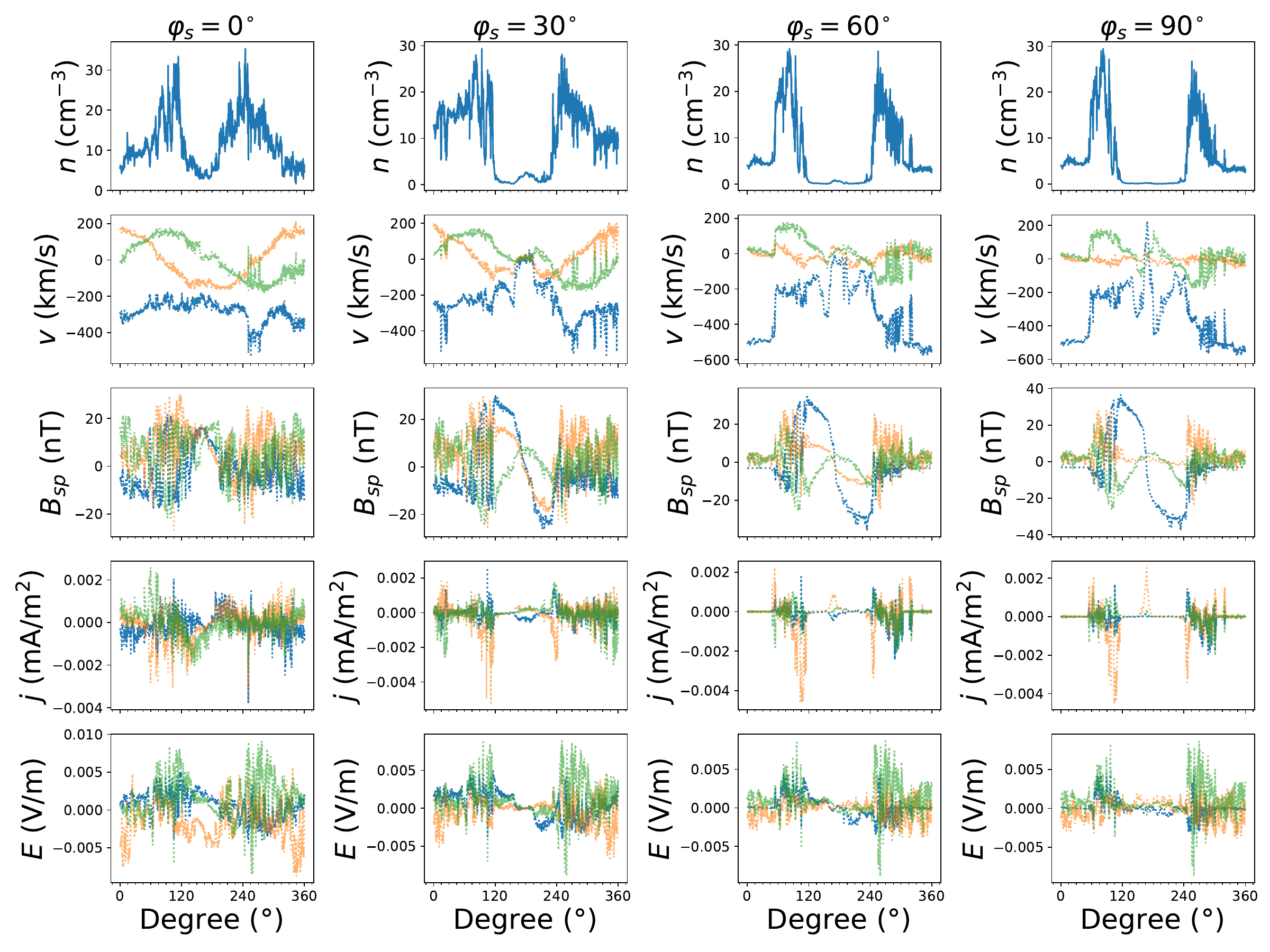}
\caption{The time series of physical quantities on the orbit of TQ that obtained by the MHD simulation. The first row is ion number density $n_{\rm i}$, the second row is bulk flow $\boldsymbol{v}$, the third row is space magnetic field $\boldsymbol{B}_{\rm sp}$, the forth row is current density $\boldsymbol{j}$, and the fifth row is the electric field $\boldsymbol{E}$ \citep{Su2020}. }
\label{fig4-para-MHD}
\end{figure}

In addition to the density around the orbit of GW detectors, solving the MHD Equations can also yield magnetic field $\boldsymbol{B}$, bulk flow speed $\boldsymbol{v}$, electric current $j$, etc. So the MHD simulation can be used not only for the study of the LPN, but also for the study of the acceleration noise due to the space magnetic field. The white curves in Figure \ref{fig-SWMF-nB} are the magnetic lines of the Earth magnetosphere obtained from the MHD simulation, and the time series of the magnetic field along the TQ's orbit during a TQ cycle is shown in Figure \ref{fig4-para-MHD}. The magnetic field around TQ's orbit is on the order of 10 nT, in the case of a solar eruption affecting the magnetosphere or magnetic storm, the magnetic field can be larger than 100 nT \citep{LiH2010}. 
The maximum of the magnetic field corresponds to the magnetosheath region, because the magnetosheath is downstream of the bow shock, and downstream of MHD shocks can enhance the magnetic field. The magnetic field in most region of the magnetosphere is larger than that of the solar wind, and in the current sheet of the magnetotail, the magnetic field tends to approach 0 nT. 

Based on Equations \eqref{eq:mag-moment-sp12345} and \eqref{eq:mag-Lorentz}, the parameters in Table \ref{tb1}, and the magnetic field on the TQ orbit obtained from the MHD simulation, the acceleration noise can be calculated.
$\boldsymbol{a}_{\mathrm{M1}}$, $\boldsymbol{a}_{\mathrm{M2}}$, $\boldsymbol{a}_{\mathrm{M3}}$, $\boldsymbol{a}_{\mathrm{M4}}$, and $\boldsymbol{a}_{\mathrm{M5}}$, are on the orders of $10^{-16} \; \mathrm{m \; s^{-2}}$, $10^{-20} \; \mathrm{m \; s^{-2}}$, $10^{-30} \; \mathrm{m \; s^{-2}}$, $10^{-32} \; \mathrm{m \; s^{-2}}$, and $10^{-23} \; \mathrm{m \; s^{-2}}$ \citep{Su2020}, $\boldsymbol{a}_{\mathrm{M1}}$ is the dominant term. In addition, $\boldsymbol{a}_{\mathrm{L}}$ is on the order of $10^{-17} \; \mathrm{m \; s^{-2}}$ \citep{Su2020}. 

Since MHD simulations require extensive computational resources, using MHD to calculate the acceleration noise caused by space magnetic fields throughout an entire solar cycle would take more than hundreds of years. To quickly assess the acceleration noise caused by space magnetic fields, a faster method is needed.

% \begin{figure}[h]
% \centering
% \includegraphics[width=0.9\textwidth]{A-t-MHD-TQ-xi10.pdf}
% \caption{asdf}
% \label{fig4-A-t-MHD}
% \end{figure}

% \begin{figure}[h]
% \centering
% \includegraphics[width=0.9\textwidth]{A-ASD-fit-req-TQ-xi10.pdf}
% \caption{asdf}
% \label{fig4-A-ASD-MHD}
% \end{figure}

\subsection{Statistics result of space magnetic acceleration noise}
\label{sec4-2}

\begin{figure}[h]
\centering
\includegraphics[width=0.86\textwidth]{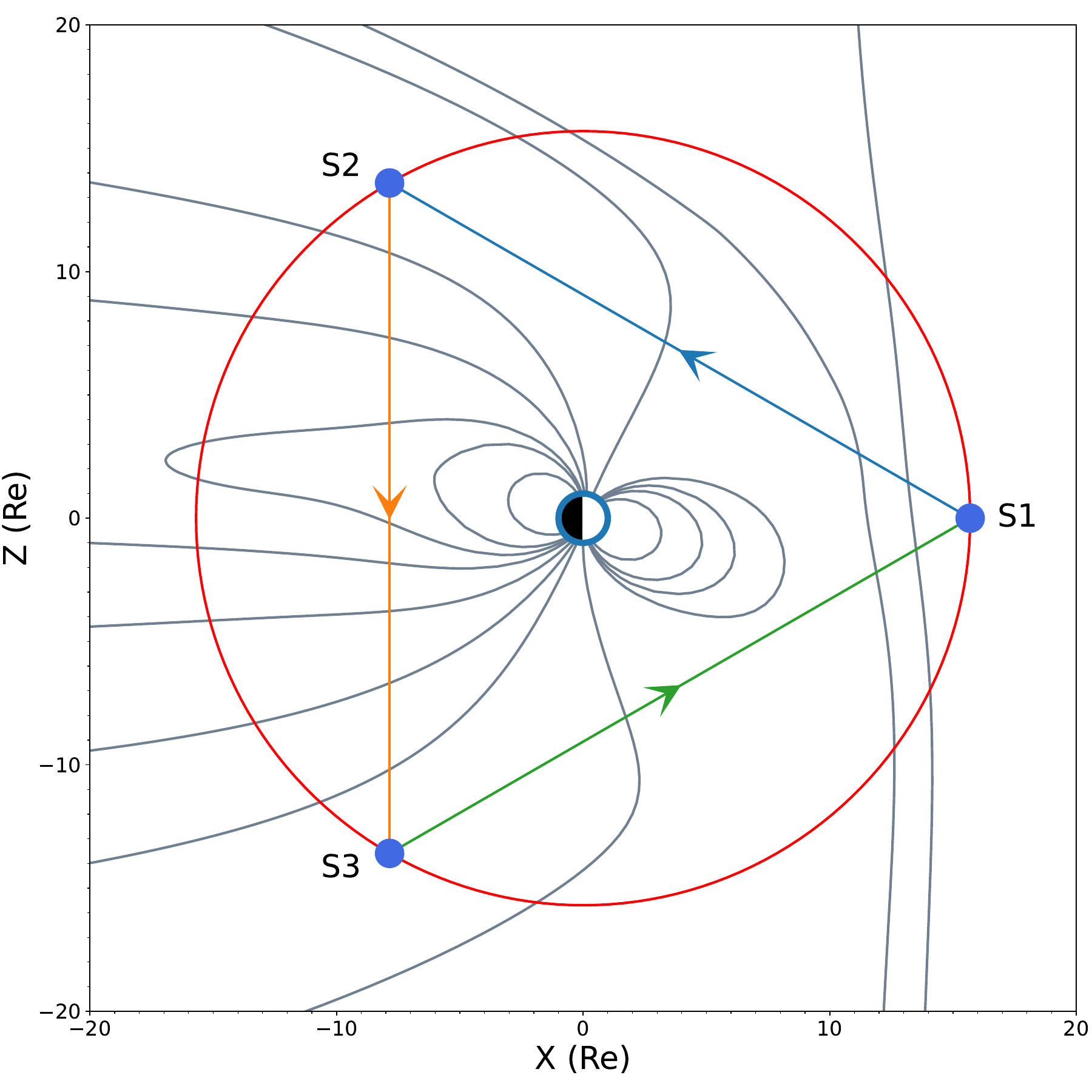}
\caption{The space magnetic field around the orbit of TQ that obtained by Tsyganenko model \citep{Su2023}. }
\label{fig4-B-2D-TS}
\end{figure}

The Tsyganenko model is a data-driven empirical model that can yield the space magnetic field in and around the Earth's magnetosphere. Figure \ref{fig4-B-2D-TS} shows a schematic diagram of space magnetic field on the TQ orbital plane that calculated by the TS model, where the grey curves are the magnetic lines calculated by Tsyganenko model, and the red circle is the orbit of TQ. Compared to the MHD simulation, it can calculate the space magnetic field much faster, and it is suitable for statistical analysis of acceleration noise due to space magnetic field for geocentric GW detectors.

% \begin{figure}[h]
% \centering
% \includegraphics[width=0.96\textwidth]{A-t-B-TS.pdf}
% \caption{asdf}
% \label{fig4-B-t-TS}
% \end{figure}

% \begin{figure}[h]
% \centering
% \includegraphics[width=0.86\textwidth]{A-ASD-Statistics-LISA-TQ.pdf}
% \caption{asdf}
% \label{fig4-B-ASD-statistics}
% \end{figure}

% \begin{figure}[h]
% \centering
% \includegraphics[width=0.96\textwidth]{A-CDF-TQ-LISA.pdf}
% \caption{asdf}
% \label{fig4-B-CDF}
% \end{figure}

\begin{figure}[h]
\centering
\includegraphics[width=0.96\textwidth]{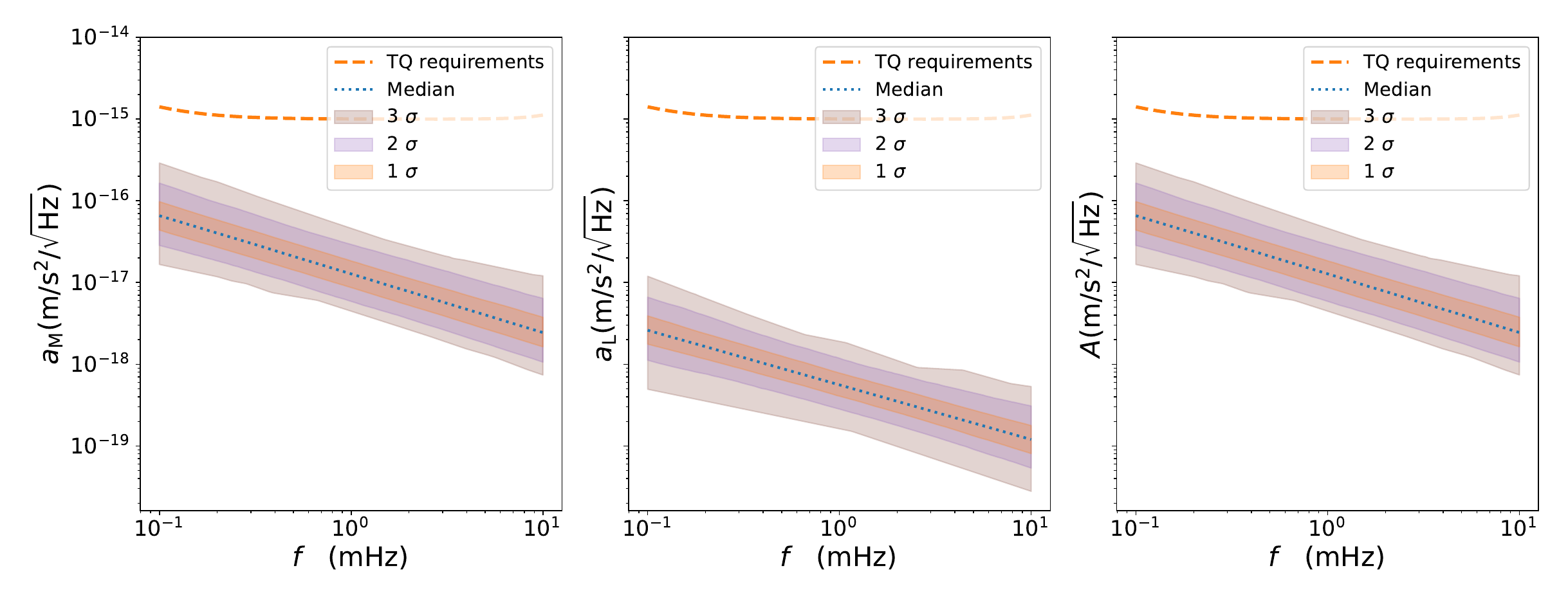}
\caption{The ASDs of $\boldsymbol{a}_{\rm m}$, $\boldsymbol{a}_{\rm L}$, and $\boldsymbol{A}$ for TQ \citep{Peng2025}. The orange curve is the acceleration requirement of TQ. }
\label{fig4-ASD-TQ}
\end{figure}

% Using the OMNI solar wind data more than 2 solar cycles as input, 
% Using 23 years (more than 2 solar cycles) of OMNI solar wind data from 1998-01-01 to 2020-12-31 as input, 
Using 23 years (more than 2 solar cycles) of OMNI solar wind data from 1998-01- 01 to 2020-12-31 as input, the Tsyganenko model was applied to calculate the space magnetic field of 2300 TQ orbital cycles over 23 years \citep{Su2023}.
Based on the Tsyganenko model, and Equations \eqref{eq:mag-moment-sp12345} and \eqref{eq:mag-Lorentz}, the space magnetic acceleration noise of a total of TQ orbital cycles is obtained \citep{Su2023}. As $\boldsymbol{a}_{\mathrm{M1}}$ is the dominant term, $\boldsymbol{a}_{\mathrm{M1}}$ is denoted by $\boldsymbol{a}_{\mathrm{M}}$ in the following, and the total space magnetic field acceleration is denoted as $\boldsymbol{A} = \sqrt{\boldsymbol{a}_{\mathrm{M}}^2 + \boldsymbol{a}_{\mathrm{L}}^2 }$. 
In this review, the ASDs of $\boldsymbol{a}_{\mathrm{M}}$ and $\boldsymbol{a}_{\mathrm{L}}$ within the 2500 TQ orbital cycles (from 1998-01-01 to 2022-12-31) are updated \citep{Peng2025}, and the results are shown in Figure \ref{fig4-ASD-TQ} \citep{Su2023,Peng2025}. The left, middle, and right panels of Figure \ref{fig4-ASD-TQ} are the ASDs of $\boldsymbol{a}_{\mathrm{M}}$, $\boldsymbol{a}_{\mathrm{L}}$, and $\boldsymbol{A}$ for TQ, respectively. The median of the 2500 ASDs is represented as dotted blue curves. The orange, purple, and brown shadows represent 1--$\sigma$, 2--$\sigma$, and 3--$\sigma$ intervals of ASDs, respectively. And the requirement of acceleration noise for TQ, $\sqrt{S_{a-{\rm TQ}} }$, is the orange dashed line, which is updated as follow \citep{LiuYC2024},
\begin{equation}
\sqrt{S_{a-{\rm TQ}} } = 1 \times \sqrt{1 + (\frac{f_{\mathrm{c1}}}{f})^2 } \sqrt{1 + (\frac{f}{f_{\mathrm{c2}}})^4} \quad \rm{fm/(s^2\cdot Hz^{1/2})}
\end{equation}
here, $f_{c1}$ and $f_{c2}$ are transfer frequencies with the value of 0.1 and 14 mHz. 
The mean value of $a_{\rm M}$, $a_{\rm L}$, and $A$ at 1 mHz for TQ are $(1.410 \pm 0.621) \times 10^{-17}$, $(6.203 \pm 2.447) \times 10^{-19}$ and $(1.412 \pm 0.621)\times 10^{-17}$ m s$^2$, respectively. 
The spectra indices of the ASDs for $a_{\rm M}$, $a_{\rm L}$, and $A$ are about $-0.5$ to $-1$, indicating that the acceleration noise due to space magnetic field is color noise, and that the amplitude of the space magnetic noise decreases with increasing frequency.

As Tsyganenko model does not contain the magnetosheath region, a hybrid modeling approach that combines the Tsyganenko-Sitnov (TS05) \citep{Tsyganenko2005} model with the Romashets-Vandas (RV) model \citep{Jelnek2012,Romashets2019} to improve accuracy in simulating the magnetic field along TQ’s orbit \citep{Low2024}. Based on the hybrid model, a careful study of the space magnetic acceleration noise in the regions such as magnetopause, magnetotail, magnetosheath, and transregion and solar wind is investigated. 
% The acceleration noise is shown in Figure xx, 
The results show that the transregion with abrupt magnetic field changes has the highest acceleration noise, the magnetosheath with strong turbulence has the middle acceleration noise, and the magnetotail with a more stable fluctuation has the lowest acceleration noise \citep{Low2024}. And the solar wind fluctuations can significantly increase the space magnetic field acceleration noise \citep{Low2024}.

% CDF TQ

\begin{figure}[h]
\centering
\includegraphics[width=0.96\textwidth]{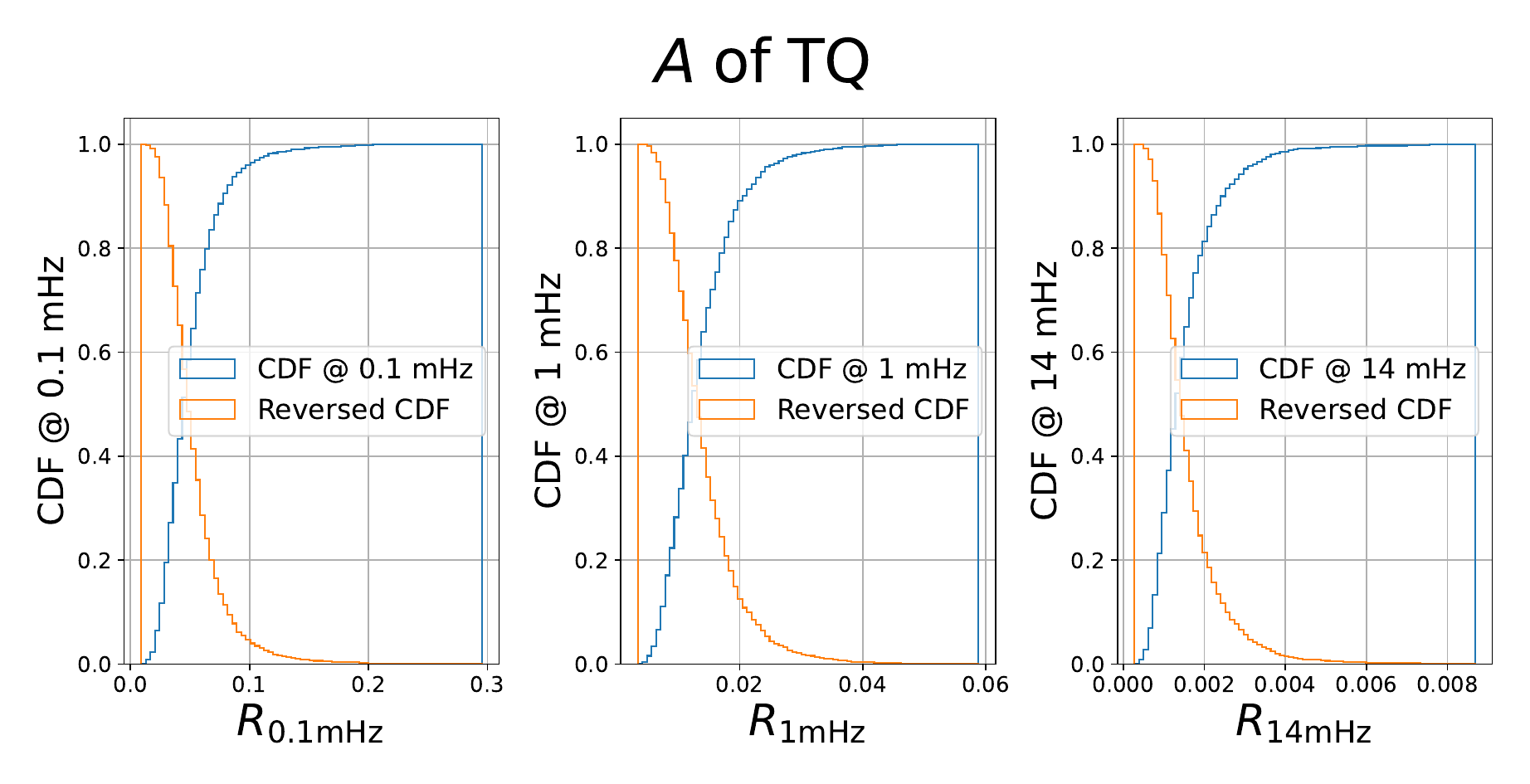}
\caption{The CDF and reversed CDF of $\boldsymbol{A}$ for TQ at 0.1, 1, and 14 mHz \citep{Peng2025}. }
\label{fig4-CDF-TQ}
\end{figure}

According to the ASDs of space magnetic field acceleration noise of TQ over 2 solar activity cycles, the ratio between the ASDs of space magnetic field acceleration noise and the acceleration requirement of TQ ($A/\sqrt{S_{a-{\rm TQ}} }$) is calculated. And the cumulative distribution function (CDF) and reversed CDF $A/\sqrt{S_{a-{\rm TQ}} }$ at the frequencies of 0.1, 1, and 14 mHz are obtained and shown in Figure \ref{fig4-CDF-TQ}. The results show that the occurrence probability of space magnetic field acceleration exceeding the TQ's requirement by 10\% at 0.1 mHz is 4.70\%, that exceeding the TQ's requirement by 10\% at 1 mHz and 14 mHz does not occur, and that the probability of exceeding the TQ's requirement by 2\% at 1 mHz is 0.13\%. It suggests that the TQ's space magnetic acceleration noise in the low frequency band ($\lesssim 1$ mHz) needs to be taken into account for TQ, and that in the high frequency band ($\gtrsim 1$ mHz) of the system are not important.

\begin{figure}[h]
\centering
\includegraphics[width=0.96\textwidth]{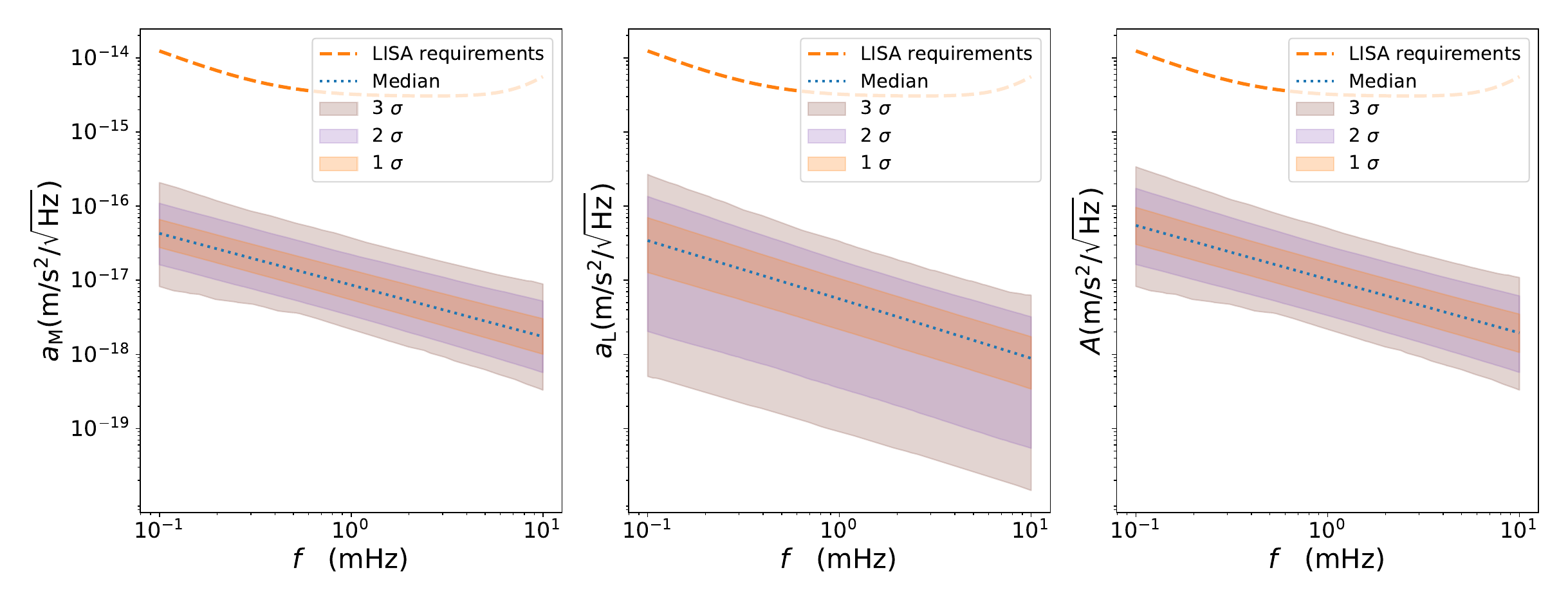}
\caption{The ASDs of $\boldsymbol{a}_{\rm m}$, $\boldsymbol{a}_{\rm L}$, and $\boldsymbol{A}$ for TQ \citep{Peng2025}. The orange curve is the acceleration requirement of LISA. }
\label{fig4-ASD-LISA}
\end{figure}

According to LISA's orbit, the LISA satellites are fully immersed in the solar wind and are approximately 1 AU from the Sun. Thus, the space magnetic field around the orbit of LISA can be considered similar to that at the Sun-Earth L1 point. Here, the OMNI dataset, which is based on in-situ solar wind observations at L1, is used to obtain space magnetic field data over 25 years from 1998-01-01 to 2022-12-31 (more than 2 solar cycles). Using the space magnetic field data from OMNI, and Equations \eqref{eq:mag-moment-sp12345} and \eqref{eq:mag-Lorentz}, the space mangetic acceleration noise $\boldsymbol{a}_{\mathrm{M}}$ and $\boldsymbol{a}_{\mathrm{L}}$ for LISA over these 9000+ days are calculated. 
And the ASDs of $\boldsymbol{a}_{\mathrm{M}}$ and $\boldsymbol{a}_{\mathrm{L}}$ within 9000+ days for LISA are calculated, and the results are shown in Figure \ref{fig4-ASD-LISA}. The left, middle, and right panels of Figure \ref{fig4-ASD-LISA} are the ASDs of $\boldsymbol{a}_{\mathrm{M}}$, $\boldsymbol{a}_{\mathrm{L}}$, and $\boldsymbol{A}$ for LISA, respectively. The shadows and curves in Figure \ref{fig4-ASD-LISA} represent the same meaning as in Figure \ref{fig4-ASD-TQ}. The LISA's acceleration noise requirement curve in Figure \ref{fig4-ASD-TQ} is as follow \citep{LISA2017},
\begin{equation}
    S_{a}^{1/2} \le 3 \times 10^{-15} \frac{\mathrm{m s^{-2}} }{\sqrt[]{\mathrm{Hz}} } \cdot \sqrt[]{1 + (\frac{0.4  \mathrm{mHz} }{f})^{2} } \cdot \sqrt[]{1 + (\frac{ f }{8 \mathrm{mHz} })^{4} }
\end{equation}
The median of $a_{\rm M}$, $a_{\rm L}$, and $A$ at 1 mHz for LISA are $9.686 \times 10^{-18}$, $6.560 \times 10^{-18}$ and $1.208 \times 10^{-17}$ m s$^2$, respectively.
Similarly to TQ, the spectra indices of the ASDs of the space magnetic accelerations are color noise for LISA, and the amplitude of the space magnetic noise decreases with increasing frequency. This is because the magnetic field in different regions of the heliosphere, such as the earth magnetosphere and the solar wind, follow the same self-similarity \citep{Sanchez2018}.

\begin{figure}[h]
\centering
\includegraphics[width=0.96\textwidth]{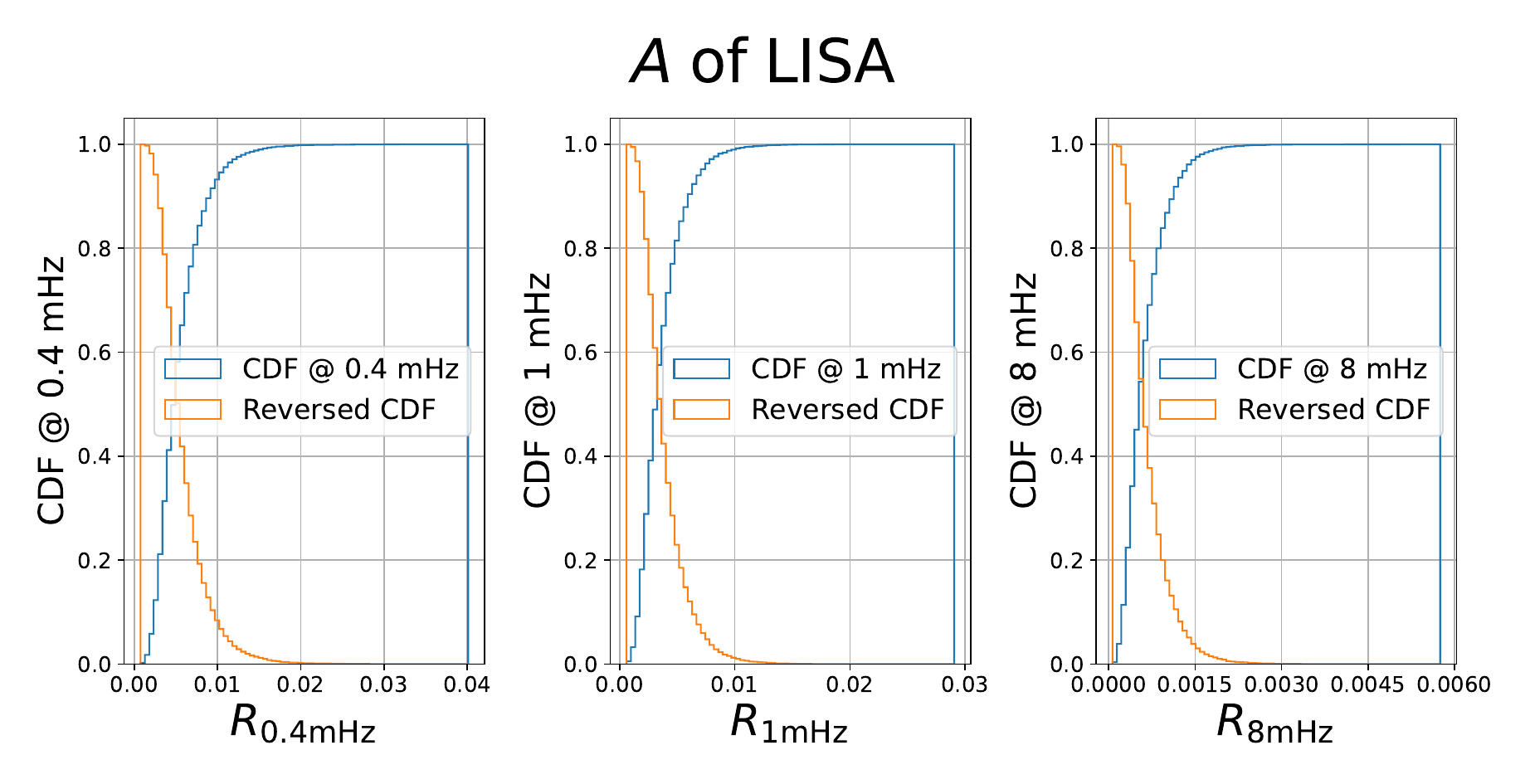}
\caption{The CDF and reversed CDF of $\boldsymbol{A}$ for LISA at 0.4, 1, and 8 mHz \citep{Peng2025}. }
\label{fig4-CDF-LISA}
\end{figure}

% According to the ASDs of space magnetic field acceleration noise of TQ over 2 solar activity cycles, 
Similarly the ratio between the ASDs of space magnetic field acceleration noise and the acceleration requirement of LISA ($A/\sqrt{S_{a-{\rm LISA}} }$) is calculated \citep{Peng2025}. And according to the ASDs of space magnetic field acceleration for LISA over 2 solar cycles, the CDFs and reversed CDFs $A/\sqrt{S_{a-{\rm LISA}} }$ at 0.4, 1, and 8 mHz are calculated, and the results are shown in Figure \ref{fig4-CDF-LISA}. The occurrence probability of space magnetic field acceleration exceeding the LISA's requirement by 10\% does not occur, that exceeding the LISA's requirement by 1\% at 0.4 mHz and 1 mHz are 2.35\% and 0.27\%, respectively, and that the probability of exceeding the TQ requirement by 1\% at 8 mHz does not occur. It suggests that the space magnetic environment of LISA is better than that of TQ.

\begin{figure}[h]
\centering
\includegraphics[width=0.96\textwidth]{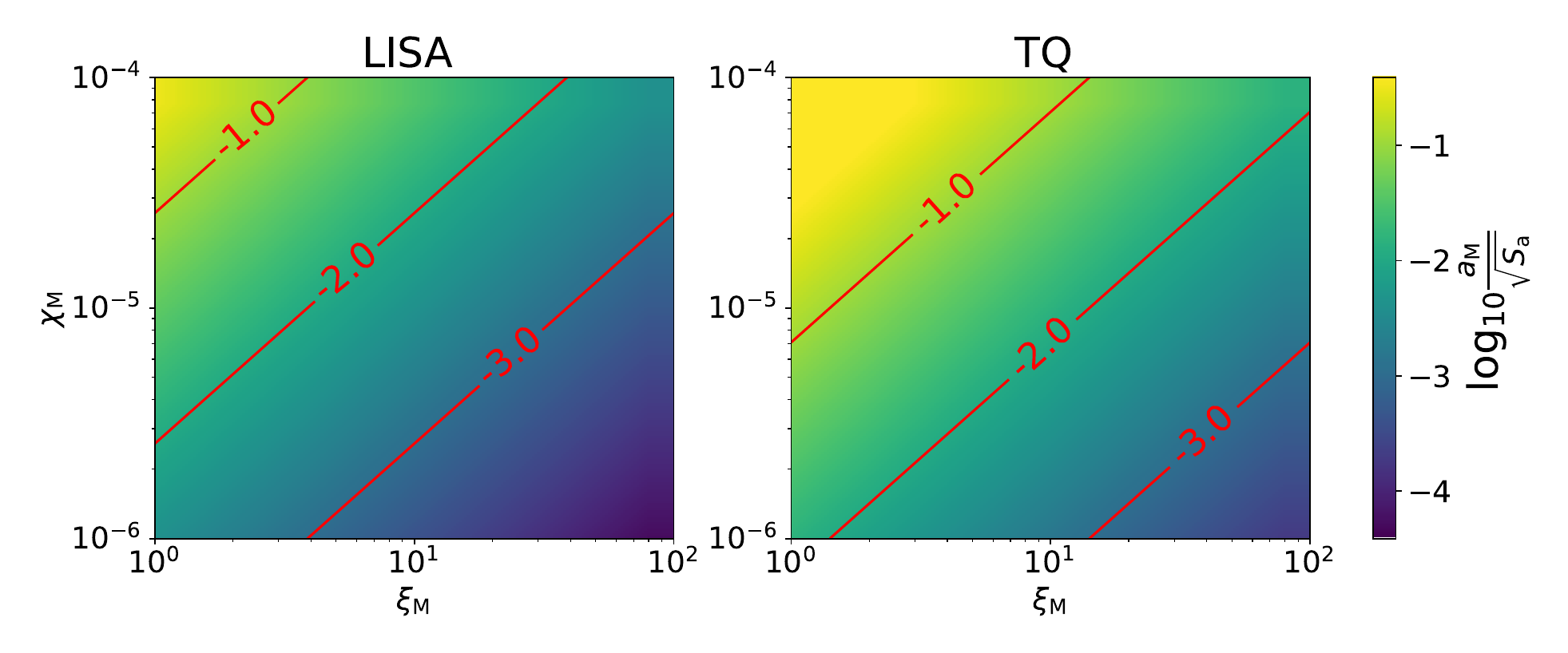}
\caption{The left and right panels are the space of $\chi_{\rm m}$--$\xi_{\rm m}$ at 1 mHz for LISA and TQ, respectively. 
$\boldsymbol{a}_{\rm m} / \sqrt{S_{\rm a}}$ is represented as the color-bar. 
The contours lines of -1, -2, and -3 are the contours of 
$\boldsymbol{a}_{\rm m} / \sqrt{S_{\rm a}} = 10^{-1} $, $10^{-1} $, and $10^{-1} $, respectively \citep{Peng2025}.
}
\label{fig4-B-para}
\end{figure}

Based on the statistical study of space magnetic acceleration noise over 2 solar cycles for LISA and TQ, the design parameters of the TM can be provided. As shown in Figure \ref{fig4-B-para}, the left and right panels are $\chi_{\rm m}$--$\xi_{\rm m}$ design spaces for LISA and TQ, respectively. The results show that TQ has more stringent requirements of $\chi_{\rm m}$--$\xi{\rm m}$ than that of LISA. This is due to the fact that the acceleration noise requirement of TQ are more stringent than that of LISA, and that the magnetic field in the Earth's magnetosphere is stronger than that in the solar wind.
Even so, with the current technical,
the space magnetic acceleration noise of TQ ($A/\sqrt{S_{a-{\rm TQ}} } \lesssim 10\%$) is lower than the total magnetic acceleration noise budget ($A/\sqrt{S_{a-{\rm TQ}} } \leq 24\%$). 
% Since recent studies have shown that $\chi_{\rm m}$ of the TM can be reduced by an order of magnitude to $10^{-6}$, and
Considering that the launch time of GW detectors at about 2035 is still a decade away, the acceleration noise due to the space magnetic field can be expected reduced to much less than 10\% of the total acceleration noise budget in 2035 for both LISA and TQ.

\section{Discussion}\label{sec12}

For laser propagation noise, studies on both heliocentric and geocentric GW detection suggest that LPN do not exceed the respective displacement noise requirements. The study of LPN for heliocentric GW detection is based on in-situ observations, while the LPN for geocentric one is derived from MHD simulations.
The advantage of using in-situ observation of $n_{\rm e}$ lies in its high reliability. But the limitation is the lack of spatial resolution.
% while LPN depends on the spatial distribution of $n_{\rm e}$. 
The theories that derive the spatial distribution of $n_{\rm e}$ along the laser link from in-situ data require validation through observations or numerical simulations.

Comparing with local observations, LPN studies based on MHD simulations can obtain the spatial distribution along the laser link. However, the discrepancy between MHD simulations and real observations makes the distribution of $n_{\rm e}$ obtained by MHD are only approximate.
Additionally, MHD simulations require enormous computational resources, making it infeasible to simulate LPN evolution over an entire solar cycle with current computational capabilities. A compromise is to study LPN noise under different space weather parameters and find the linear correlation between the space weather parameters and LPN, then estimate of LPN throughout the solar cycle. This method introduces errors.
Furthermore, the temporal resolution of MHD simulations for LPN studies is 60 s, much larger than 1 s, meaning the frequency ($\gtrsim 0.01$ Hz) do not cover the high-frequency band of space-borne GW detection. Current estimations assume that $n_{\rm e}$ follows the same spectra index in the high-frequency range, the in-situ observation around the space-borne GW detectors can further validate the spectra index in the high-frequency range.
Regarding spatial resolution, since GW detection satellites travel at speeds of the order of 10 km/s, the finest spatial resolution is required to be about 10 km ($10~\mathrm{km~s^{-1}}/ 1~\mathrm{Hz}$). However, the spatial resolution of the existing MHD simulations is still 2 orders of magnitude short of the demand.
Future research needs higher temporal and spatial resolution MHD simulations to improve the LPN studies.

Due to the influence of magnetic fields, anisotropy is one of the fundamental differences between plasma and fluid \citep{Wu2023}. 
Until now, in studies of the effects of space plasma and magnetic fields on space-borne GW detection, the anisotropy is neglected, the effects caused by anisotropy are worth further investigation.
The impact of anisotropy needs to be considered in future work for heliocentric and geocentric GW detection.
Both the dual-frequency scheme and the time delay interferometry scheme can suppress LPN. The dual-frequency scheme increases the complexity of the GW detection system, while the TDI scheme does not require additional hardware, making it more cost-effective \citep{Tinto2021}. But the TDI is not as effective in suppressing the noise at high frequency band as it is at low frequency band. The final choice between these two schemes requires further discussion.

For acceleration noise caused by space magnetic fields, current results indicate that both heliocentric and geocentric orbit GW detectors meet their respective acceleration noise requirements. The acceleration noise performance of heliocentric GW detectors is better than that of geocentric ones. 
The difference arises from the different space environments around the orbits. Heliocentric GW detectors are fully immersed in the solar wind, whereas geocentric GW detectors are primarily within Earth's magnetosphere. Roughly, the solar wind is dominated by plasma dynamic pressure, while the Earth's magnetosphere is dominated by magnetic pressure \citep{Parker1958PoF}. Consequently, the magnetic field in the solar wind is weaker than that in the magnetosphere \citep{Toffoletto2003}. It results in TQ having a more stringent parameters ($\chi_{\rm m}$--$\xi_{\rm m}$) design requirements compared to LISA.
Additionally, the space magnetic acceleration noise analysis for geocentric GW detectors is based on empirical and MHD models, and further validation through observation-based studies is necessary.

\section{Conclusion}\label{sec13}

% Conclusions may be used to restate your hypothesis or research question, restate your major findings, explain the relevance and the added value of your work, highlight any limitations of your study, describe future directions for research and recommendations. 

% In some disciplines use of Discussion or 'Conclusion' is interchangeable. It is not mandatory to use both. Please refer to Journal-level guidance for any specific requirements.

Laser propagation effects in the space plasma and acceleration noise due to space magnetic fields are critical challenges in space-based GW detections. The optical path noise induced by laser propagation in space plasma is only about one order of magnitude below the displacement requirement of GW detectors, and it can fully consume the noise budget during strong solar eruptions. The dual-frequency scheme can deduct the LPN. 
And TDI can suppress the LPN in low frequency band ($\lesssim 10$ mHz) effectively without adding a new laser beam with different frequency.
Acceleration noise due to space magnetic fields is about 1–2 orders of magnitude below the acceleration requirement for GW detections. Note that the result depends on specific satellite parameters such as $\chi_{\rm m}$ and $\xi_{\rm m}$, the design of the TM's parameters space can be constrainted based on space magnetic field acceleration noise analysis.
Both LPN and acceleration noise studies are based on models and assumptions, such as the Kolmogorov spectrum of space electron number density $n_{\rm e}$, Tsyganenko models, and MHD simulations of SWMF. Due to model limitations, some results, such as the occurrence rate of LPN over solar cycles, remain estimation. Current findings require further validation through space satellites observations, and more detailed studies on LPN and acceleration noise are needed.

\backmatter

% \bmhead{Supplementary information}

% If your article has accompanying supplementary file/s please state so here. 

% Authors reporting data from electrophoretic gels and blots should supply the full unprocessed scans for key as part of their Supplementary information. This may be requested by the editorial team/s if it is missing.

% Please refer to Journal-level guidance for any specific requirements.

\bmhead{Acknowledgements}

S.W. is supported by the National Key R\&D Program of China (No. 2020YFC2201200), NSFC (grant No. 12473060 and 122261131504).

% Acknowledgements are not compulsory. Where included they should be brief. Grant or contribution numbers may be acknowledged.

% Please refer to Journal-level guidance for any specific requirements.

\section*{Declarations}

\bmhead{Conflict of interest}

The author declares no conflict of interest.

% \begin{appendices}

% \section{Section title of first appendix}\label{secA1}

% An appendix contains supplementary information that is not an essential part of the text itself but which may be helpful in providing a more comprehensive understanding of the research problem or it is information that is too cumbersome to be included in the body of the paper.




% \end{appendices}


\bibliography{sn-bibliography}% common bib file

\end{document}